\pdfoutput=1

\documentclass[epj,nopacs]{svjour}

\usepackage[T1]{fontenc}                                  
\usepackage{lmodern}                                  
\usepackage{calc}                                 
\usepackage{subfig}
\usepackage{graphicx}                           
\usepackage{booktabs}                 
\usepackage{textcomp}                                 
\usepackage{xspace}                                   
\usepackage{relsize}                                   
\usepackage{amssymb}                               
\usepackage{amsmath}                                  
\usepackage{listings}                                 
\usepackage{microtype}                                  
\usepackage{multirow}                                   
\usepackage{tabularx}                                       
\usepackage{array}                                         
\usepackage{placeins}                                    
\usepackage{cuted}                                              
\usepackage{fixltx2e} 
\usepackage[latin1]{inputenc}
\usepackage{graphics}
\usepackage{longtable}
\usepackage{color}
\usepackage{pslatex}
\usepackage{hyperref}
\hypersetup{colorlinks,linkcolor={blue},citecolor={blue},urlcolor={cyan}} 
\usepackage{cite}
\usepackage[normalem]{ulem}

\usepackage[absolute,overlay]{textpos}

\begin{document}
%%%%%%%%%%%%%%%%%%%%%%%%%%%%%%%%%%%%%%%%%%%%%%%%%%%%%%%%

\title{Evaluation of {\tt Photos} Monte Carlo ambiguities in case of four fermion final states}

\author{
A.~Kusina{\thanks{\email{aleksander.kusina@ifj.edu.pl}}}
\and
Z.~Was{\thanks{\email{zbigniew.was@ifj.edu.pl}}}
}
\institute{%
Institute of Nuclear Physics Polish Academy of Sciences, PL-31342 Krakow, Poland
}

%\author*[1]{\fnm{Aleksander} \sur{Kusina}}\email{aleksander.kusina@ifj.edu.pl}
%
%\author[1]{\fnm{Z.} \sur{Was}}\email{zbigniew.was@ifj.edu.pl}
%
%
%\affil*[1]{\orgname{Institute of Nuclear Physics Polish Academy of Sciences}, \orgaddress{\street{Street}, \city{Krakow}, \postcode{PL-31342}, \country{Poland}}}

\date{Received: date / Revised version: date}

\abstract{
With the increasing precision requirements and growing spectrum of applications
of Monte Carlo simulations the evaluation of different components of such
simulations and their systematic ambiguities become of utmost interest.
In the following, we will address the question of systematic errors for {\tt Photos}
Monte Carlo for simulation of bremsstrahlung corrections in final states,
which can not, in principle, be identified as a decay of resonances.
It is possible, because the program
features explicit and exact parametrization of phase space for multi-body plus
multi-photon final states.
The {\tt Photos} emission kernel for some processes consist of complete matrix
element, in the remaining cases appropriate approximation is used.
Comparisons with results of simulations, from generators based on exact phase
space and exact fixed order matrix elements, can be used.
For the purpose of such validations {\tt Photos} provides an option to restrict
emissions to single photon only.
In the current work we concentrate on final state bremsstrahlung in 
$q \bar{q}(e^+e^-) \to l^+l^- l^+l^- \gamma$ and $\gamma\gamma \to l^+l^- \gamma$
processes. The reference distributions used as a cross-check are obtained from
the fixed-order {\tt MadGraph} Monte Carlo simulations.
For the purpose of validation we concentrate on those phase space regions where
{\tt Photos} is not expected to work on the basis of its design alone.
These phase space regions of hard, non-collinear photons,
do not contribute to large logarithmic terms.
We find that in these phase space regions the differences between {\tt Photos}
and {\tt MadGraph} results do not surpass a few percent and these regions,
in turn, contribute about 10\% to the observed process rates.
This is encouraging in view of the possible ambiguities for precise calculation
of realistic observables.
}
%
%\PACS{
%{12.38.Bx}{Perturbative calculations} \and
%{12.38.-t}{Quantum chromodynamics}  \and
%{24.10.Lx}{Monte Carlo simulations}
%{12.20.-m}{Quantum electrodynamics}  \and
%}
%\keywords{keyword1, Keyword2, Keyword3, Keyword4}

%
\begin{textblock*}{55mm}(15.5cm,2.2cm)
{\small {\bf IFJPAN-IV-2022-11}}
\end{textblock*}

\maketitle
\setcounter{tocdepth}{3}
\tableofcontents

%==================================
\section{Introduction}
\label{sec:intro}
%==================================
Phenomenology of High Energy Physics experiments require careful comparison of
experimental results with theoretical predictions. An agreement between the two
represents confirmation of the current theory and calculational schemes,
a discrepancy on the other hand can be an indicator
of New Physics phenomena or point to inadequacy of the
applied approximations. For that purpose, all elements of theoretical
predictions and detector responses, need to be reviewed whenever new
threshold of sophistication is reached. Each element of the predictions
as well as the strategy of combining the individual parts need to be validated anew.

It is generally believed that Monte Carlo simulations offer feasible solution
whenever all theoretical and experimental effects need to be taken into account
simultaneously. Electroweak effects can be defined as separate part
of such simulation systems. Recently, we have evaluated numerically if such
separation scheme developed in LEP times still holds in phenomenology of Dell-Yan
processes for present day applications~\cite{Richter-Was:2020jlt}.
See there for further references, in particular, on the fundamental result enabling
the separation of the electroweak effects into separate parts.
Here we concentrate on
QED final state radiations in processes of four-lepton final states produced
in high energy $e^+e^-$ or $pp$ collisions as well as $\gamma\gamma \to l^+l^-$
hard processes.
In such cases precision requirements are lower, nonetheless recently
an interest for such predictions arise, see e.g.~\cite{Gutschow:2020cug}.
The separation of the QED effects from the complete electroweak effects does not
seem to be the precision obstacle, nor dividing the QED contributions into parts;
one of them the final state radiation.

In the present paper, we address the adequacy of {\tt Photos} Monte Carlo
\cite{Barberio:1994qi,Davidson:2010ew} for final state photon radiation in 
$e^+e^- (q \bar q) \to l^+l^- l^+l^- (n\gamma)$ and $\gamma\gamma \to l^+l^- (n\gamma)$ processes.
These processes are outside of the default {\tt Photos} applicability domain.
However, predictions for them can be obtained from different
Standard Model calculations\cite{Alwall:2014hca,Campbell:2011bn,Campbell:2019dru,Melia:2011tj,Alioli:2010xd,Cascioli:2013gfa}. Therefore, for the validation of {\tt Photos} in this region
one does not need to rely on comparisons with the data.
At the same time {\tt Photos} is very valuable as it can be combined
with calculations/simulations enabling higher order corrections and flexible acceptance
cuts providing complete calculations assuring
control of technical aspects, resummations, technical cuts used to separate
phase space regions of singularities.
This can be done because of its design and algorithm modifying events stored already
in event records.
Fortunately the lowest order processes and processes with added photons,
are implemented e.g. in {\tt MadGraph}~\cite{Alwall:2014hca} and can be used for
phase space regions where predictions of {\tt Photos} require validation.

Crude level {\tt Photos} Monte Carlo algorithm is non-Markovian.
The event to which photons may be added, is first read from the event record
produced by other program,
then the momenta coordinates are calculated back, using specifically chosen
parametrization. These variables are used for parametrization
of phase space slots were additional photons are added.
The number of photon candidates is generated from Poissonian
(or binomial) distribution. For each photon, to be constructed,
the variables are generated. They are used to complete
phase space parametrization for the new configuration with additional photons.
At this step, no energy momentum
conservation is enforced, Jacobians of phase space parametrization  are absent
as well as emission kernels representing approximated or interpolated matrix
elements. They are introduced later with weights and rejection of photon candidates through
iterative algorithm.%
    \footnote{This sometimes leads to confusion and misinterpretation
      of the algorithm design as Markovian of shower type which it is not.}
If construction of the photon
is rejected, previous kinematic configuration  is retained.
For rejection  we rely on Kinoshita-Lee-Nauenberg theorem~\cite{Lee:1964is,Kinoshita:1962ur}.
Only numerically minor, process dependent,  corrections need 
further care if complete first order QED matrix element is used. 
The details of event generation in {\tt Photos} are explained
in \cite{Golonka:2005pn,Davidson:2010ew}, phase space
parametrization is possibly best described in \cite{Nanava:2006vv}.
What is important is that the algorithm covers the full multi-photon phase space
and parametrization is exact, whenever it is necessary.
This means that approximations for the phase space always match those of matrix element.%
    \footnote{When testing algorithm for $Z$ decays, using $Z\to l^+l^- n\gamma$
      matrix elements, it was found that phase space approximations related to the
      combination of generation of parallel presamplers and implementation of
      interference weight should match those of matrix element~\cite{RichterWas:1994ep}.}

For the sake of universality, since ref.~\cite{Golonka:2005pn}, simplified
kernel with respect to the exact first order matrix element was used for
all processes and multi-photon radiations. However, already then,
interference effects for emissions from all final state charges were
introduced and good agreement with (matrix element based) reference
simulations was achieved. This, distinct from parton shower approach
enabled rigorous introduction of exact matrix elements
which was done in certain two-body decays including: $Z/\gamma^*$~\cite{Golonka:2006tw},
$W$,$\gamma^*$ \cite{Nanava:2010xvz} and $K^*,B^*$ \cite{Nanava:2006vv}.

Now we turn our attention to final state photon radiation in $q\bar{q} \to l^+l^- l^+l^- (n\gamma)$
and $\gamma \gamma \to l^+l^-$ processes and specially to {\tt Photos} ambiguity for these cases.

The remaining part of the paper is organized as follows.
In Sec.~\ref{sec:cut} we provide details on the samples used for tests
with certain more technical details delegated to appendix~\ref{app:MGinput}.
Section~\ref{sec:ZZ} is devoted to validation of {\tt Photos} for
$q\bar{q}(e^+e^-) \to l^+l^- l^+l^- (n\gamma)$ process and Sec.~\ref{sec:gamgam}
for the $\gamma \gamma \to l^+l^-$ process.
Section~\ref{sec:pheno} discusses {\tt Photos} usecases and its
interfacing with other programs.
The obtained results are summarized in Sec.~\ref{sec:summary}.
Additionally, appendix~\ref{sec:mc-tester} collects numerical
results automatically obtained with the help of the {\tt MC-tester}
program~\cite{Davidson:2008ma}.

%==================================
\section{Details of samples used for validation}
\label{sec:cut}
%==================================

In order to perform validation of {\tt Photos} in the new kinematic
domain discussed earlier we use the following approach. For each of
the considered processes (e.g. $q \bar q \to l^+l^- l^+l^-$) we use
{\tt MadGraph} to generate two samples: first for a base process
without any photons in the final state, and second for an analogous
process but with one additional photon in the final state. Then the
first sample is supplemented with additional photon generated using
{\tt Photos}. This is possible because of the special single photon
emission mode of {\tt Photos}.
Typically we use samples (after selection cuts) of around 50000
events with photons.
In the next step distributions calculated
from the two samples ({\tt MadGraph} with additional photon and
{\tt MadGraph} supplemented by photon from {\tt Photos}) are compared.
Histograms of all possible invariant masses which can be build from the
final state momenta are constructed and then compared. That simplifies
the first step of tests. For the actual study of ambiguity for observables
of phenomenological merit the appropriate selection cuts, which are as
close as possible to the realistic ones, should be used.
Only then systematic ambiguity can be obtained with sufficient certainty.

The kinematic selection cuts play here an additional more technical role.
The event simulation with {\tt MadGraph} is best suited for generation
of configurations, where there is a separation between the outgoing particles.
That is why cuts on distance, $\Delta R$, between final state particles,
and minimal energy of the final state photon, $E_\gamma$, are always used.
In this way non-interesting for tests of {\tt Photos}, but difficult for
{\tt MadGraph}, phase space region of infrared and/or collinear singularities
is avoided. Fortunately in those collinear and infrared regions {\tt Photos}
does not need to be re-tested.

In the following we discuss separately the details of samples generated
for the two considered types of processes.

%++++++++++++++++++
\subsection{$q \bar{q} \to l^+l^- l^+l^- (\gamma)$ process}
\label{subsec:cuts4l}
%++++++++++++++++++

%-----------------
\begin{figure*}[h!]
  \begin{center}
    \includegraphics[width=0.4\textwidth]{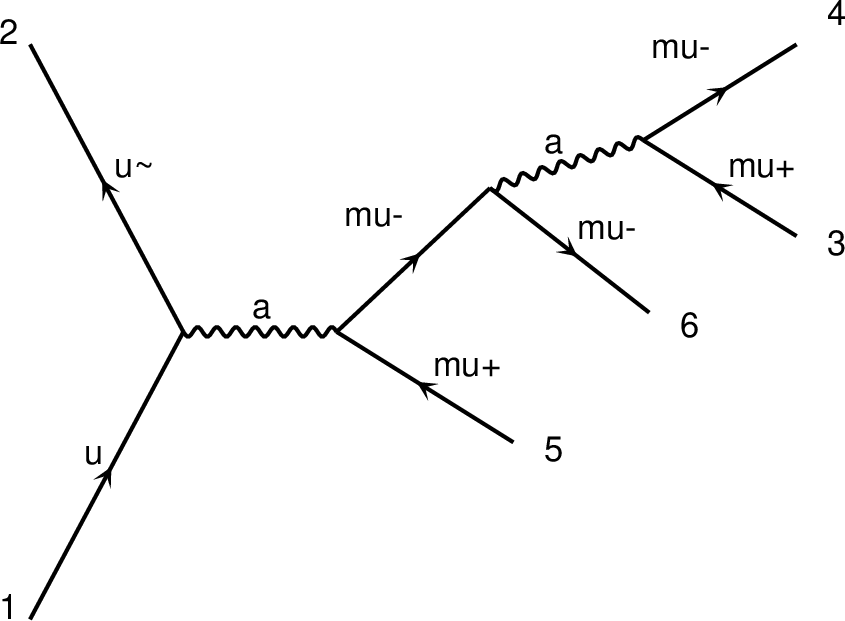} \qquad\qquad
    \includegraphics[width=0.4\textwidth]{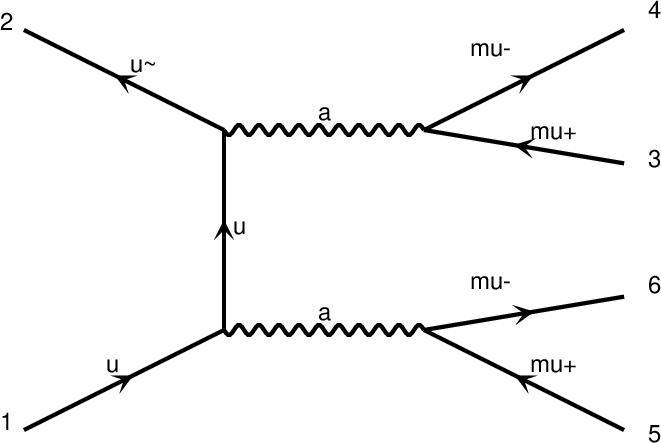} 
    \caption{Typical diagrams contributing to the amplitude of
      $q \bar{q} \to \mu^+\mu^-\mu^+\mu^-$ production.
      Note that $a$ stands here for either photon or $Z$ boson.}
    \label{fig:mad24}
  \end{center}
\end{figure*}
%-----------------

\begin{figure*}[htb!]
  \begin{center}
    \includegraphics[width=0.4\textwidth]{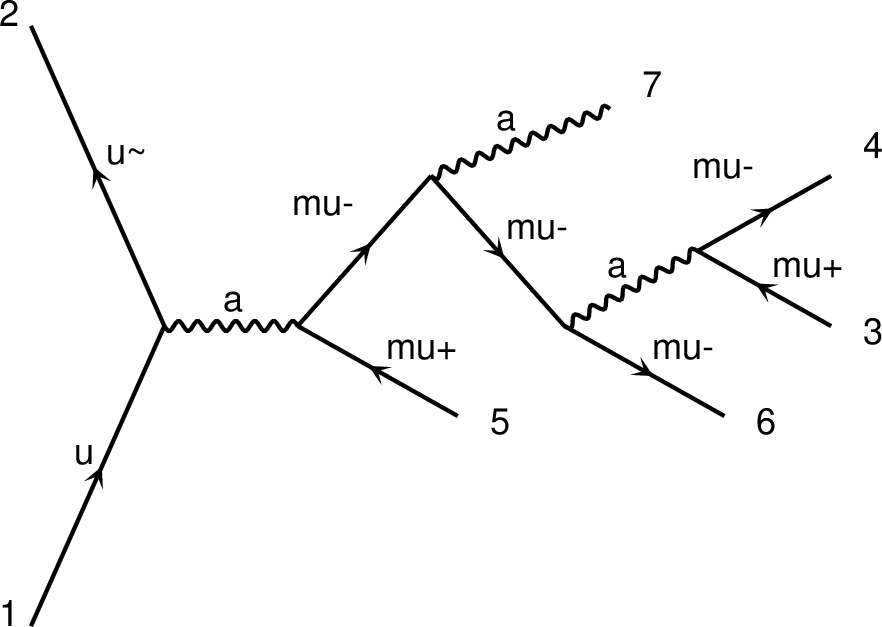} \qquad\qquad
    \includegraphics[width=0.4\textwidth]{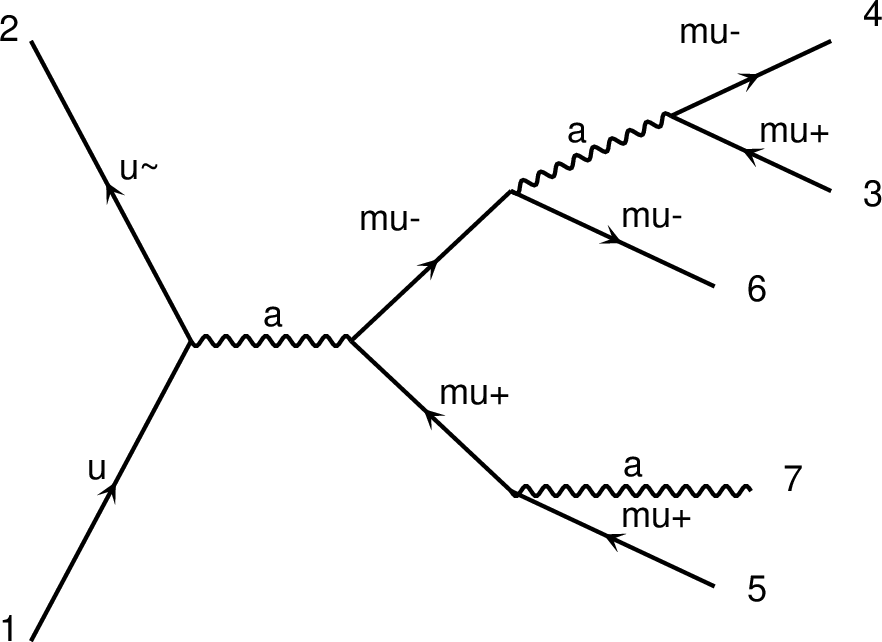} \\
    \includegraphics[width=0.4\textwidth]{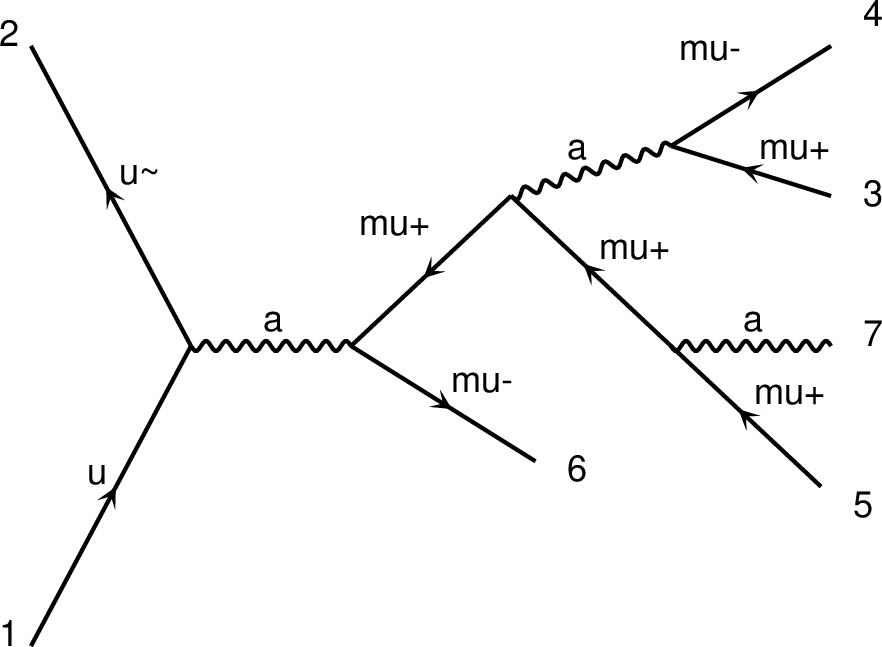} \qquad\qquad
    \includegraphics[width=0.4\textwidth]{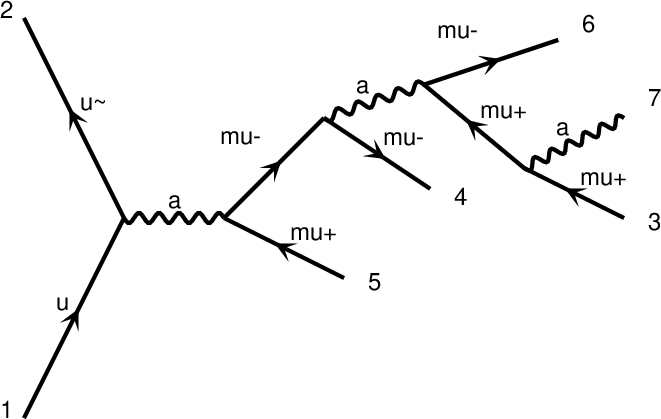} \\
    \includegraphics[width=0.4\textwidth]{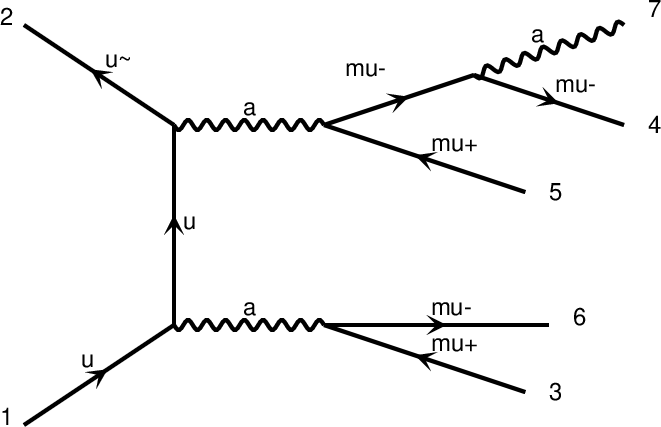}  \qquad\qquad
    \includegraphics[width=0.4\textwidth]{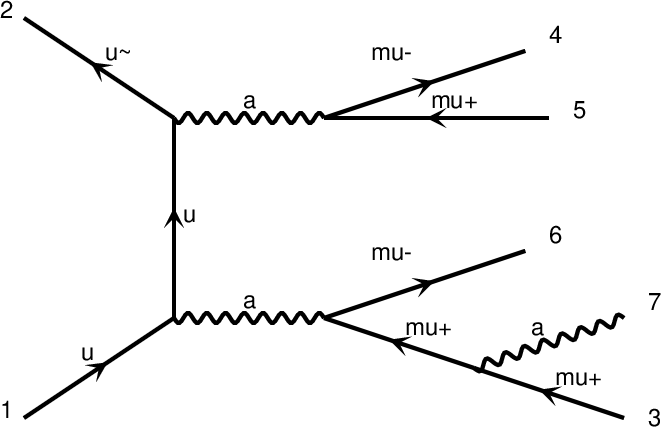} 
    \caption{Typical diagrams contributing to the amplitude of
      $q \bar{q} \to \mu^+\mu^-\mu^+\mu^-\gamma$ production.
      Note that $a$ stands here for either photon or $Z$ boson.}
    \label{fig:mad24g}
  \end{center}
\end{figure*}

\begin{figure*}[h!]
  \begin{center}
    \includegraphics[width=0.4\textwidth]{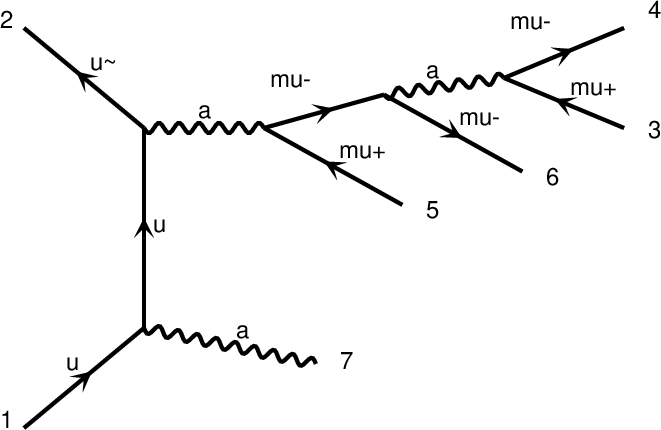} \qquad\qquad
    \includegraphics[width=0.4\textwidth]{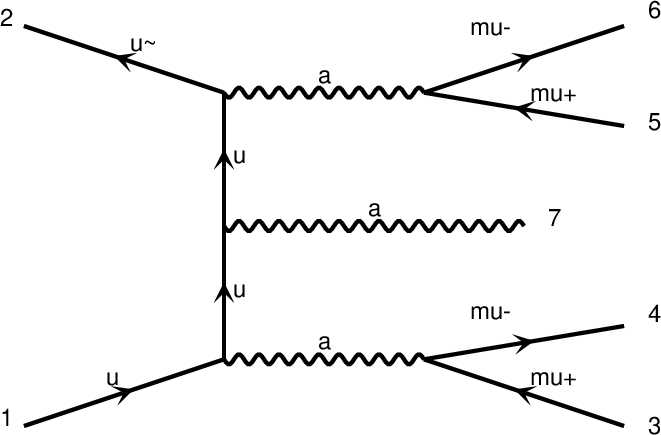} 
    \caption{Typical diagrams with photon radiation in the initial state
      or from the $t$-channel which were removed from {\tt MadGraph}
      calculations of $q \bar{q} \to \mu^+\mu^-\mu^+\mu^-\gamma$ process.
      Note that $a$ stands here for either photon or $Z$ boson.}
    \label{fig:madRMg}
  \end{center}
\end{figure*}

For the $q \bar{q} \to l^+l^- l^+l^- (\gamma)$ process we have generated
three sets of samples for different center of mass energies:
$\sqrt{s}=\{125, 150, 240\}$ GeV (each set consisted of a process with and
without a photon in the final state).
The selection of Feynman diagrams entering such calculations are shown in
Figs.~\ref{fig:mad24} and~\ref{fig:mad24g} correspondingly for process
without and with the final state photon.
Additionally, one should note one more important detail. When generating
{\tt MadGraph} samples for processes with additional final state photon we
wanted to add the photon only in the final state. As a consequence we needed
to remove some of the diagrams from {\tt MadGraph} calculations. Specifically,
we removed diagrams where photon was radiated from the initial state quarks
as well from the $t$-channel intermediate quark, example diagrams of this
kind are displayed in Fig.~\ref{fig:madRMg}. In this way we were left with
a gauge invariant subset of diagrams yielding a sensible result which in
turn could be directly compared with the {\tt MadGraph} sample supplemented
with photon added from {\tt Photos}.%
\footnote{On the technical level this was achieved by invoking a user defined
  {\tt diagram\_filter} function. We would like to thank Richard Ruiz and
  Olivier Mattelaer for help in doing this.}

For all these samples the kinematic
selection cuts used in {\tt MadGraph} were always the same (with additional
isolation cuts on the final state photon if present). We tried to
keep them as wide as possible but taking care not to spoil the convergence
of the calculation. That way we allowed ourselves an option to further
restrict the cuts later on after including additional photon from {\tt Photos}.
This can be useful as more restrictive cuts on the {\tt MadGraph} sample
without the final state photon could, due to the kinematics, restrict the
possibility of generating the additional photon with {\tt Photos}.

%-----------------
%-----------------

%-----------------
%-----------------

The specific values used for kinematic selection cuts are listed below,
we also provide an example {\tt MadGraph} input file (run card)
in App.~\ref{app:MGinput}.
One should note that the same cuts for the photons were applied independently
of whether the photon was generated from {\tt MadGraph} or added later with
the help of {\tt Photos}. Also the electroweak initialization parameters
for the compared two cases were taken the same.
\begin{enumerate}
\item Maximal rapidity of individual charged leptons: $|\eta_l<3|$.
\item Minimal invariant mass of same flavor and opposite charge lepton pairs: $m_{l^+l^-}>9$ GeV.
\item Minimal distance between final state leptons: $\Delta R_{ll}>0.4$.
\item Maximal rapidity of final state photons: $|\eta_\gamma|<3$.
\item Minimal energy of final state photons: $E_\gamma>5$ GeV.
\item Distance between final state photon and each lepton: $\Delta R_{l\gamma}>0.4$.
\item Transverse energy:
  $E_t < \frac{1-\cos(\Delta R_{\gamma l_i})}{1-r_0^{\gamma}} \sqrt{(p_{\gamma}^1)^2+(p_{\gamma}^2)^2}$,
  with $r_0^{\gamma}=0.4$.
\end{enumerate}
The definitions of kinematic variables used for constructing the above
kinematic cuts are provided in App.~\ref{app:MGinput}.

%++++++++++++++++++
\subsection{$\gamma \gamma \to l^+l^- (\gamma)$ process}
\label{subsec:cutsGG2l}
%++++++++++++++++++

\begin{figure*}[htb!]
  \begin{center}
    \includegraphics[width=0.4\textwidth]{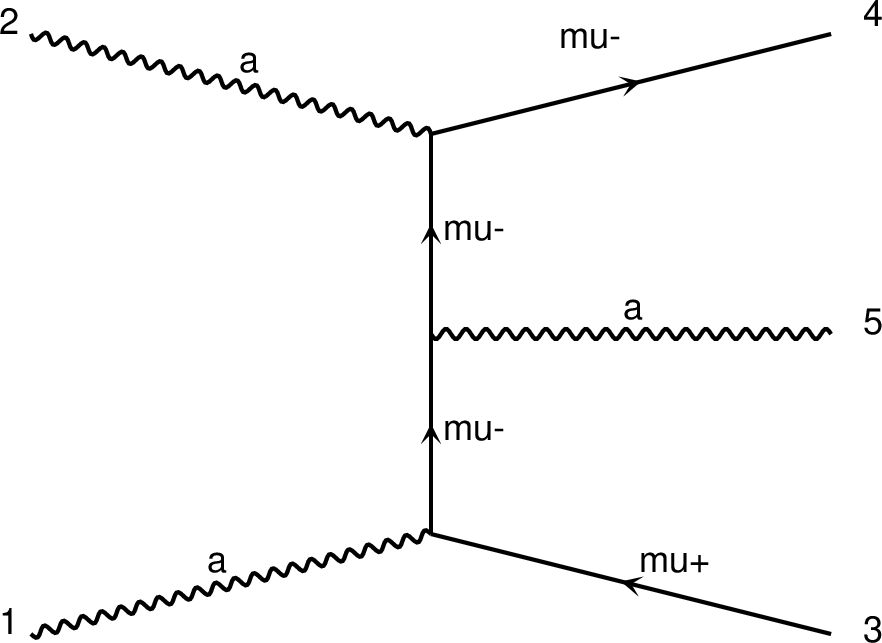} \qquad\qquad
    \includegraphics[width=0.4\textwidth]{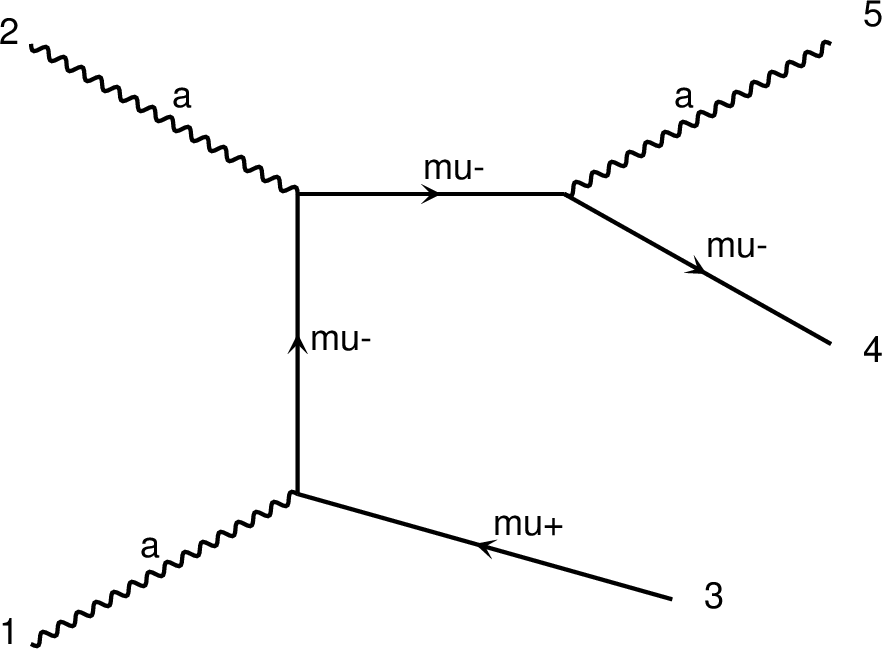}
    \caption{Diagrams contributing to amplitude of $\gamma\gamma \to \mu^+\mu^-\gamma$ production.
    Note that $a$ stands here for photon only.}
    \label{fig:MEgg}
  \end{center}
\end{figure*}

For the $\gamma \gamma \to l^+l^- (\gamma)$ process we have generated
two {\tt MadGraph} samples: one for the process without the photon in
the final state and one with the additional photon, both samples were
generated at center of mass energy of $\sqrt{s}=125$ GeV;
both comprise of 40000 events.
Example Feynman diagrams contributing to the calculation of the
$\gamma \gamma \to l^+l^- \gamma$ process are displayed in Fig.~\ref{fig:MEgg}.

The kinematic cuts used for these processes were the same as for the
process $q \bar{q} \to l^+l^- l^+l^- (\gamma)$ with the simplification
that there was no need to subselect diagrams that enter {\tt MadGraph}
calculation with extra photon. Again the {\tt MadGraph} sample without
final state photon was supplemented with a photon generated using
{\tt Photos} and the two were compared.

%-----------------
%-----------------

%==================================
\section{Results}
\label{sec:results}
%==================================

For the purpose of validation we have chosen two sets of processes,
first the $q \bar{q} \to l^+l^- l^+l^- (\gamma)$ and later the
$\gamma \gamma \to l^+l^- (\gamma)$ which turned out to be simpler.
We should note that the results shown for the
$q \bar{q} \to l^+l^- l^+l^- (\gamma)$ process holds with nearly
no alteration for the process where quarks are replaced by electrons
$e^+e^- \to  l^+l^- l^+l^- (\gamma)$. Such process was also checked
but we will not show explicit results for it here.

In order to simplify the comparisons and also to obtain a better understanding
we have staged the performed tests. This is especially true for the case of
four lepton production process. In what follows we will discuss the different
stages but will present explicit numerical results only for the most important
parts.

We note that options of {\tt Photos} initializations, such as activation
of interference effects, single or multiple photon mode of operation are
carefully explained in the program manual, ref.~\cite{Davidson:2010ew}.

%++++++++++++++++++
\subsection{Case of four lepton final states}
\label{sec:ZZ}
%++++++++++++++++++

The algorithm of {\tt Photos} use explicit phase space parametrization, also
part of the emission kernel originating from matrix element is explicitly
coded. It is thus possible to install exact, single photon emission matrix
element and improve quality of generation. That was implemented for several
two body final states.

For the four fermions final state like  $l^+l^- l^+l^-$,
until now, there were no matrix elements installed into public version of
{\tt Photos} emission kernel. In principle, it is possible and in fact
such kernel, was temporarily installed for emission of lepton pairs in case of
$Z \to l^+l^- (l^+l^-)$ that is for lepton pair emission instead of
bremsstrahlung photons, see
refs.~\cite{Antropov:2017bed,Antropov:2020noh}. In that case four body
phase space was elaborated and exact parametrization was used.
It turned
out, that approximate form of matrix element used for emission kernel,
was sufficient for quite impressive precision. An  approximation
better than the eikonal one of ref.~\cite{Jadach:1993wk} was developed.
Now we investigate if the same level of accuracy can be achieved
for final state bremsstrahlung and processes
like $q \bar{q} \to l^+l^- l^+l^- \gamma$.
One could think of such process as production and
decay of the $Z$-boson pair, but this weakens assumptions, especially for energies
which are insufficient for the on shell two-boson state formation. In case
of four (same flavor) fermion final states,
more than two resonant configuration may  simultaneously contribute and 
complicate  the input for {\tt Photos}. For such processes tests
are required. The case is at the edge of program applicability domain.

When testing {\tt Photos} for the case of process with four lepton
final states we did the following steps:
\begin{enumerate}
\item Consider final states where two pairs of leptons carry different flavors.
\item Require certain number of intermediate $Z$-bosons.
\item Test different initialization options + interferences.
\item Consider different center of mass energies.
\item Investigate the most complex production of four (same flavor) leptons
  including the $\gamma/Z$ interferences.
\end{enumerate}

%%%%%%%%%%%%%%%%%
\subsubsection{$q \bar{q} \to \mu^+\mu^- \tau^+\tau^- (\gamma)$}
%%%%%%%%%%%%%%%%%

We have started our tests with a simple case of 
$q \bar{q} \to Z Z \to  \mu^+\mu^- \tau^+\tau^- (\gamma)$
process. Also the mass of $\tau$ lepton was reduced to the one
of the muon. That way the number of interferences and contributing diagrams was
much smaller than for the four muon production. 
Once the intermediate bosons were explicitly written into the event
record, the agreement between {\tt MadGraph} and {\tt Photos} was perfect.%
    \footnote{Since by default {\tt Photos} assumes nearly flat distributions
      the efficiency of the algorithm drops when encountering resonances. Because
      of this we need to inform {\tt Photos} about the presence of the resonance
      which is done by introducing such intermediate states into the event record
      produced with {\tt MadGraph}.}
Comparisons were prepared with help of the {\tt MC-tester} \cite{Davidson:2008ma}
program. All possible invariant masses which can be formed from momenta
of the outgoing particles were monitored, but not multidimensional
distributions.

For the $\mu^+\mu^- \tau^+\tau^-$ (later also for the 
four muon final states) several options of {\tt Photos} initialization settings
were investigated.
Interference effects in {\tt Photos} between emissions from $\mu^+\mu^-$ and $ \tau^+\tau^-$ pairs
were switched on and off. For that purpose, intermediate
$Z$ bosons (alternatively none or just one)
were explicitly present/absent in the event record before call to {\tt Photos}.
It was found that at very high energies the interference between emission amplitudes
for two $Z$ bosons forming lepton pairs are small. However, at lower energies
when $Z$ bosons are not ultra-relativistic, emission interferences are not so much
reduced by the directions of leptons and the resulting separation of the emission dipoles.
As a result the $Z$ peak constraint has to be taken into account explicitly.

In the first run of tests, the final state fermions were of different flavor
to minimize number of diagrams contributing to {\tt MadGraph} amplitudes.
 At 240 GeV agreement was reasonable, but no $s$-channel photon exchange
was taken into account. Production takes place above $ZZ$ threshold so clearly
two $Z$ resonant contribution dominates and such intermediate states were written
into event record to simplify the task for {\tt Photos} algorithm. Two
independent, two body $Z$ decays were processed. Then, second round of
tests were performed. The only modification was that center of mass energy
was reduced to 150 GeV. Thus one of the $Z$ bosons must have been off the
resonance peak. Still no $s$-channel photon exchange was taken into account.
Further reduction of center of mass energy did not introduce any changes or
concerns.

In our following tests with center-of-mass system energy in 125-240 GeV range
the agreement with {\tt MadGraph} simulations, seem to be optimal when lepton
pair of virtuality closest to the $Z$ mass was written in the event record
as originating from the intermediate $Z$ boson (written explicitly into event
record). If both possible lepton pairs virtualities were more
than 3-4 GeV away from the $Z$ virtuality, the intermediate bosons were not
written into the event record.
Similar  arrangements for small lepton pair virtualities (when 
exchange of virtual photon was taken into account) was not helpful/necessary, possibly because
of the applied cuts, excluding such phase space regions.

Finally we have found, that 
if the interference effects were active in {\tt Photos}  and intermediate bosons were explicitly written
the agreement was reasonable for all numerical results of standard tests
for all  one-dimensional invariant mass distributions.

%%%%%%%%%%%%%%%%%
\subsubsection{$q \bar{q} \to \mu^+\mu^- \mu^+\mu^- (\gamma)$}
%%%%%%%%%%%%%%%%%
Encouraged by the positive results of the tests described above,
we have moved to the next step: the four muon final states.
We have taken into account the $s$-channel photon exchange too.%
    \footnote{Depending on the choice, the Feynman diagrams with
    $s$-channel photon exchange were taken into account, or not,
    both in four lepton and four lepton plus final state photon
    samples, generated by {\tt MadGraph}.}
Typical diagrams contributing to this process are collected in
Sec.~\ref{subsec:cuts4l}

\begin{figure*}[htb!]
  \begin{center}
    \includegraphics[width=0.6\textwidth]{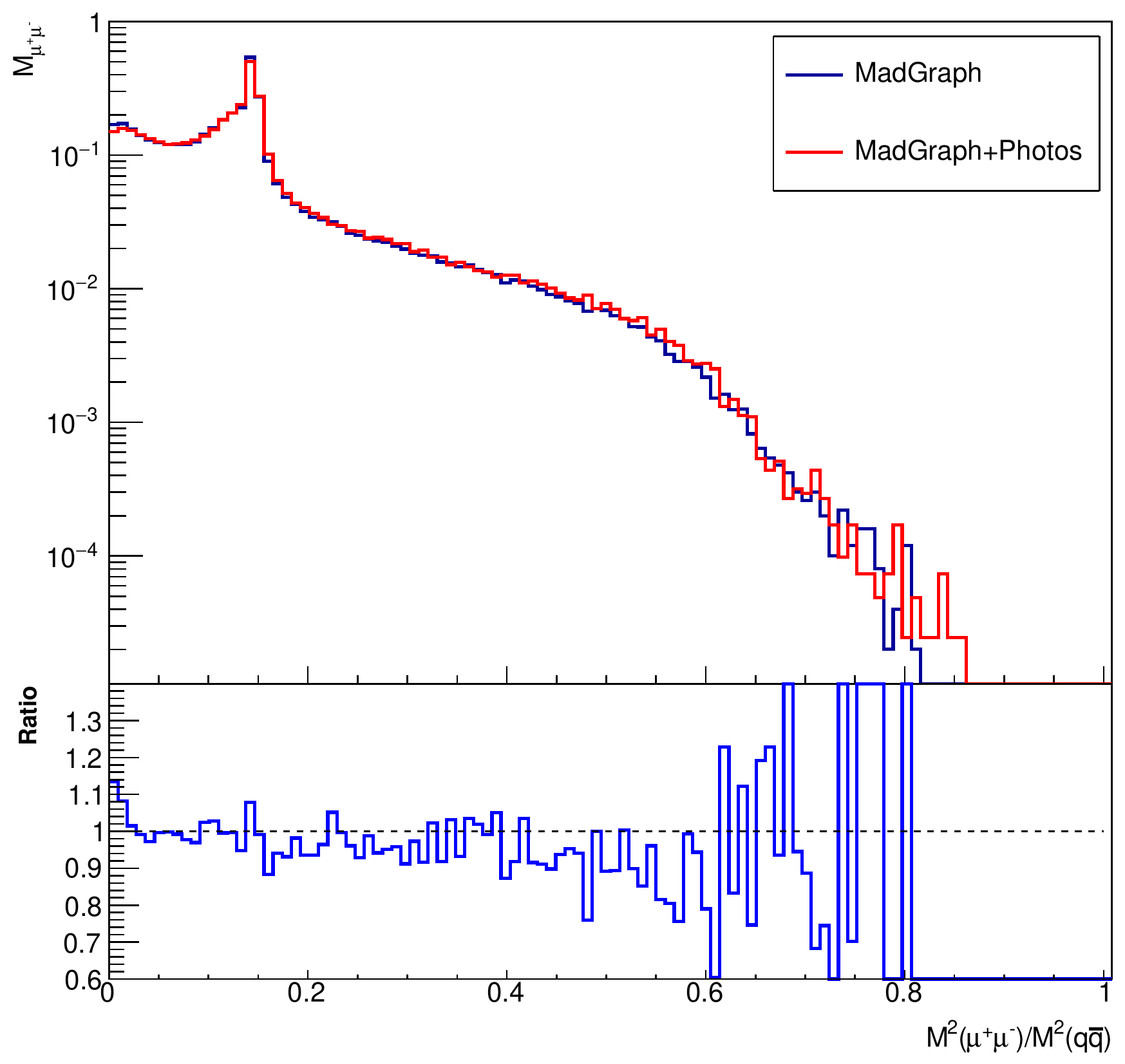}
    \caption{The distribution of $\mu^+\mu^-$ invariant mass
      for $\sqrt{s}=240$ GeV
      with all possible $\mu^+\mu^-$ pairs contributing
      (for each event four entries/combinations are added in the histogram).
      The upper panel shows the absolute distributions from the two generators,
      the lower panel shows the ratio of the two.
      The agreement is much better than what could have been deduced
      from Appendix \ref{sec:mc-tester240}, where technical bias due to
      identical particles partial ordering was present.}
    \label{fig:symetriz}
\end{center}
\end{figure*}

\begin{figure*}[h!]
  \begin{center}
    \includegraphics[width=0.6\textwidth]{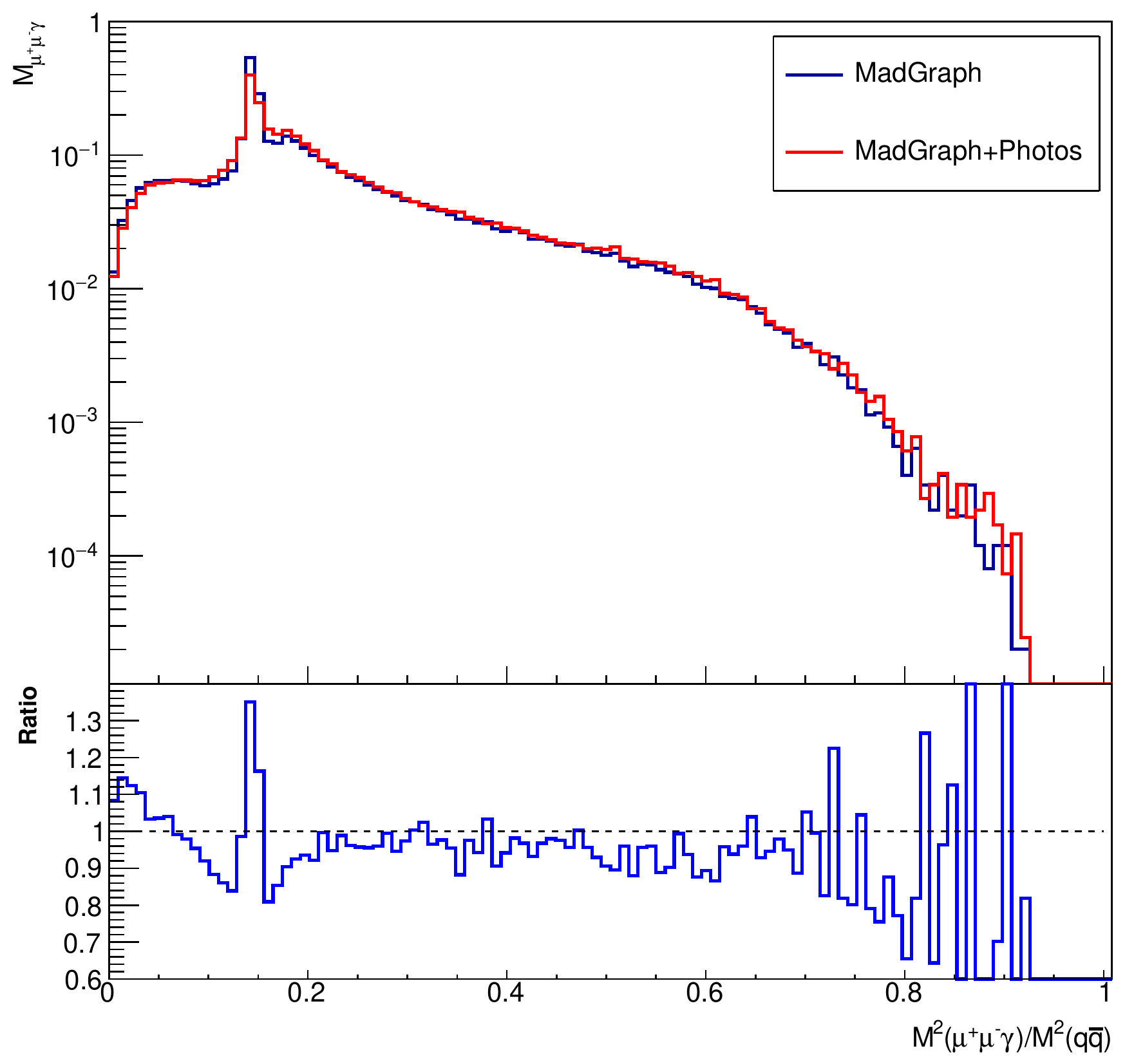}
    \caption{The distribution of $\mu^+\mu^-\gamma$ invariant mass
      for $\sqrt{s}=240$ GeV
      with all possible $\mu^+\mu^-$ pairs contributing
      (for each event four entries/combinations are added in the histogram).
      The upper panel shows the absolute distributions from the two generators,
      the lower panel shows the ratio of the two.
      The agreement is much better than what could have been deduced
      from Appendix \ref{sec:mc-tester240}, where technical bias due to
      identical particles partial ordering was present.
      Note events migration from the $Z$ peak to side bands. This is due to
      approximations in {\tt Photos} kernel, for details see the text.}
    \label{fig:symetriz1}
\end{center}
\end{figure*}

The results of these technical tests are collected in Appendices
\ref{sec:mc-tester125}-\ref{sec:mc-tester240} for center of mass
energies of \{125, 150, 240\} GeV. In these cases the obtained agreement may
look not that impressive, as the differences of the normalized area under
the one dimensional histograms of up to 5\% were found.
However, one should not forget that these histograms were
constructed from events where only hard, non-collinear, photons were taken,
that is from about 10\% of the accepted events.
For 90\% of events, either no bremsstrahlung photons were present, or they
were soft and/or collinear. For such cases {\tt Photos} algorithm performs
accurately thanks to its design, that is why, such regions of the phase space
were not included in the test distributions as they do not contribute
to ambiguities.
Naively this would point to a tiny ambiguity of the order $\sim0.005$.
In reality the differences may be even smaller, as the use of {\tt Photos}
distorts original order of particles and the first and second occurrence of
e.g. $\mu^+$ may feature different distribution.%
    \footnote{For the case of $\mu^+\mu^+\; \mu^- \mu^-$ final state
    pair of intermediate $Z$ bosons contribute. There are two interfering
    diagrams depending on how identical muon pairs are attributed to bosons.
    This has consequences not only for {\tt Photos} interface and its
    ambiguities but also for the technicality of the tests. In principle,
    we could have symmetrize randomly position of same charge muons before
    our testing program {\tt MC-tester}~\cite{Davidson:2008ma} is invoked.
    {\tt MC-tester} is sensitive to how identical particles are ordered
    in the event record. This artefact of our tests slightly increase the
    differences seen in the test results. However, since results are anyway
    sufficiently good, we have not investigated at this time, how large is
    the resulting increase of the ambiguity.}
This may be observed from
our Fig.~\ref{fig:symetriz} for $\mu^+\mu^-$ invariant mass
and Fig.~\ref{fig:symetriz1} for $\mu^+\mu^-\gamma$.%
    \footnote{The $y$-axis of Fig.~\ref{fig:symetriz} (and later plots)
    shows the number of events normalized in such a way that the area
    under the curve is equal to 1.}

%-----------------
%-----------------

%-----------------
%-----------------

Now, the $s$-channel photon exchange is taken into account. Also, the four muons
in the final state imply, that the $Z$ resonance can be formed from four combinations
of leptons. The pattern of  interferences and intermediate resonances, may be more
challenging for the {\tt Photos} algorithm to include.

For the highest center of mass energy of 240 GeV the virtuality of the system
is above the threshold for the pair of $Z$ boson production. As a consequence,
in order to achieve agreement, we have to call {\tt Photos} for the $Z$ decays
separately. This prevents the algorithm to ignore the $Z$ peak constraint on
distributions. Additionally, we have to order momenta, prior to invoking
{\tt Photos}, so the first pair of muons would be closest to the $Z$ peak.
If the pair could be within $\pm 3$ GeV from the $Z$ peak, we write the $Z$
into the event record prior to invoking {\tt Photos}. As a consequence its algorithm
will not deform the shape of the intermediate resonance peak, but at the same
time interferences of emissions from all four muons will be reduced
to interferences within the two separate $\mu^+\mu^-$ pairs. This seems not to
be physical if none of the muon pairs is originating from the resonance peak.
Such interface to {\tt Photos} clearly leads to temporary ordering in the events
with identical muons. This ordering could make the interpretation of the comparisons
produced with {\tt MC-tester} difficult.
That is why before it is invoked, the order of the muons in the event record is
returned to its prior state before the call to the {\tt Photos} interface.
This has to be done to adopt to the
{\tt MC-tester} which recognizes the order of identical particles in the event record.
The residual traces of the above arrangement are visible in the plots from appendices,
but less so in Figs.~\ref{fig:symetriz} and~\ref{fig:symetriz1}
where the sum of the contributions from all $\mu^+\mu^-$ pairs is taken.
The separation of phase space (with $\pm 3$ GeV cut) into off- and on- $Z$ peak regions
helps but still migration of events from the $Z$ peak to its side bands takes place.
This exhibits as a peak in the ratio plot of Fig.~\ref{fig:symetriz1}.
{\tt Photos} algorithm can not simultaneously assure interferences of all
Fig.~\ref{fig:mad24g} diagrams and $Z$ peak phase space constraint.
For that, complete
4-fermion and photon matrix element in emission kernel would be needed.

In order to answer the question how this translates into the ambiguities
of observables used for the precision measurements requires use of observables
that are (semi-)realistic from the detection point of view.
To do it the prepared event samples (which were used for producing the presented plots), can be used.

Finally one may wonder about multi-photon effects. These can be implemented
with {\tt Photos} too and no further ambiguity should be expected, because the same emission kernel is then used.

%%%%%%%%%%%%%%%%%
\subsubsection{$q \bar{q} \to \tau^+\tau^- \tau^+\tau^- (\gamma)$ at $\sqrt{s}=10.5$ GeV}
%%%%%%%%%%%%%%%%%

Finally, we have performed comparison between {\tt MadGraph} and
{\tt Photos} final state bremsstrahlung for the process
$q \bar{q} \to \tau^+\tau^- \tau^+\tau^-$ which will work in exactly
the same way for the $e^+e^- \to \tau^+\tau^+\tau^-\tau^-$ process
but is easier to initialize in {\tt MadGraph}.
In this case, $\tau$ leptons are moderately relativistic and terms
proportional to $\tau$ mass are of importance, on the other hand
contribution of the $Z$ boson interaction is negligible. The study
of {\tt Photos} performance in such very different regime can be
instructive. It may be of potential interest for Belle II
phenomenology~\cite{Belle-II:2022cgf}, that is why we have chosen
the center of mass energy to be 10.5 GeV.

\begin{figure*}[htb!]
  \begin{center}
    \includegraphics[width=0.47\textwidth]{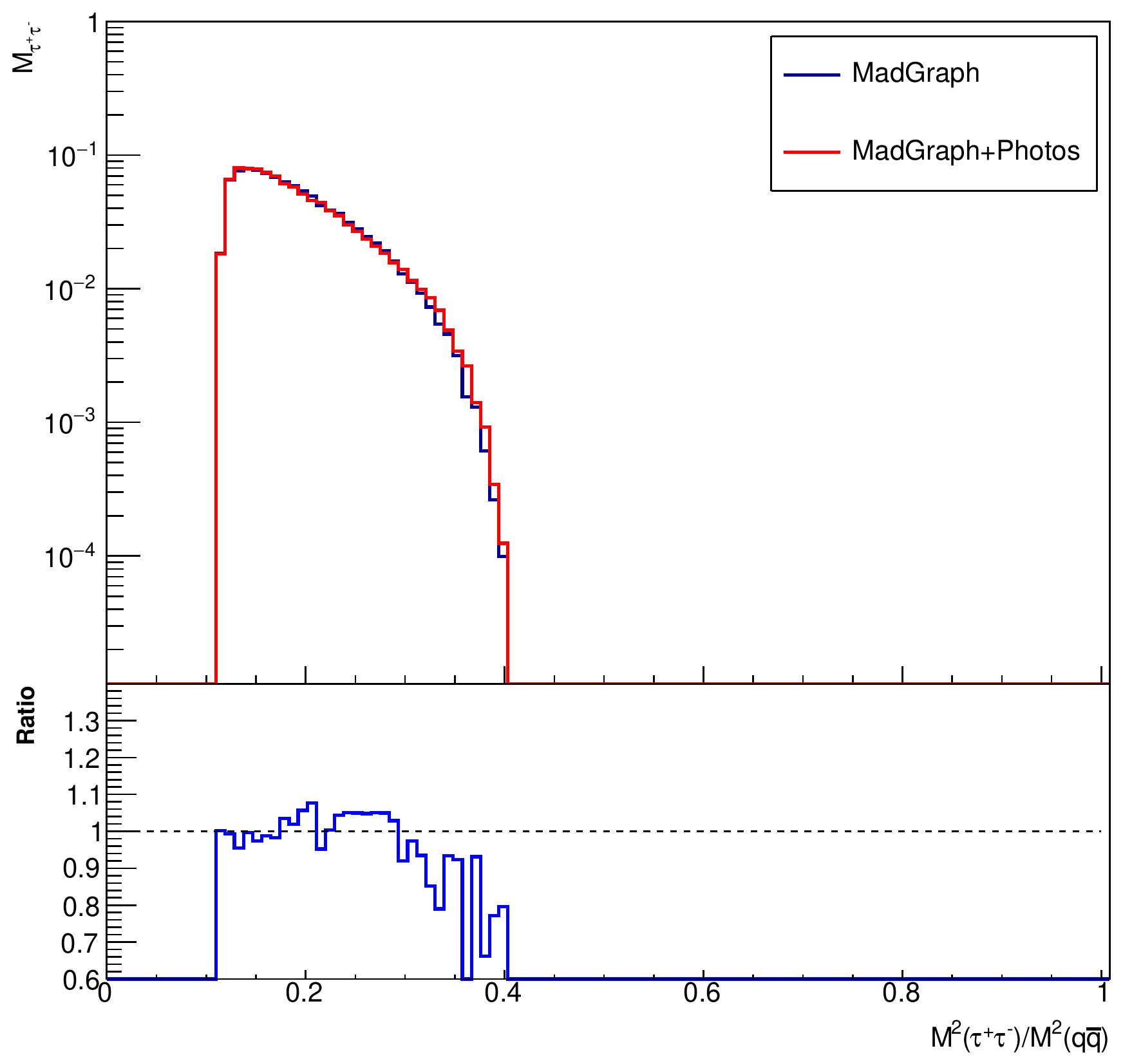}\quad
    \includegraphics[width=0.47\textwidth]{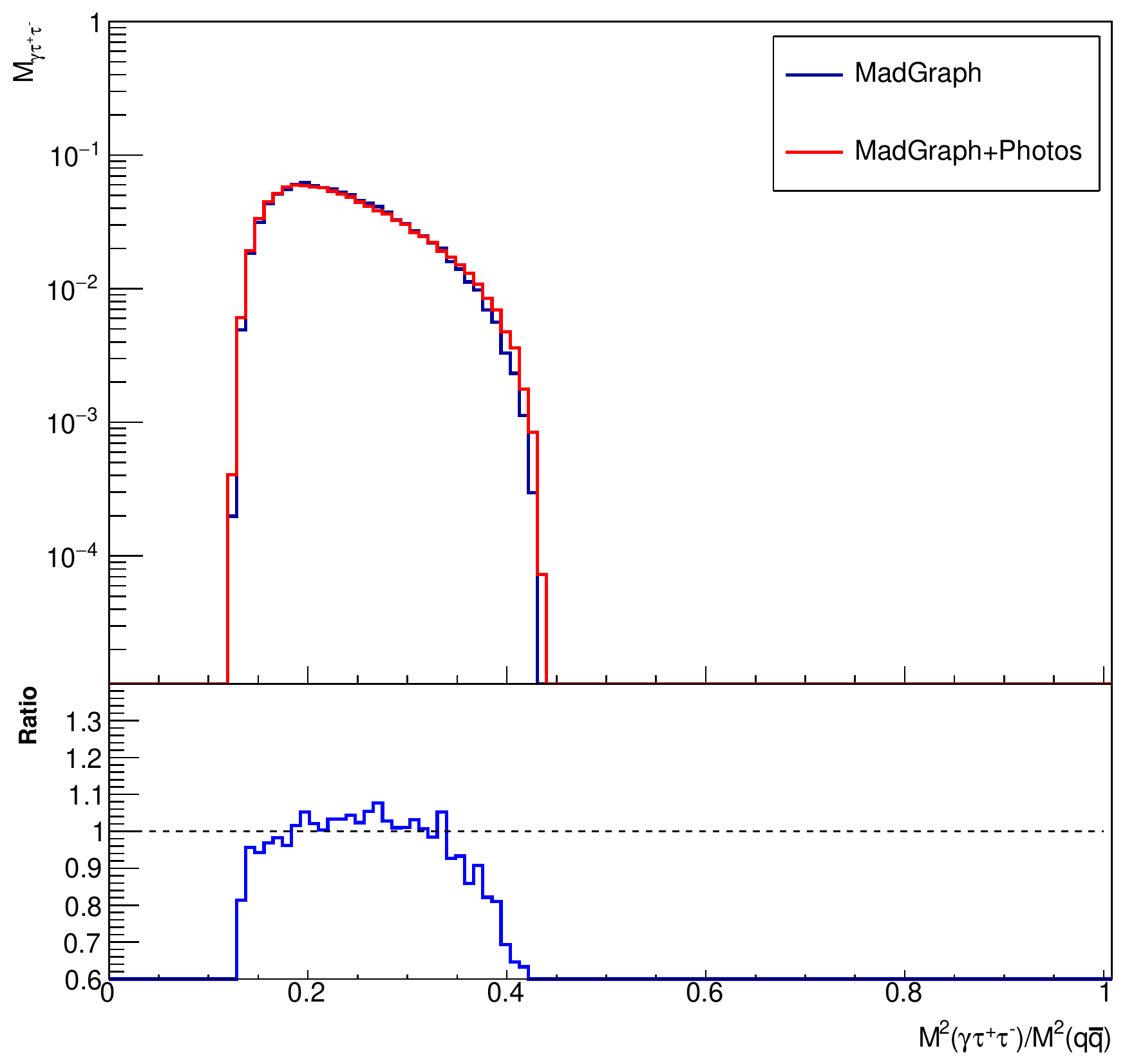}
    \caption{The distributions of invariant mass of $\tau^+ \tau^-$ (left panel)
      and $\tau^+ \tau^-\gamma $ (right panel) for the
      $q\bar{q} \to \tau^+\tau^- \tau^+\tau^- \gamma$ process
      at $\sqrt{s}=10.5$ GeV with $E_{\gamma}>0.2$ GeV.
      The upper panel shows the absolute distributions from the two samples
      generated using {\tt MadGraph} and {\tt MadGraph} + {\tt Photos};      
      the lower panel shows the ratio of the two.}
    \label{fig:Belle}
\end{center}
\end{figure*}

With the default version of {\tt Photos} the agreement at the level
of a factor of two was obtained. Clearly the mass terms of the 5-body
system for $\tau$ leptons and photon final state are needed.
It was relatively easy to introduce the necessary adjustment with the
additional factor $(1 -4m_\tau/\sqrt{s} -E_\gamma/\sqrt{s})^2$
for the internal {\tt Photos}  weight, resulting in softening of the
photon energy spectrum. Such an ad~hoc factor, can not replace a study
of the matrix element but it can be assumed as an educated guess that
represents its dominant, missing in this case, part. In fact, that is
supported by the results of Fig.~\ref{fig:Belle}
which shows a very good agreement with the {\tt MadGraph}
calculation using exact matrix elements.
Even though the obtained
numerical results are encouraging such hypothesis require more rigorous
study. In case of plots from Fig.~\ref{fig:Belle} no kinematic cuts,
except minimal photon energy of 0.2 GeV, were applied.
In Fig.~\ref{fig:Belle} we can see that,
compared to the analogous distributions for the four muon case at higher
energies (e.g. in Fig.~\ref{fig:symetriz1} and figures from
Apps.~\ref{sec:mc-tester125}-\ref{sec:mc-tester240}),
the mass of $\tau$ lepton provides a clear cut-off at low $M^2(\gamma\tau\tau)$.
Also we no longer see
any remnants of the $Z$-peak.
%-----------------
%-----------------

%++++++++++++++++++
\subsection{Case of $\gamma\gamma \to l^+l^- (\gamma)$}
\label{sec:gamgam}
%++++++++++++++++++

Let us turn our attention to $\gamma\gamma \to e^+e^- (\gamma)$
that is a building block for some $pp$ simulation programs.
From technical side this case is simpler because incoming partons are not charged,
and if one works in  $\gamma\gamma$ collision frame, the $\gamma\gamma \to e^+e^-$ angular distribution
does not peak. 
 This may not be the case once strong
boost to lab frame takes place.

The consequence of boost requires attention, especially for the
cases when there are initial-state high $p_T$ jets present.
That may require detailed explanation and broader windows of
generations of $\gamma\gamma \to l^+l^-$ samples than acceptance,
as well as checks of the impact from the cuts in combinations with
boosts. Note that {\tt Photos} will ``kick'' some of the events
into acceptance region and this need to be checked. It may be
especially important for the multi-photon bremsstrahlung configurations.

\begin{figure*}[h]
  \begin{center}
    \includegraphics[width=0.47\textwidth]{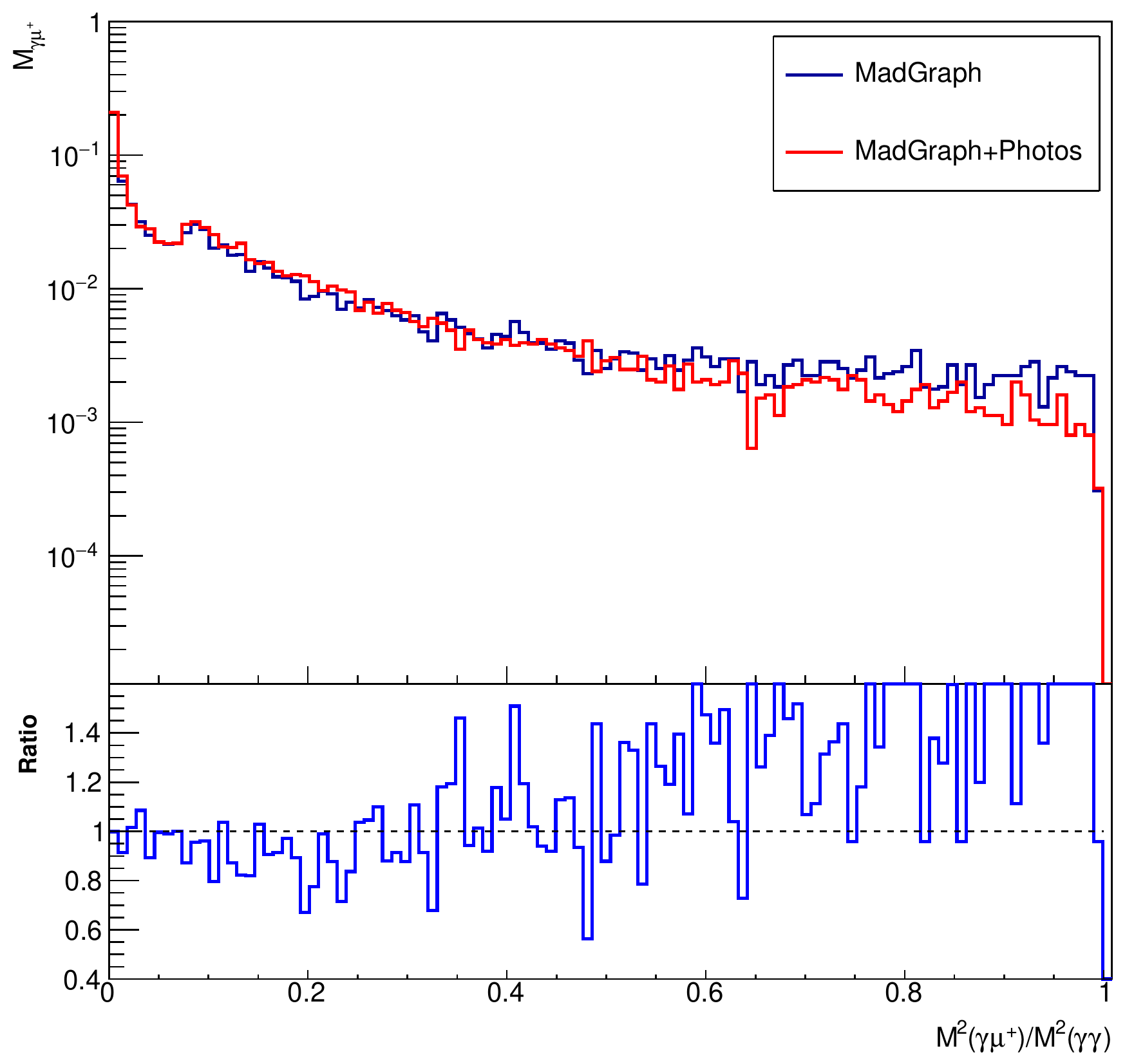}\quad
    \includegraphics[width=0.47\textwidth]{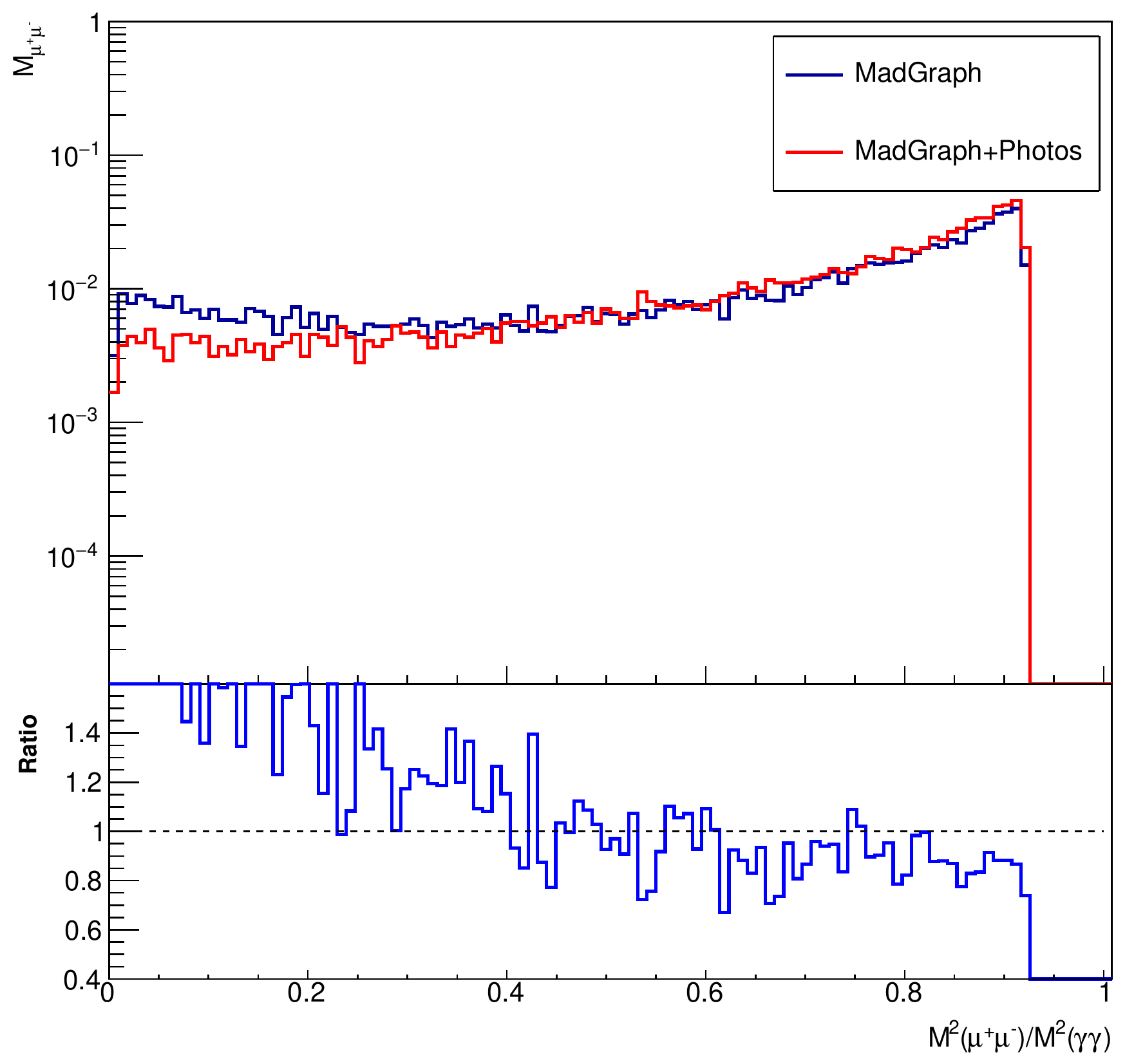}
    \caption{The distribution of invariant mass of $\mu^+ \gamma $ (left panel) and 
      $\mu^-\mu^+ $ (right panel) for the $\gamma \gamma \to \mu^-\mu^+ \gamma$ process.
      Comparison between {\tt Photos} and {\tt MadGraph}. For details
      of selection cuts see the text. Collinear and soft photon phase space regions
      are excluded, but all hard non-collinear photon configurations are taken into
      account. That is the phase space region where {\tt Photos} is not guaranteed
      to work well due to its design. Still, only for hardest emission sub-region
      which is scarcely populated, differences may approach 70\%.
      That should be expected, as the kernel in {\tt Photos} is not improved with
      matrix element for this case.}
    \label{fig:gamgam}
\end{center}
\end{figure*}

On the other hand, nothing like that seems to be challenging from the
perspective of our test runs, see Fig.~\ref{fig:gamgam}. The obtained
distributions are smooth and thanks to the used approximations
{\tt Photos} works well.
One should remember that collinear and soft photon phase space regions
are excluded, but all hard non-collinear photon configurations are taken
into account. That is the phase space region where {\tt Photos} is not
guaranteed to work well due to its design. Still, only for hardest
emission sub-region which is scarcely populated, differences may approach 70\%.
That should be expected, as the kernel in {\tt Photos} is not improved with
matrix element for this case.
The overall agreement between {\tt Photos} and {\tt MadGraph}
is at the level of 0.01$\times$0.05, but at present this is from
simplistic/naive tests only.
The diagrams used for spin amplitudes are rather simple in this case,
see Fig.~\ref{fig:MEgg}.
%-----------------
%-----------------

To enable {\tt Photos} activation we had to introduce intermediate cluster
formed by the incoming $\gamma\gamma$ pair and then decaying to lepton pair.

%\clearpage

%==================================
%\section{Outlook for phenomenology}
\section{Usage for phenomenology}
\label{sec:pheno}
%==================================
In the previous parts of the paper we have explained the interface
of {\tt Photos} to our test environment (including {\tt MadGraph},
{\tt MC-tester} and {\tt Photos} working in a single emission mode).
This might have look complicated and hard to reproduce, but 
 in usual applications, when it is not necessary to manipulate event record content,
 it is straightforward, see
program manual~\cite{Davidson:2010ew}.
As long as the events, for which the bremsstrahlung is to be added, are stored in
HepMC~\cite{Dobbs:2001ck} or Hepevt~\cite{Altarelli:1989hv} format,
we simply feed them to {\tt Photos} and obtain back events of the same
format but possibly with additional photons added. Such events  can be further processed,
e.g. by simulation of  detector
response. That is why  {\tt Photos} can work with any generator
chain, without any burden for the user.

Our paper explains reliability tests of {\tt Photos} for events with
hard non-collinear photons where it may not be accurate. For comparison
with {\tt MadGraph} simulation where single photon configurations are
generated from exact phase space and exact matrix element, also the
{\tt Photos} single photon mode of operation has to be activated.
%The initialization variants are explained in {\tt Photos} documentation.%~\cite{Davidson:2010ew}.
In practice, {\tt Photos} is expected to be used in the multi-photon
mode covering the whole phase space also that of collinear and soft
photons. {\tt Photos} algorithm is closely related to exponentiation,
that is why, e.g. the Yennie-Frautschi-Suura $\beta_1$
part~\cite{yfs:1961,yfsww:1998,kkcpc:1999,Arbuzov:2020coe}
of the second order matrix element is taken into account.

We note that a full scale matching of {\tt Photos} with higher order
matrix element for $q \bar q \to 4l+\gamma$ like processes would require
development of a dedicated matching scheme. Preferably separating
matrix element into parts corresponding to Born times eikonal factor
and the so-called $\beta_1$ of Yennie Frautchi Suura exponentiation.
Unfortunately it require major effort. Separation of the phase space
into soft photon region, for which {\tt Photos} would be used like
in~\cite{Barze:2013fru} and matrix element generator for hard emissions
is also possible, but again would require considerable effort.
For the moment we have demonstrated that using an approximate treatment
provided by {\tt Photos} gives sufficiently good results for many
current applications.

%==================================
\section{Summary}
\label{sec:summary}
%==================================
In the paper, we have addressed the question of how reliable is using
{\tt Photos} Monte Carlo for simulation of final state bremss\-trahlung %bremsstrahlung
in $e^+e^- (q \bar q) \to l^+l^- l^+l^-$ and $\gamma\gamma \to l^+l^- (n\gamma)$
hard processes. {\tt Photos} algorithm is prepared for combination
with complete simulation chain (as used in experimental analyses)
and is expected to be used as an add up solution for bremsstrahlung
in decay of particles or resonances.
The applicability of {\tt Photos} extends beyond processes selected here.
Evaluation of ambiguities is thus of particular importance,
especially if hard and/or non-collinear photon emissions are of interest. For such
configurations comparisons with matrix element tree-level based simulation
was now performed; {\tt Photos} operation was restricted to single
photon emissions only. Then full phase space with explicit parametrization was
used and the differences with respect to reference {\tt MadGraph}
simulations are shown. In this way approximations used in {\tt Photos}
emission kernel (its simplified matrix element) are exposed.
The comparisons relied on automatically defined and collected
1-dimensional distributions of all possible invariant masses which
can be constructed from the outgoing particles.
These full results are collected in the Appendices,
whereas in Sec.~\ref{sec:results} selected results
of particular interest were recalled.
Only the events with hard non-collinear photons were
taken into account because that is the phase space region where
source of ambiguities is expected to reside. For the soft or collinear
photons {\tt Photos} algorithm is expected to work in a general case.

The obtained agreement level is encouraging.
We quantify it as the difference between area under the invariant mass
histograms, and in all considered cases such differences are below 3\%.
Note that monitored events represented up to 10\% of the samples
for the selected processes and cuts. 
In this way, we can indicate size of the ambiguities due to the approximations
used in {\tt Photos}. Similar reliability for further processes also
somewhat outside of {\tt Photos} applicability range can be expected.

Naively one could take the size of the obtained differences ($\sim3\%$)
and the fact that approximations affect only subset of actual events ($\sim10\%$)
to estimate the global effect as $3\%*10\%$ thus giving $0.003$.
For  $\gamma \gamma \to \mu^+ 2\mu^- (\gamma) $ agreement seem to be even better.
Once these tests are completed we can address
effects of multi-photon radiation. This can be taken into account with the help of
bremsstrahlung photon simulation segment of {\tt Photos}. Full
simulation chains is then enriched, without much of intervention into the software
design.
However, since we have not used realistic or even idealized observables
to evaluate the impact of approximation on observables of phenomenological
interest, we can not claim that precision with certainty.
For that, additional work using dedicated event samples and idealized observables
is needed and we are looking forward to such work.

One should stress that
{\tt Photos} was not expected to be used for processes with strong $t$-channel
transfer dependence or for processes where non-bremsstrahlung matrix elements
vary largely over the phase space, e.g. due to intermediate resonances.
This required special attention. E.g. if the intermediate $Z$-boson
resonances were explicitly written into the content of event records,
prior to invoking {\tt Photos}, allowed to reduce the differences,
especially for the cases of $2\mu^+ 2\mu^-$ and $\mu^+ \mu^- l^+ l^-$
production at 240 GeV.

%%%%%%%%%%%%%%%%%%%%%%%%%%%%%%%%%%%%%%%%%%%%%%%%
\section*{Acknowledgments}
%\centerline {\bf Acknowledgments}
This project was supported in part from funds of Polish National Science
Centre under decisions DEC-2017/27/B/ST2/01391.
Encouragement from the Polish-French collaboration
no. 10-138 within IN2P3 through LAPP Annecy and
during the years leading to  completion of this work is
also acknowledged.

%%%%%%%%%%%%%%%%%%%%%%%%%%%%%%%%%%%%%%%%%%%%%%%%
%%%%%%%%%%%%%%%%%%%%%%%%%%%%%%%%%%%%%%%%%%%%%%%%
%\bibliographystyle{unsrt}
\bibliographystyle{utphys}
\bibliography{paper-EWcorr}
%%%%%%%%%%%%%%%%%%%%%%%%%%%%%%%%%%%%%%%%%%%%%%%%

\onecolumn
%%%%%%%%%%%%%%%%%%%%%%%%%%%%%%%%%%%%%%%%%%%%%%%%
\appendix

%\clearpage
%==================================
\section{Details on running {\tt MadGraph} and kinematic cuts}
\label{app:MGinput}
%==================================
Below we list the definitions of kinematic variables used when constructing
kinematic cuts in Sec.~\ref{subsec:cuts4l} for the events in samples generated
from {\tt MadGraph} and later analyzed with addition of photon from {\tt Photos}:
\begin{itemize}
\item Rapidity: $\eta = 0.5\ln\left(\frac{p_t+p_z}{p_t-p_z}\right)$,
  with $p_t(p_z)$ being the zero (third) component of 4-momentum of given particle.
\item Invariant mass of two particles: $m_{ab}=\sqrt{(p_a+p_b)^2}$,
  with $p_a$, $p_b$ the 4-momenta of particle $a$ and $b$.
\item Distance between two particles $a$ and $b$:
  $\Delta R_{ab}=\sqrt{(\eta_a-\eta_b)^2+ \Delta\phi_{ab}^2}$,
  with $\phi$ being the orientation angle (in radians) which is defined for
  the momenta components in the $x$-$y$ plane which is perpendicular to the beam
  direction given by $z$, specifically we have:
  $\Delta\phi_{ab}^2=\arccos\left(\frac{p_{ax}p_{bx}+p_{ay}p_{by}}{\sqrt{p_{ax}^2+p_{ay}^2}\sqrt{p_{bx}^2+p_{by}^2}}\right)$.
\item Transverse energy: $E_t=\sum_{i\in leptons} \sqrt{(p_i^1)^2+(p_i^2)^2}$.
\end{itemize}

We also provide here an example input file (run card) for running {\tt MadGraph}
simulation for the $q \bar{q} \to l^+l^- l^+l^- \gamma$ process.
\begin{verbatim}
import model loop_sm-lepton_masses 
generate p p > mu+ mu- mu+ mu- a QCD=0 QED=5 / h --diagram_filter
output FCC240_qq_4muA_Inc_QED5_noISR_500k
launch FCC240_qq_4muA_Inc_QED5_noISR_500k
analysis=off
set mta 0.10566
set ymtau 0.10566
set nevents 500k
set lpp1 0
set lpp2 0
set ebeam1 120.
set ebeam2 120.
set use_syst false
set no_parton_cut
set cut_decays true
set etal 3 # max rapidity of charged lepton
set etaa 3 # max rapidity of photons
set ea   5 # minimum E for the photons
set mmll 9 # minimal invariant mass of (same flavour) lepton pairs: l+l-
set eta_max_pdg {15:3, 13:3} # max eta for mu (13:) and tau (15:)
set mxx_min_pdg {15:9, 13:9} # min invariant mass of a pair of particles X/X~ (mu, tau)
set drll 0.4 # distance between leptons
set dral 0.4 # distance between gamma and lepton
set r0gamma 0.4
set xn 1
set isoEM true
set epsgamma 1.0
done
\end{verbatim}

%\vspace{-4cm}
%\bigskip

%==================================
\section{Technical comparison plots for $q \bar q \rightarrow \gamma \mu^{+} \mu^{+} \mu^{-} \mu^{-}$ at different energies}
\label{sec:mc-tester}
%==================================
In appendices~\ref{sec:mc-tester125}-\ref{sec:mc-tester240} we present
technical comparison plots for processes with production of four leptons
at energies of 125-240~GeV that were discussed in Sec.~\ref{sec:ZZ}.
The plots display invariant mass distributions of all combinations of
the final state particles produced either with {\tt MadGraph} or with
{\tt MadGraph} plus {\tt Photos}. The plots have two axes the right one
gives the number of events and corresponds to the invariant mass
distributions (plotted in green and red), the left one corresponds to
the ratio of the two distributions (plotted in blue).
All the plots were produced automatically using the {\tt MC-tester}
program~\cite{Davidson:2008ma}. Additionally in yellow box we display
value of ``SDP'' which is the difference of the area under the normalized
histograms of the distributions from the two generators which is used
to judge the differences between the two generators. The details on the
precise definition of ``SDP'' can be found in the program
manual~\cite{Davidson:2010ew}.

%==================================
\subsection{{\tt MC-tester}: $q \bar q \rightarrow \gamma \mu^{+} \mu^{+} \mu^{-} \mu^{-}$ at $\sqrt{s}=125$ GeV}
\label{sec:mc-tester125}
%==================================
{ \resizebox*{0.49\textwidth}{!}{\includegraphics{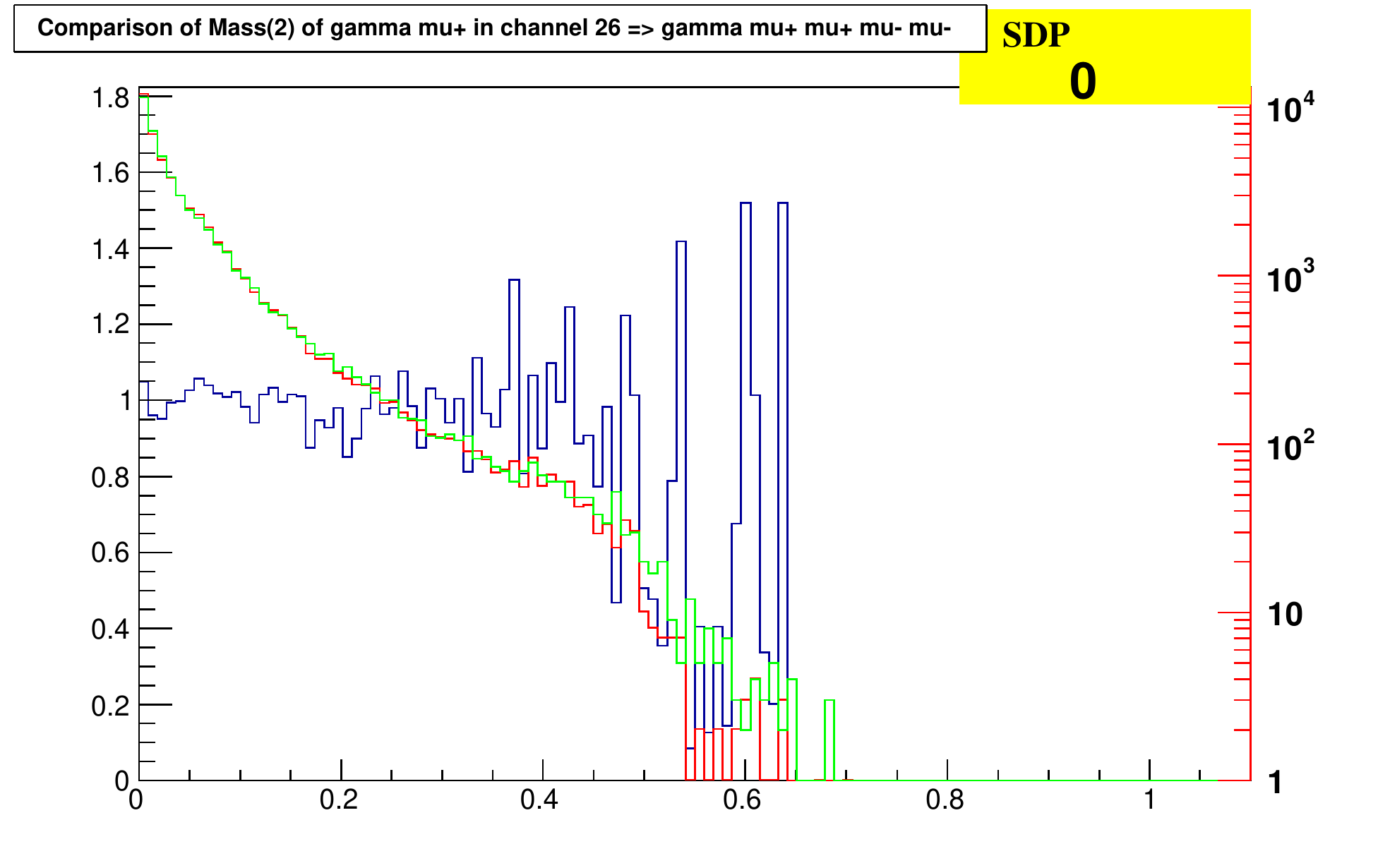}} }
{ \resizebox*{0.49\textwidth}{!}{\includegraphics{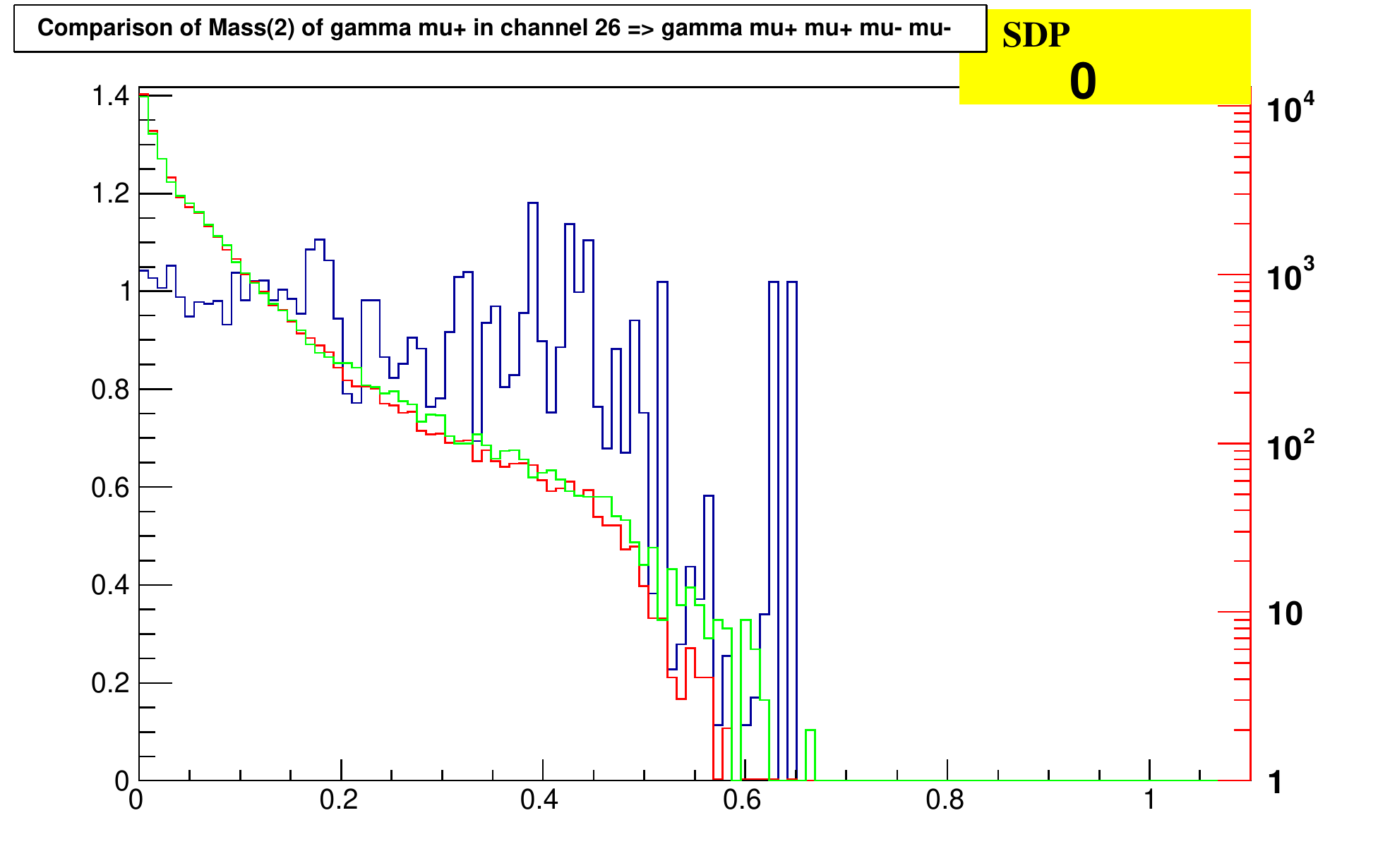}} }
{ \resizebox*{0.49\textwidth}{!}{\includegraphics{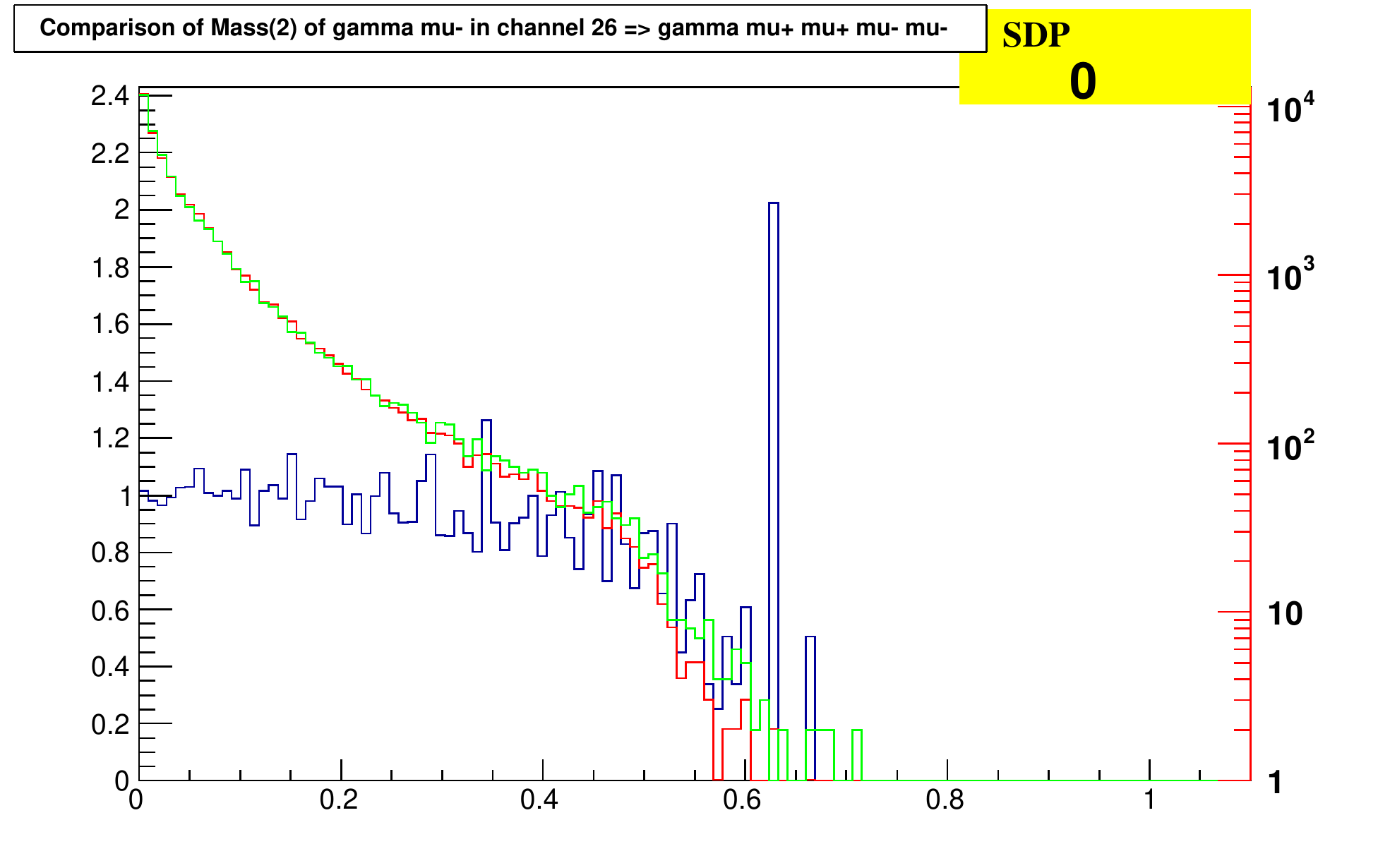}} }
{ \resizebox*{0.49\textwidth}{!}{\includegraphics{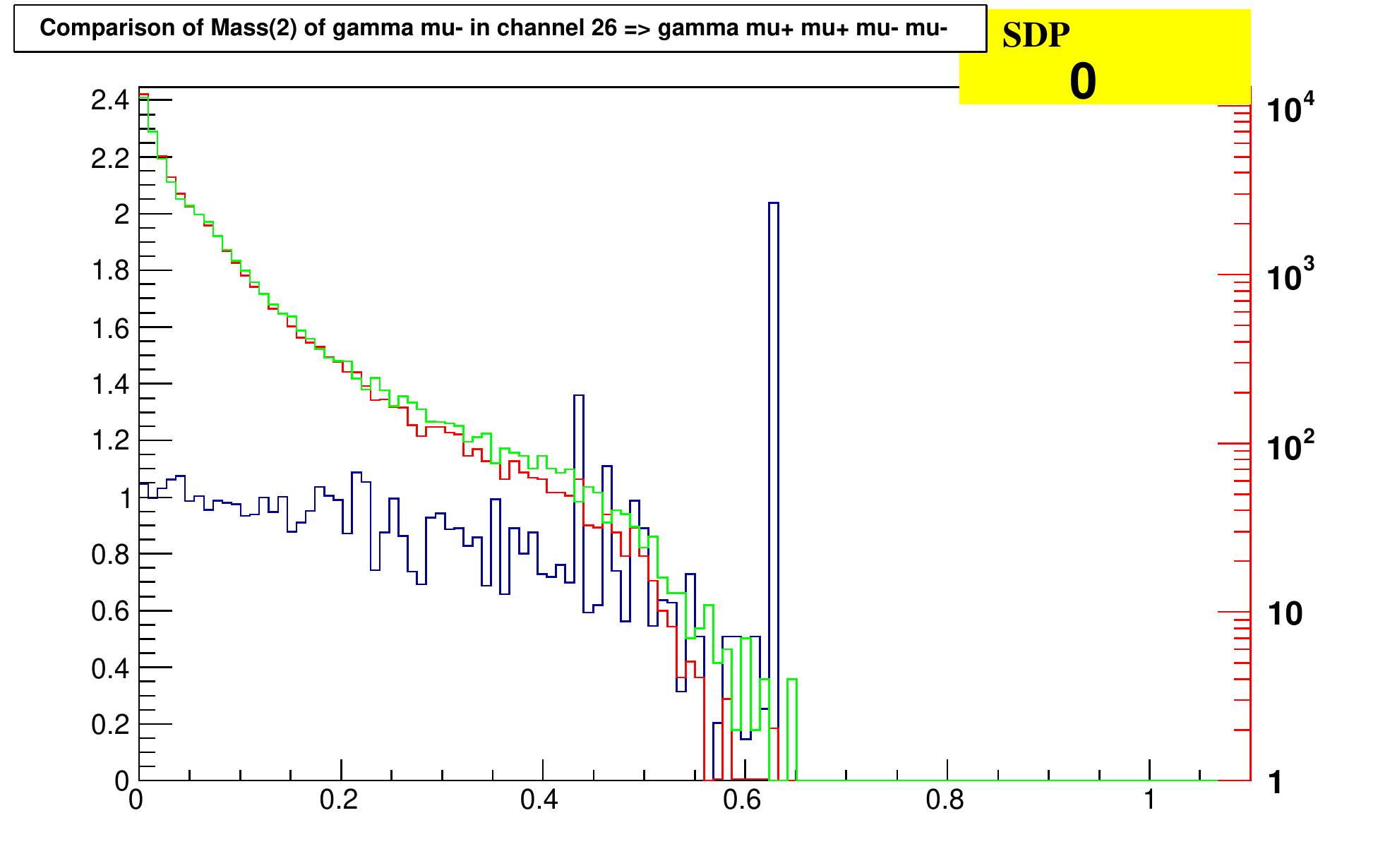}} }
{ \resizebox*{0.49\textwidth}{!}{\includegraphics{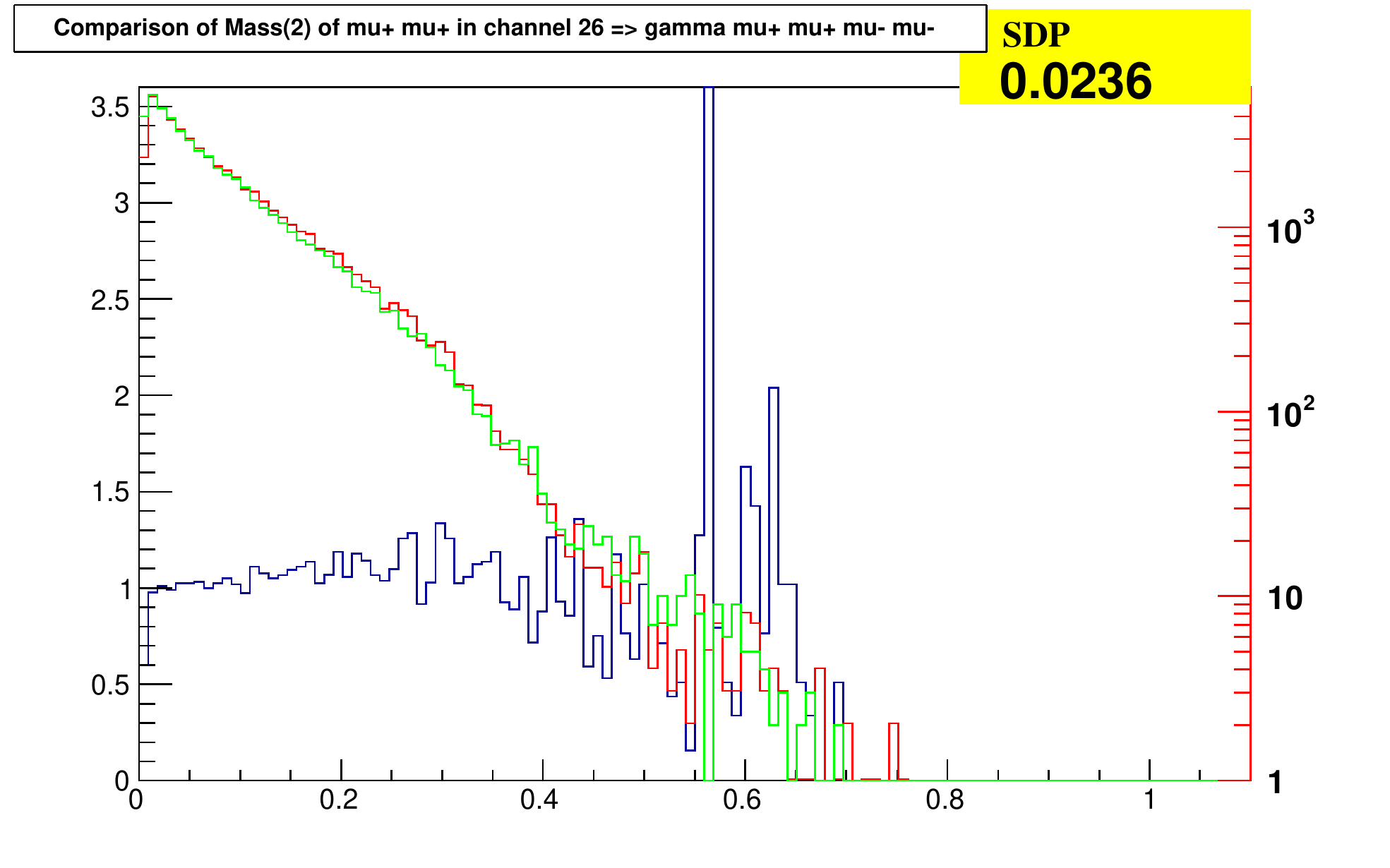}} }
{ \resizebox*{0.49\textwidth}{!}{\includegraphics{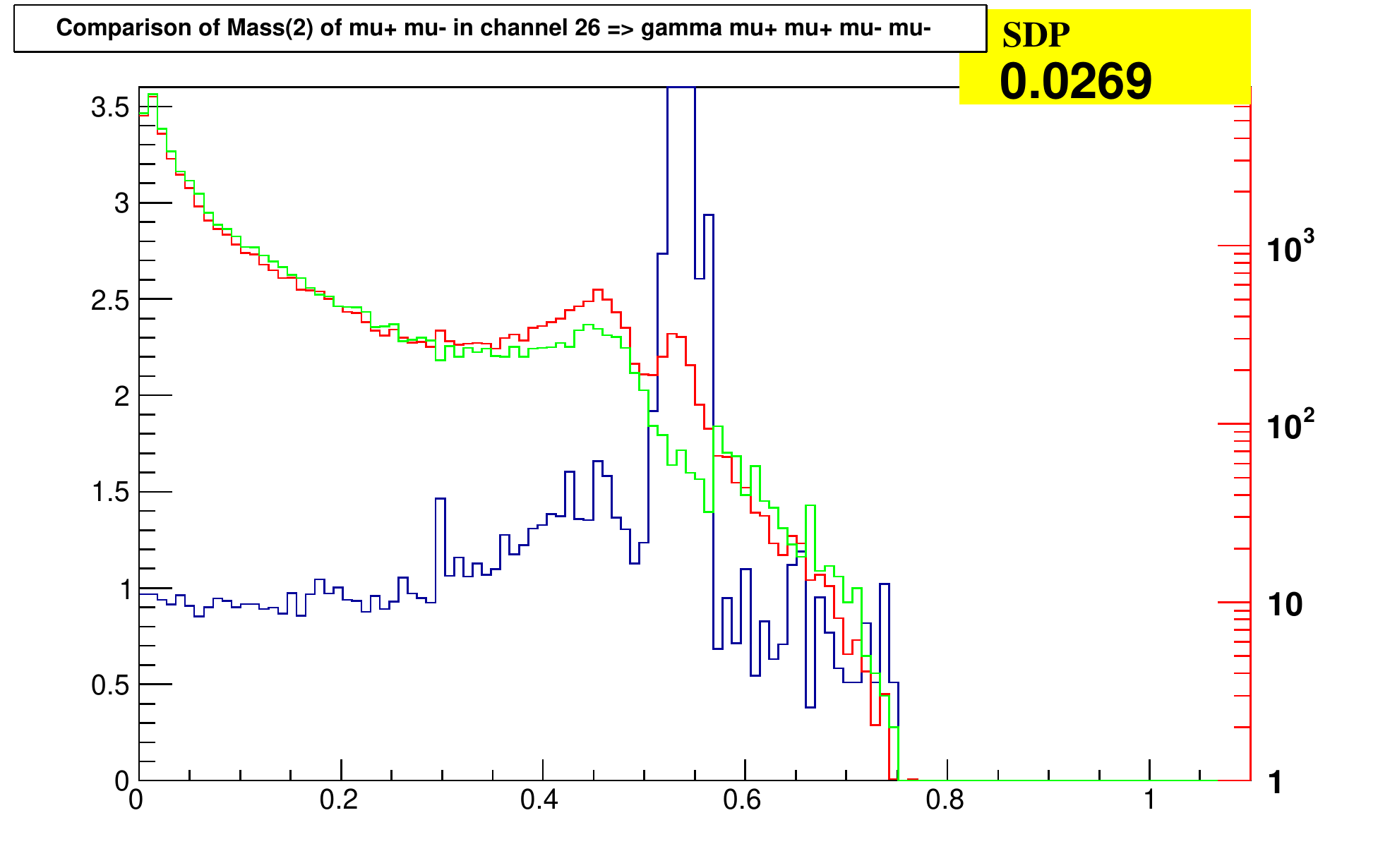}} }
{ \resizebox*{0.49\textwidth}{!}{\includegraphics{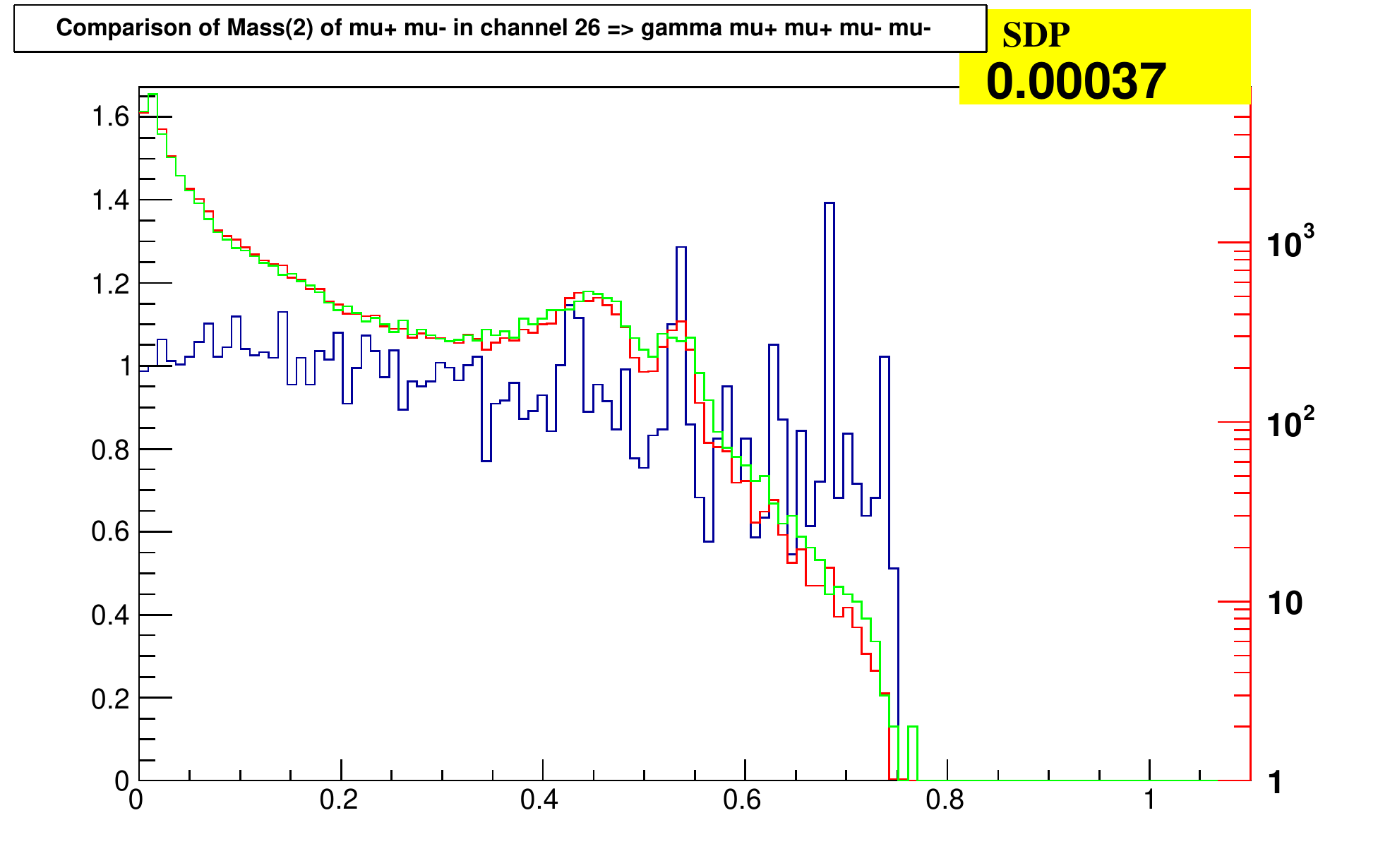}} }
{ \resizebox*{0.49\textwidth}{!}{\includegraphics{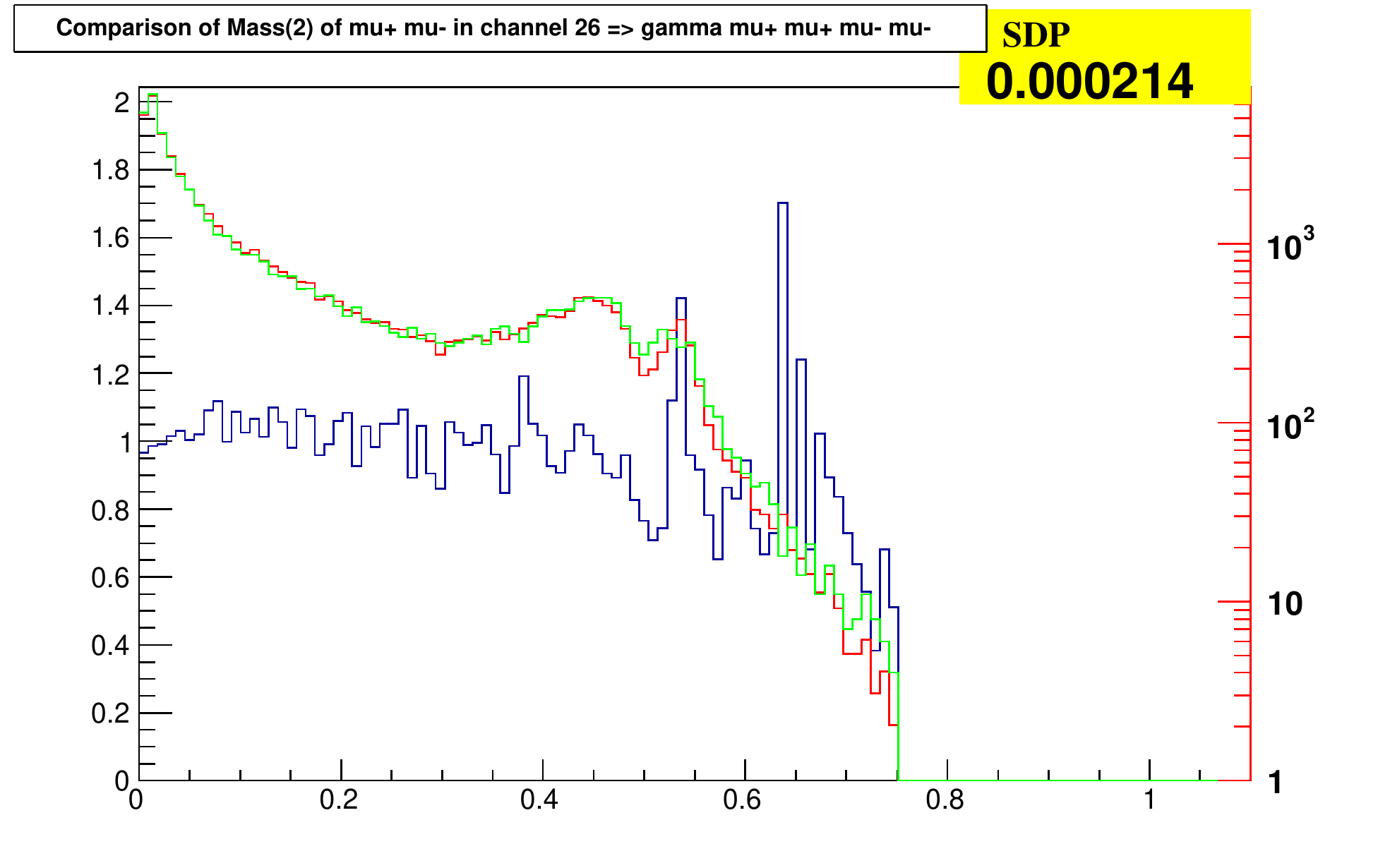}} }
{ \resizebox*{0.49\textwidth}{!}{\includegraphics{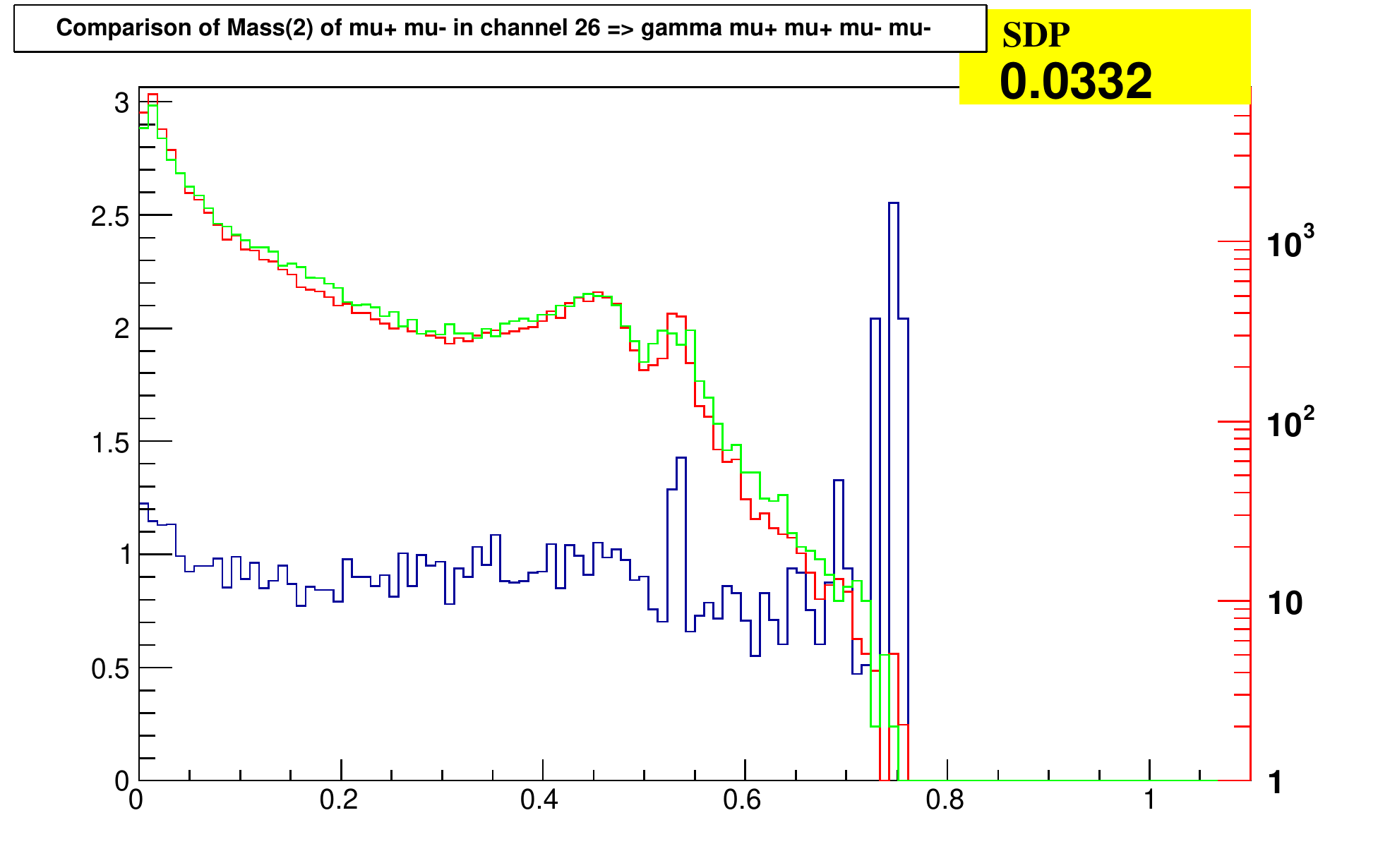}} }
{ \resizebox*{0.49\textwidth}{!}{\includegraphics{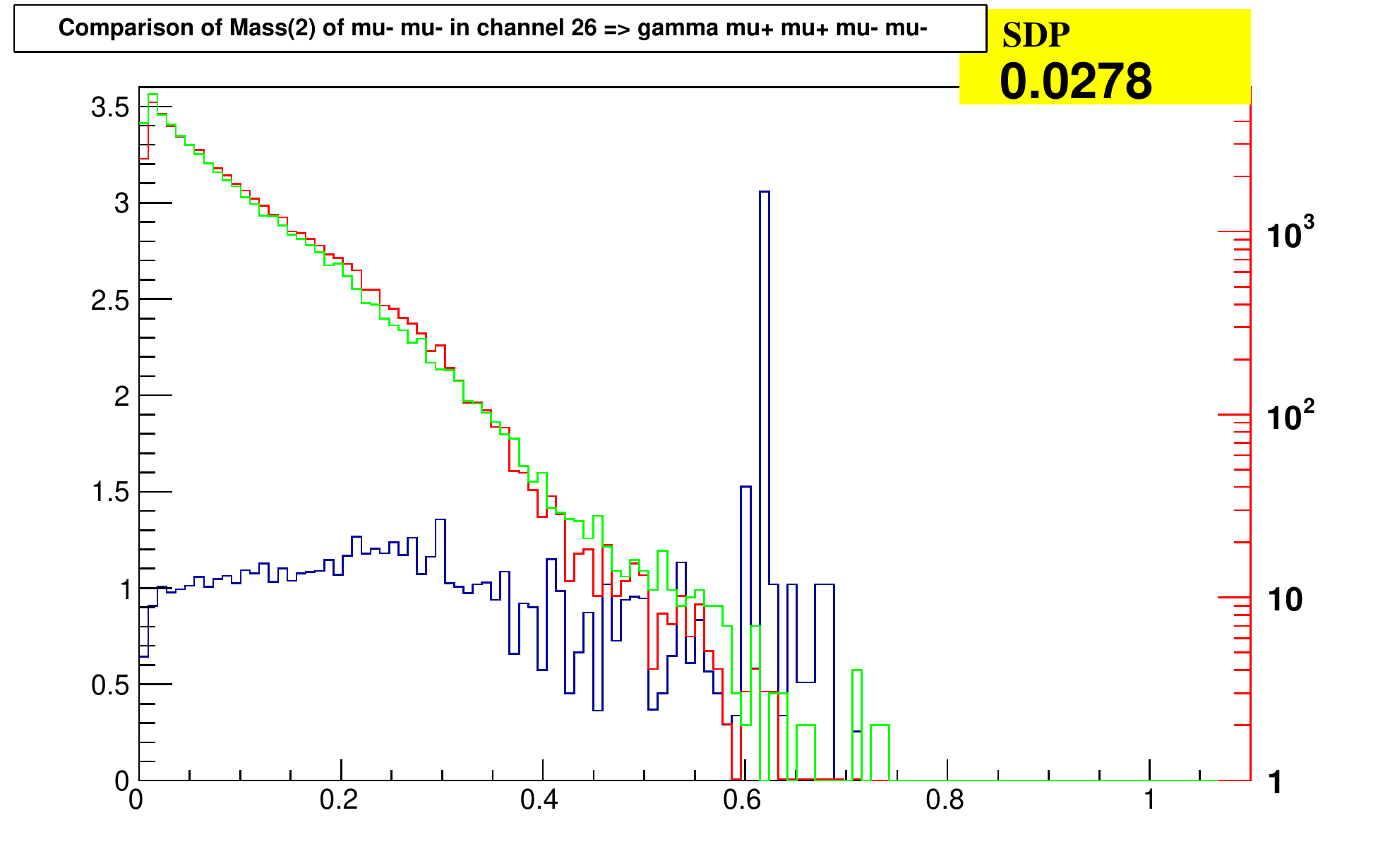}} }
{ \resizebox*{0.49\textwidth}{!}{\includegraphics{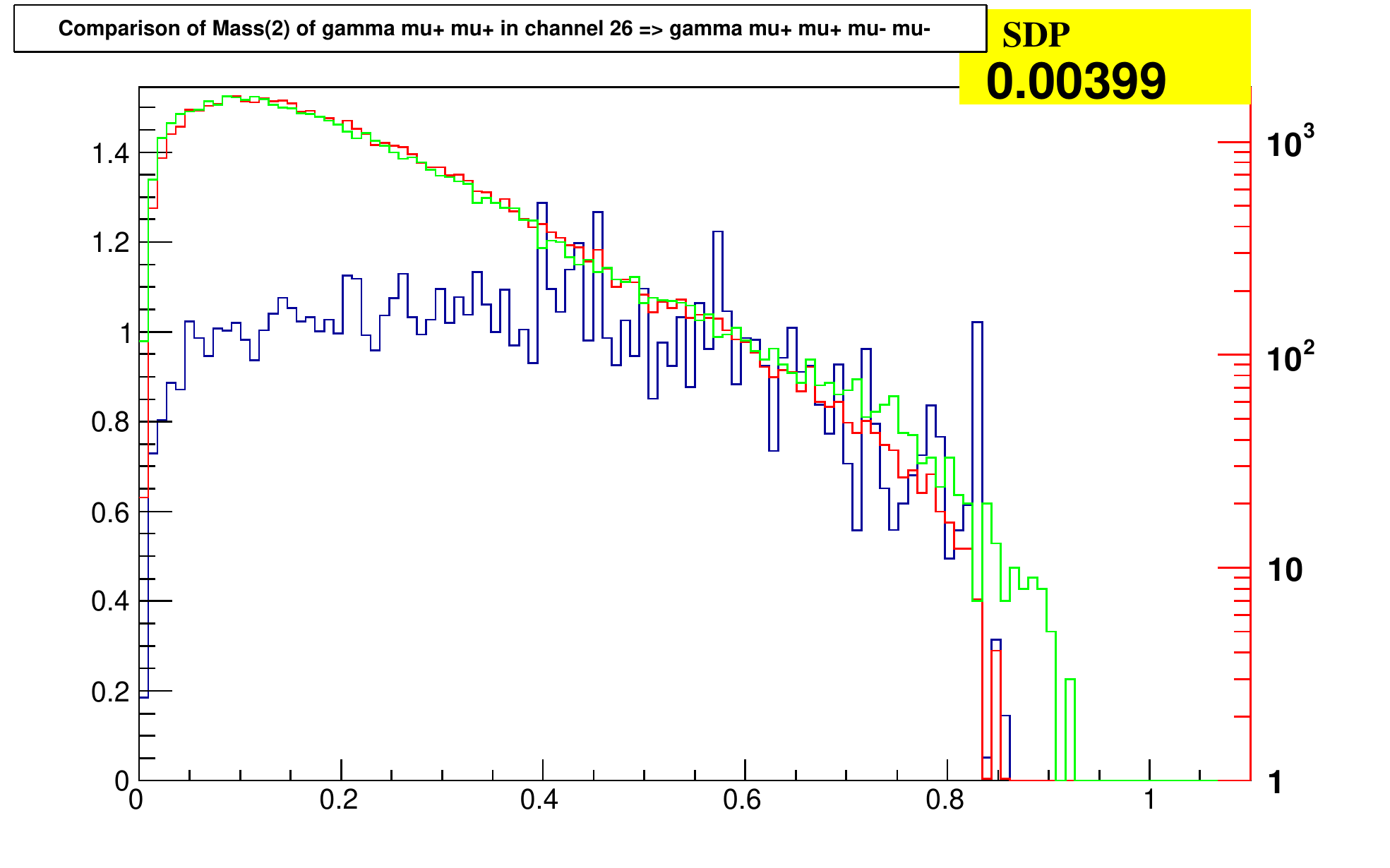}} }
{ \resizebox*{0.49\textwidth}{!}{\includegraphics{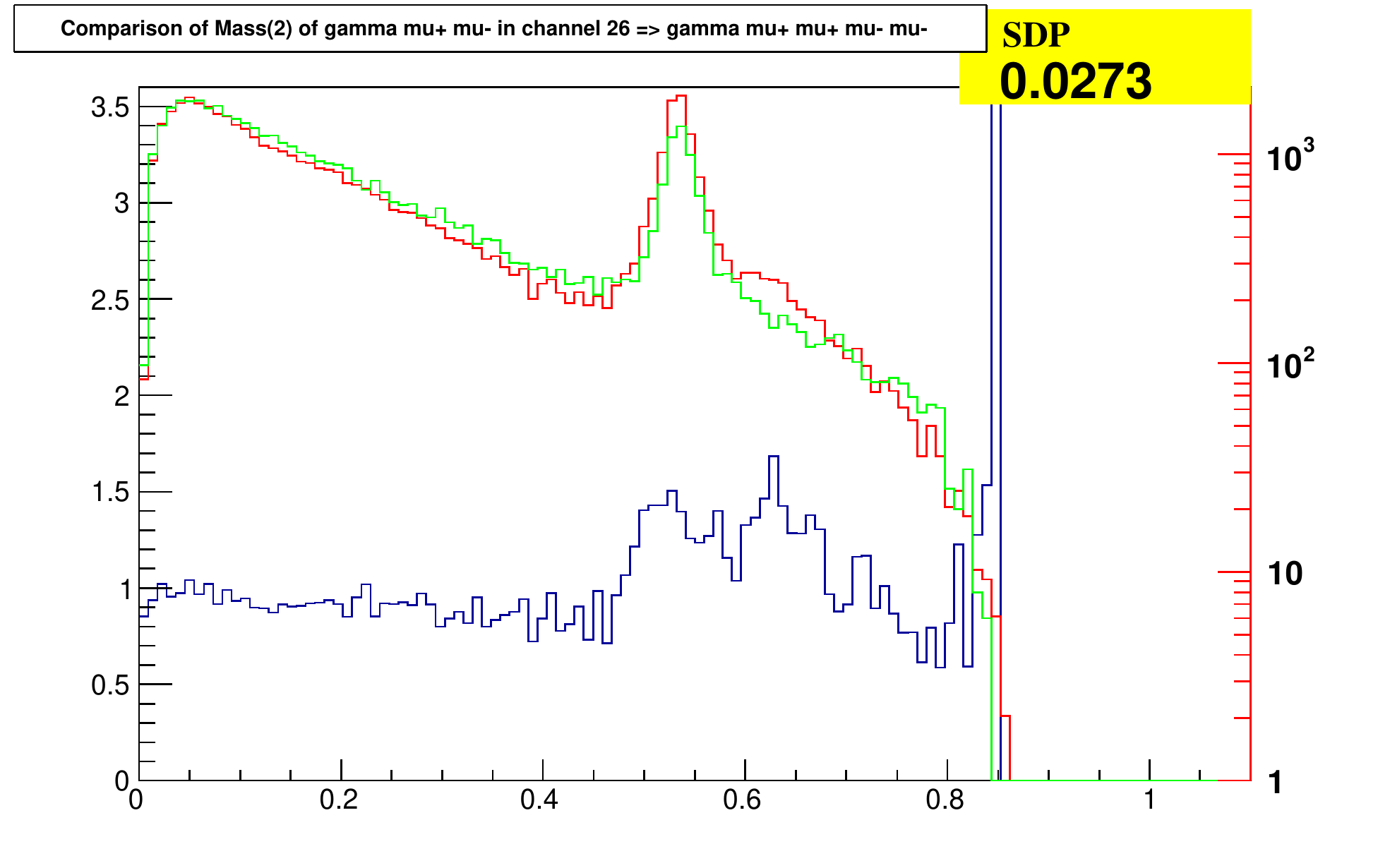}} }
{ \resizebox*{0.49\textwidth}{!}{\includegraphics{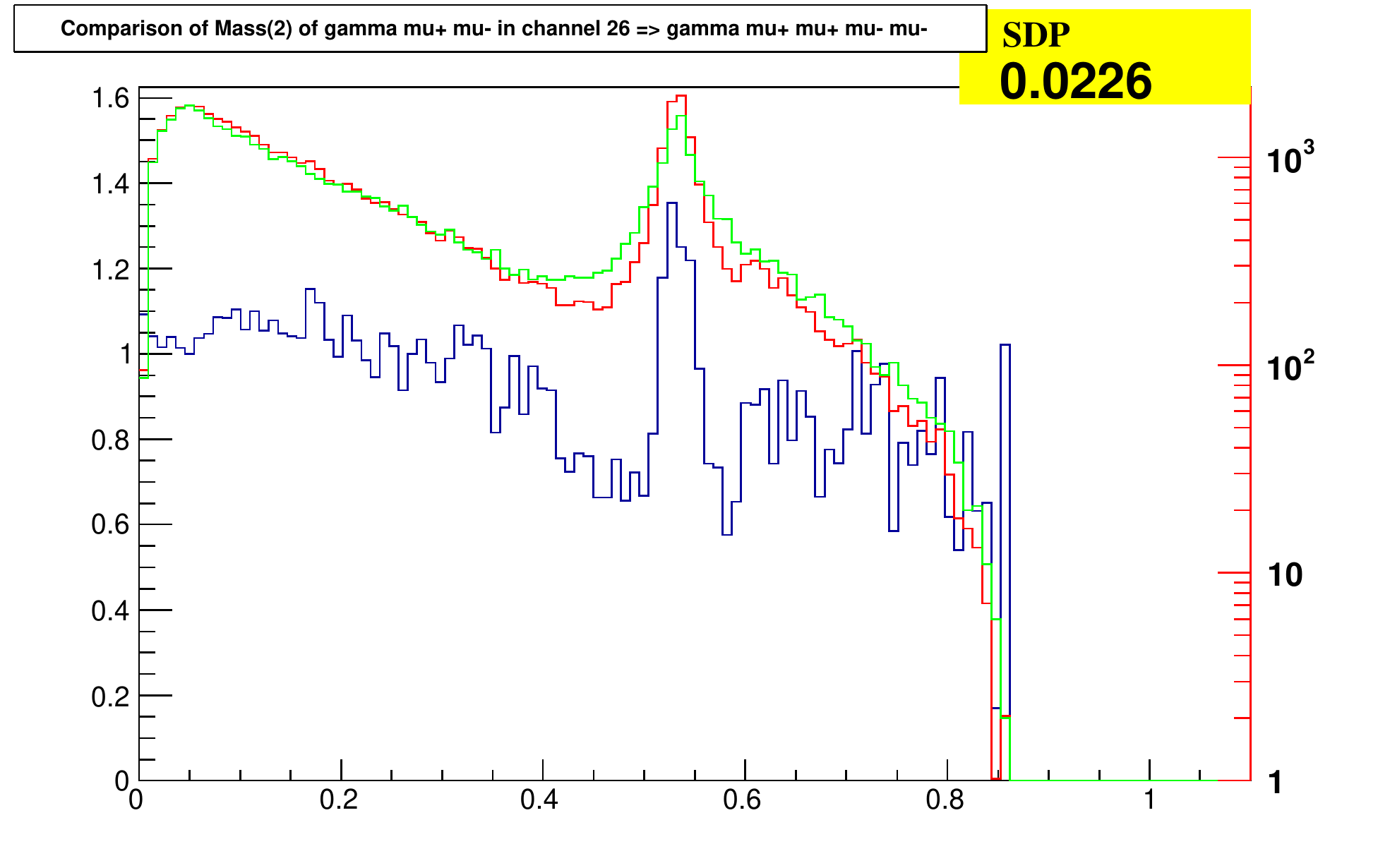}} }
{ \resizebox*{0.49\textwidth}{!}{\includegraphics{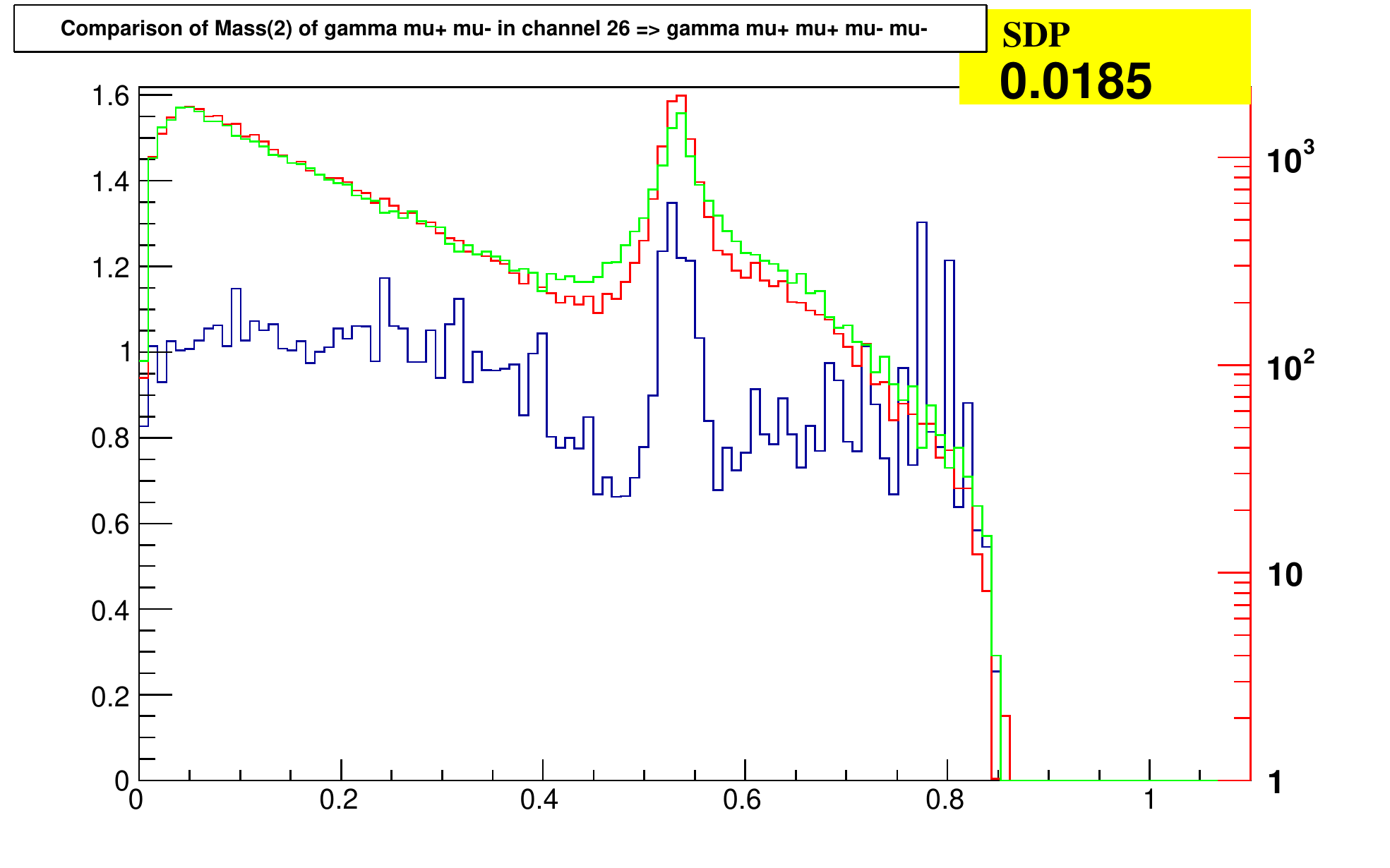}} }
{ \resizebox*{0.49\textwidth}{!}{\includegraphics{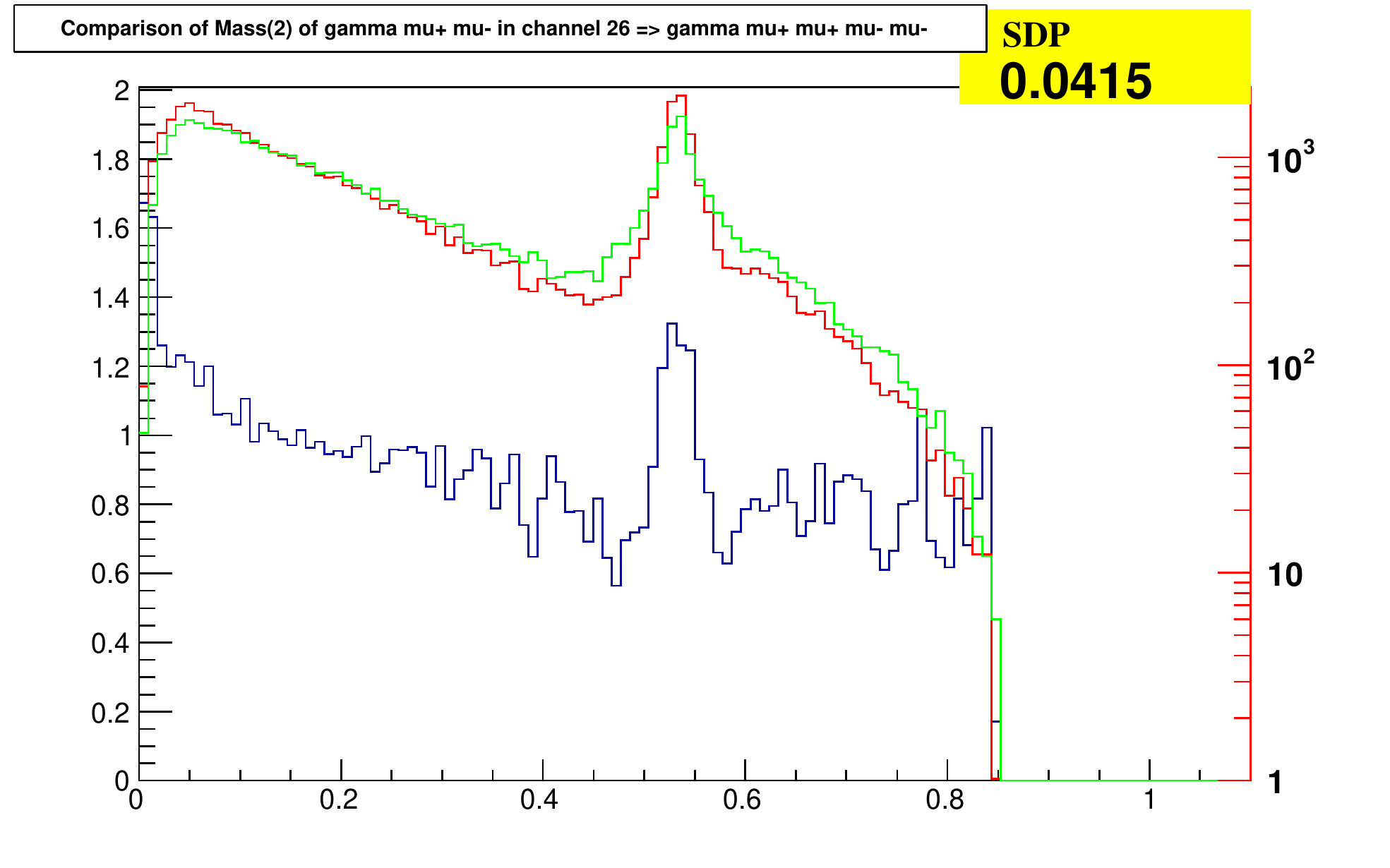}} }
{ \resizebox*{0.49\textwidth}{!}{\includegraphics{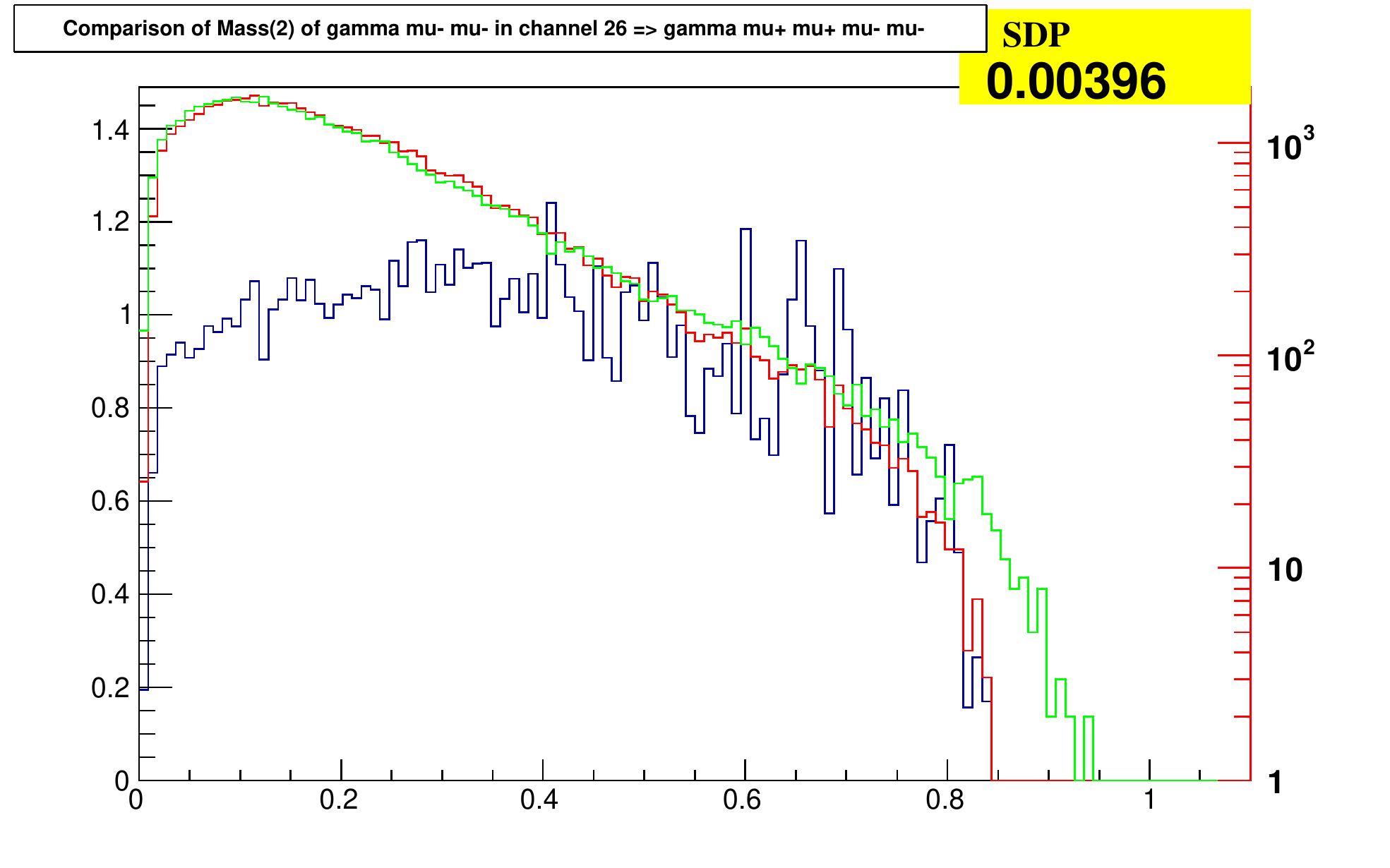}} }
{ \resizebox*{0.49\textwidth}{!}{\includegraphics{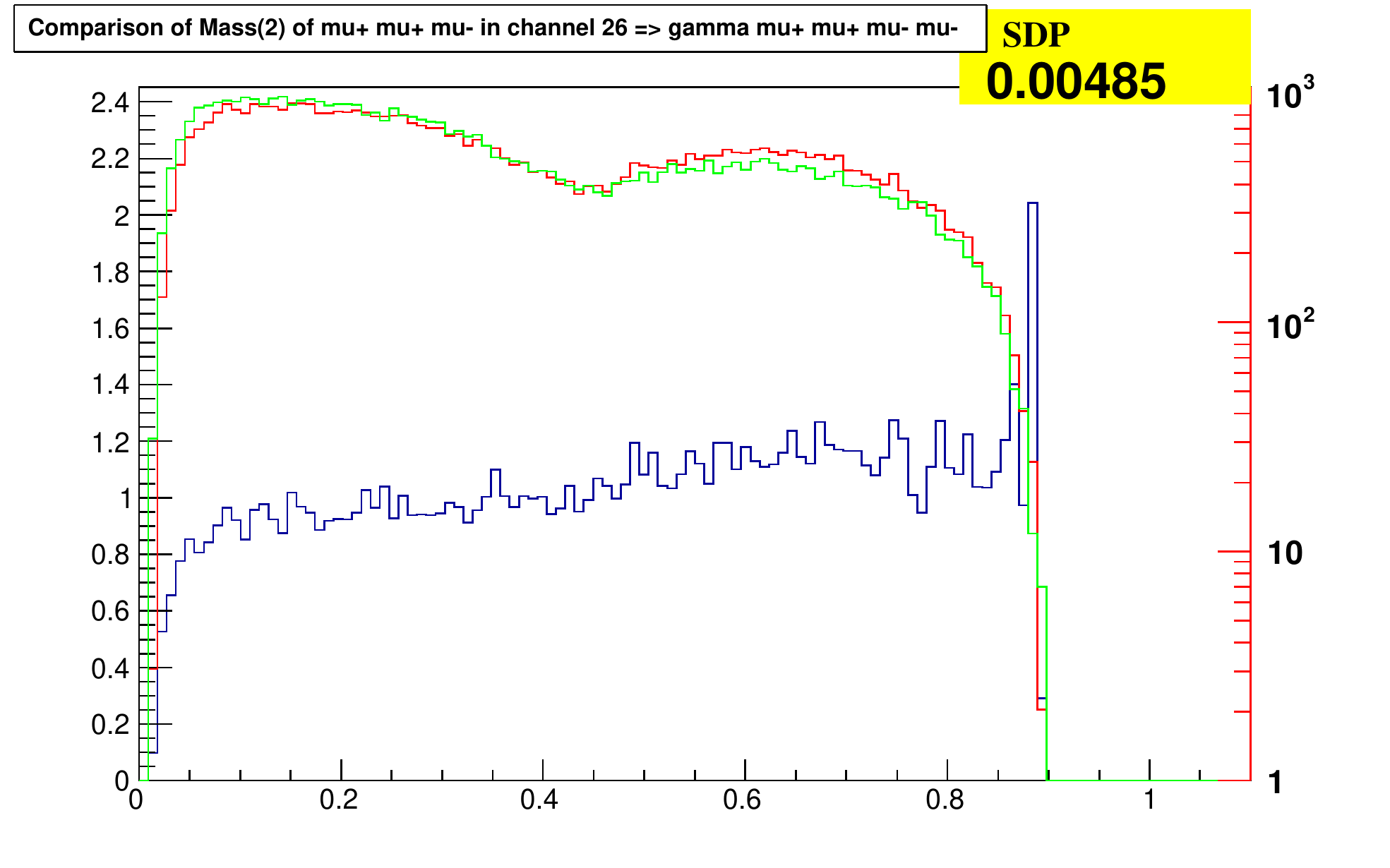}} }
{ \resizebox*{0.49\textwidth}{!}{\includegraphics{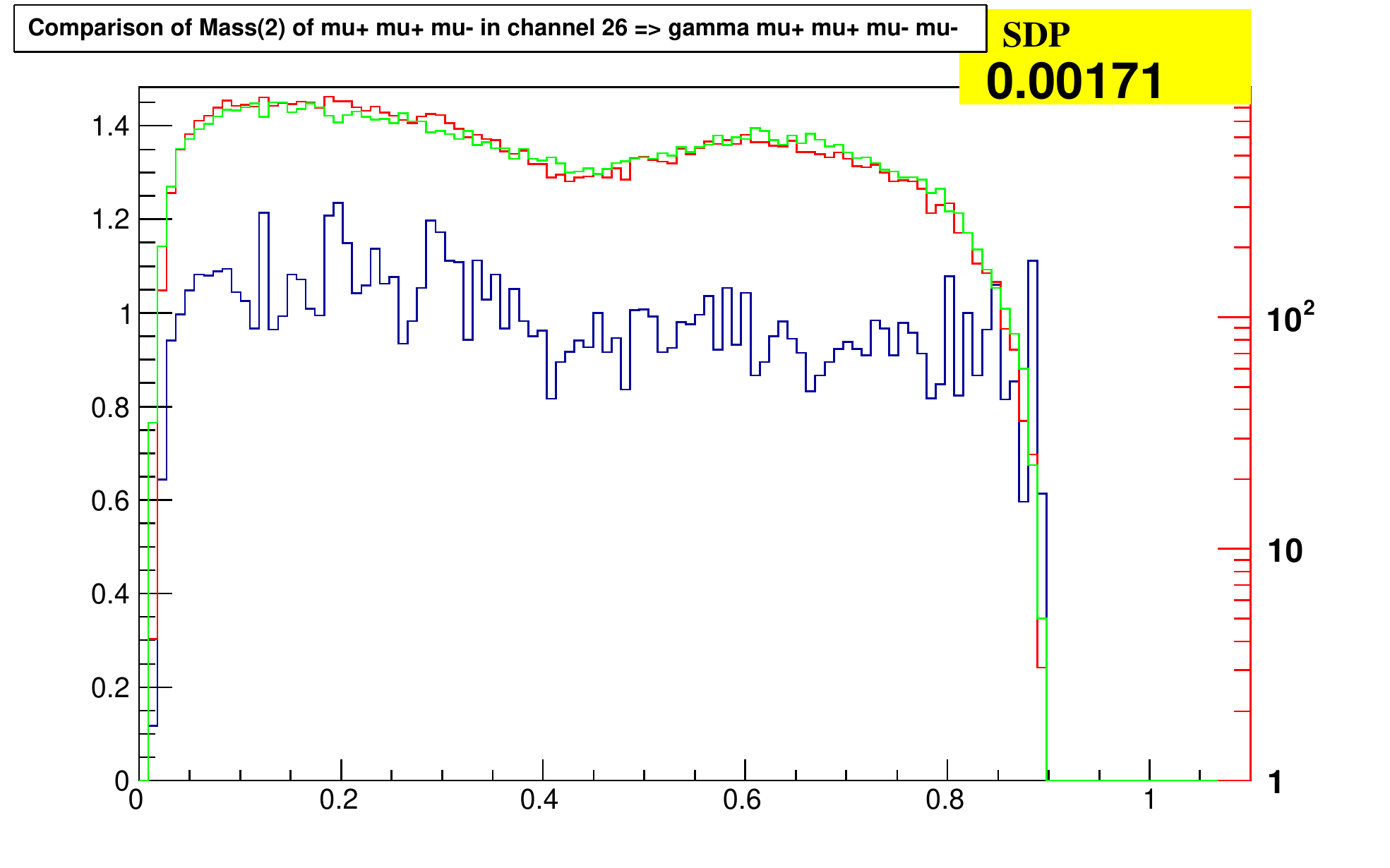}} }
{ \resizebox*{0.49\textwidth}{!}{\includegraphics{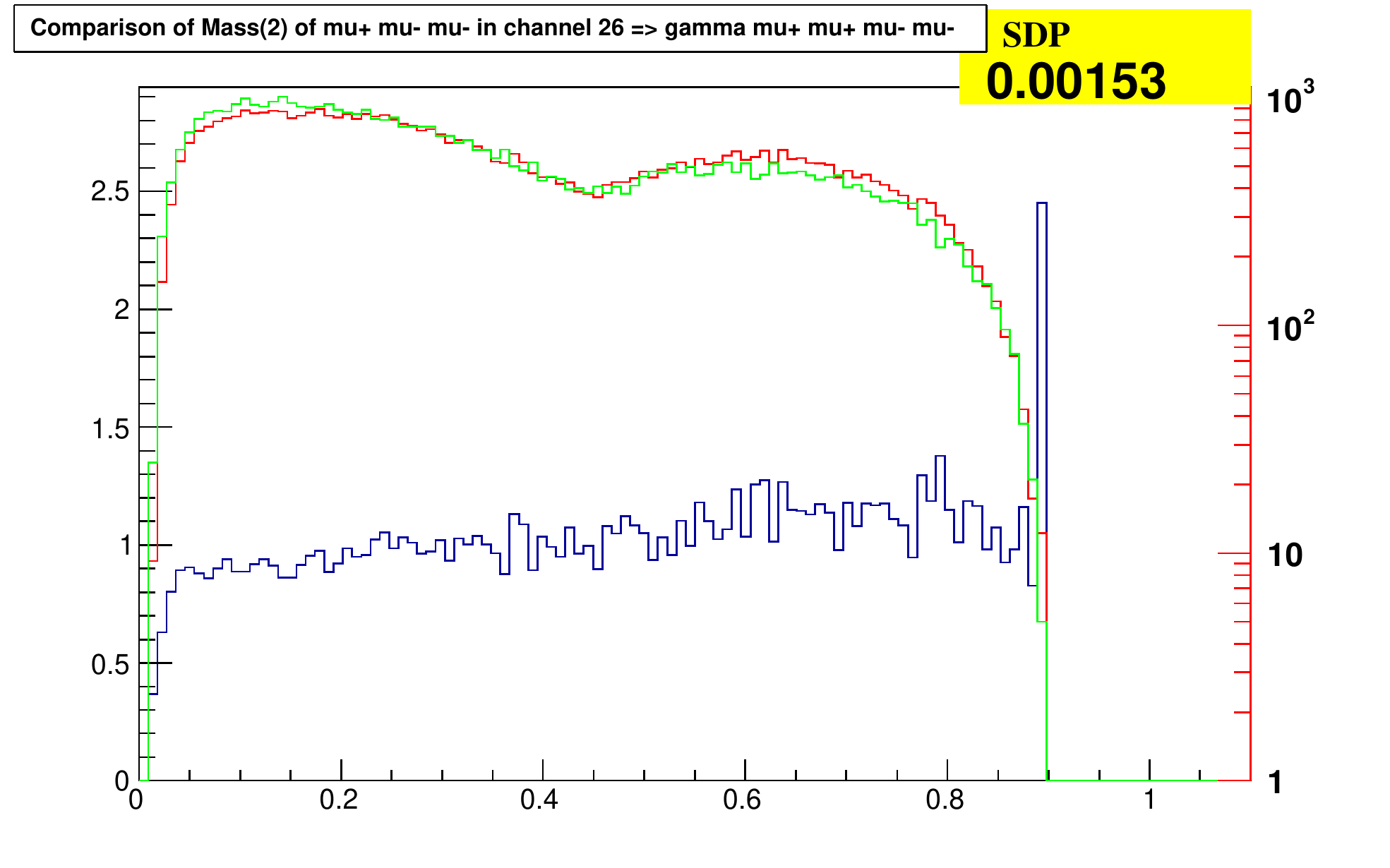}} }
{ \resizebox*{0.49\textwidth}{!}{\includegraphics{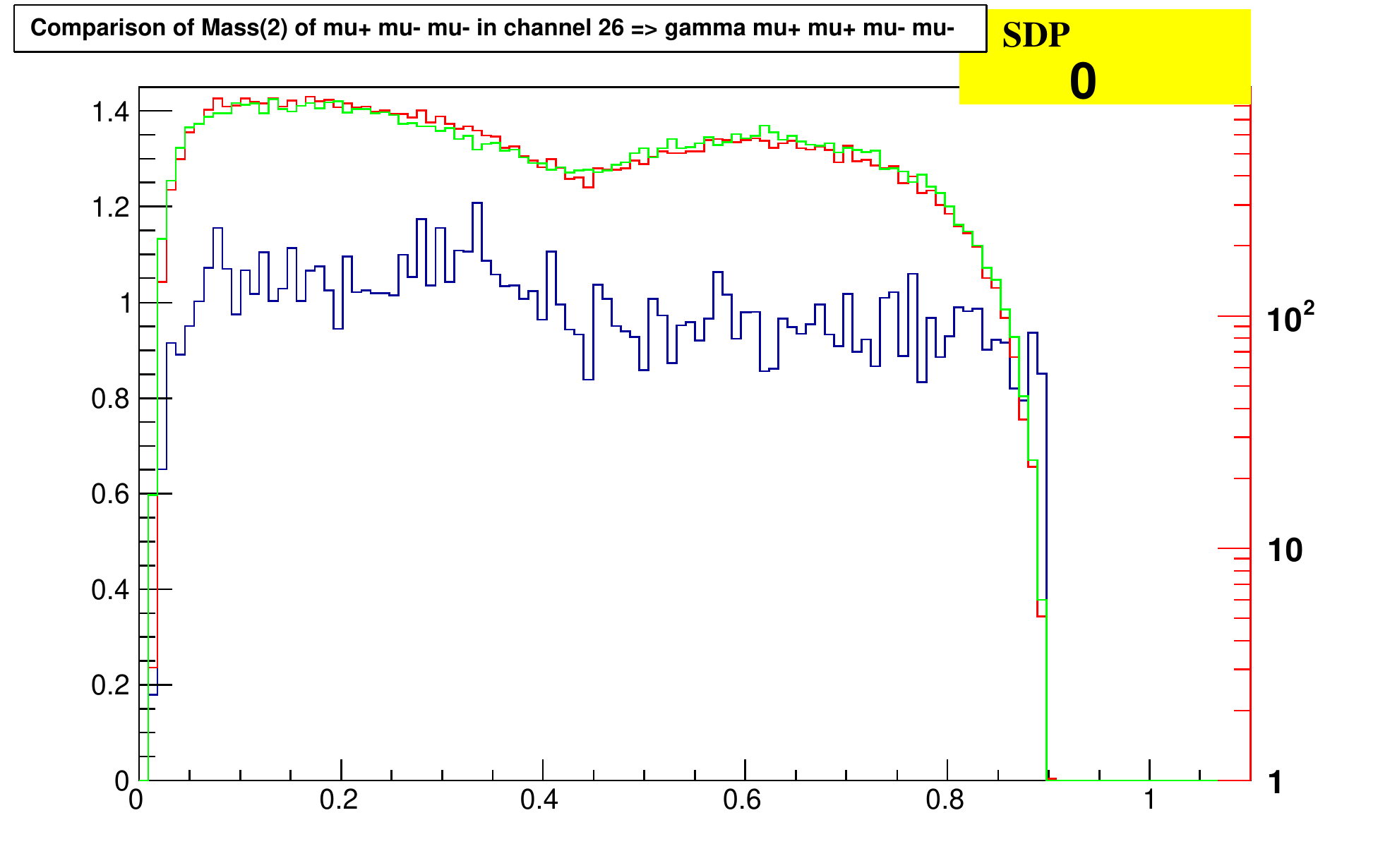}} }
{ \resizebox*{0.49\textwidth}{!}{\includegraphics{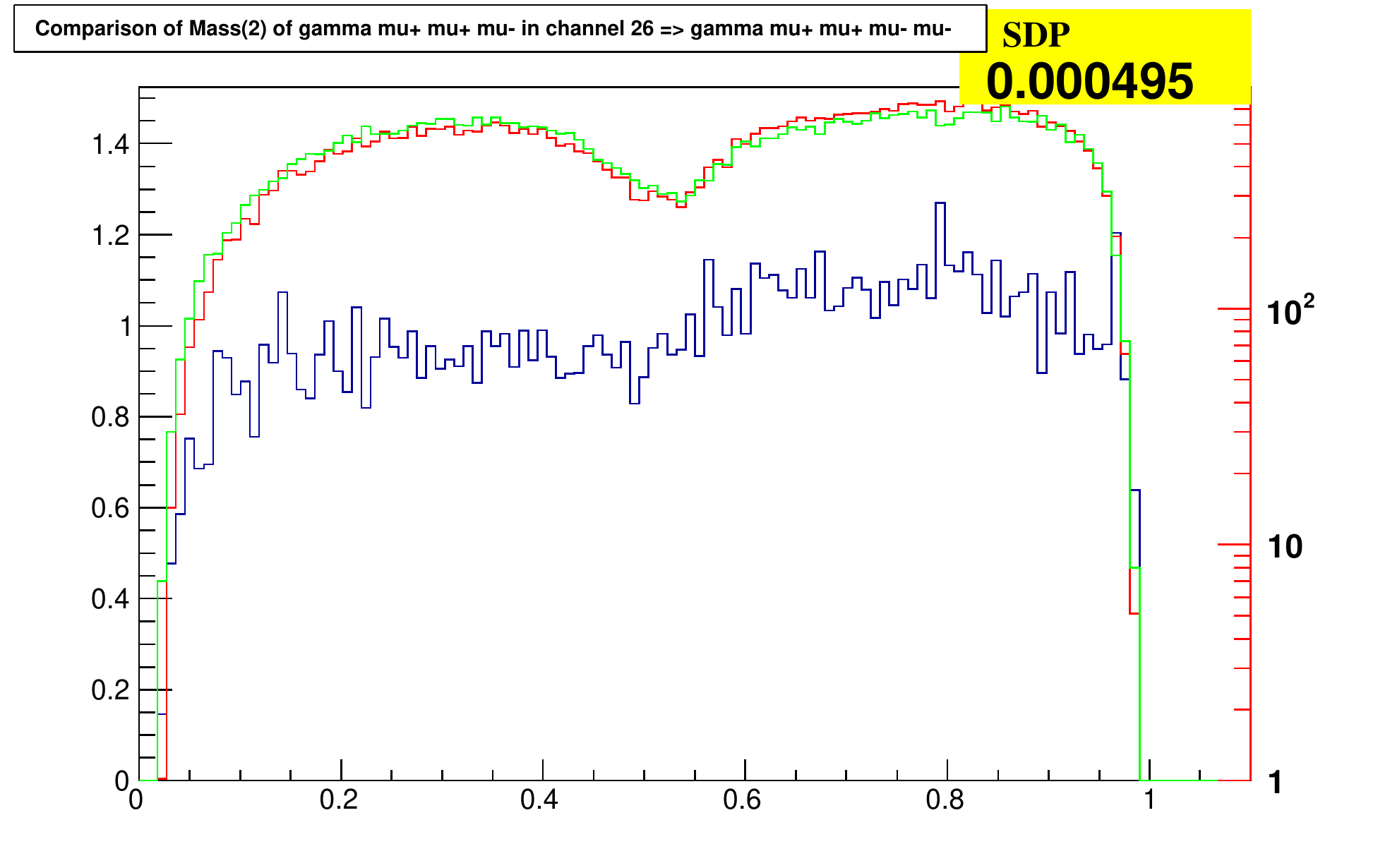}} }
{ \resizebox*{0.49\textwidth}{!}{\includegraphics{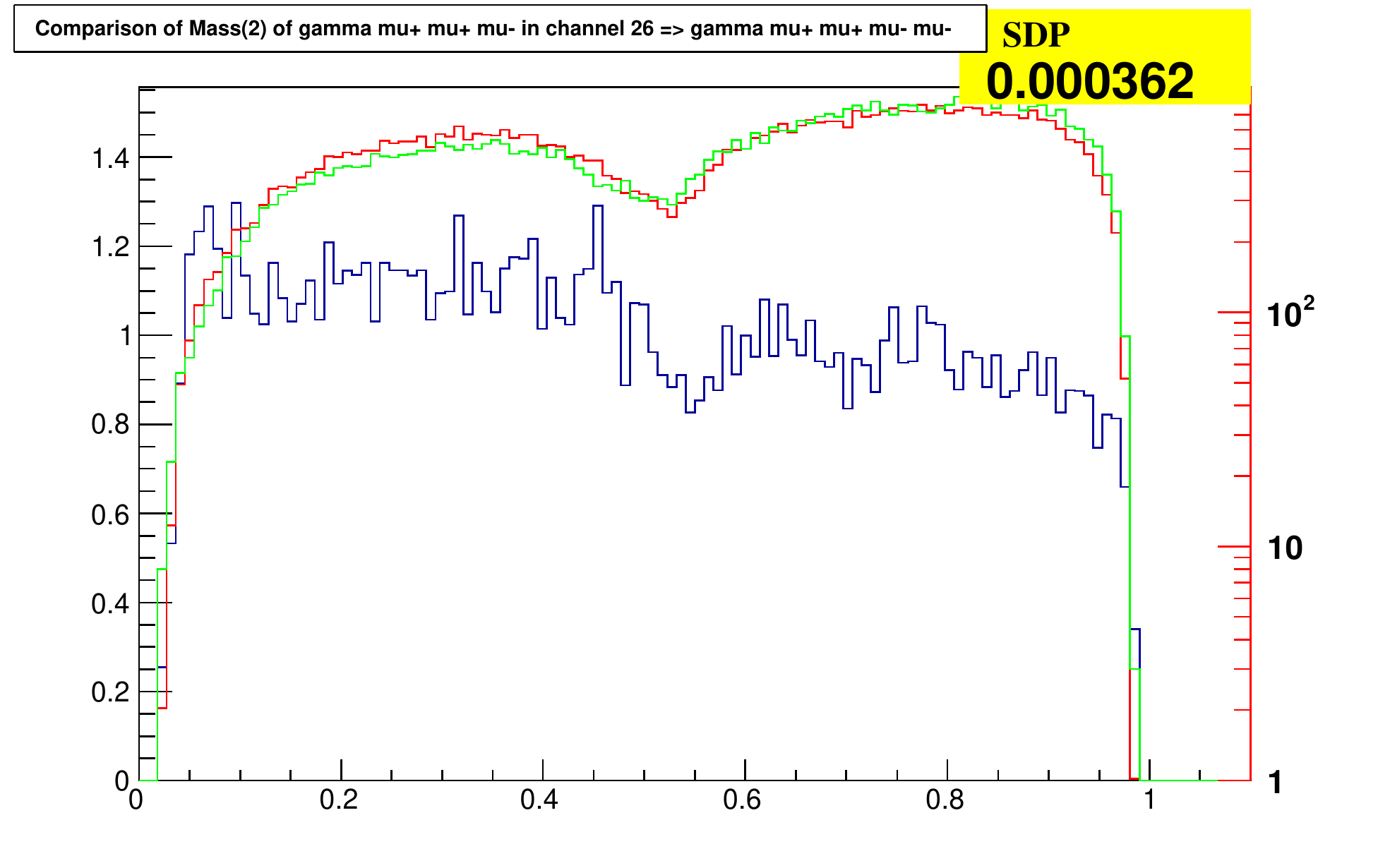}} }
{ \resizebox*{0.49\textwidth}{!}{\includegraphics{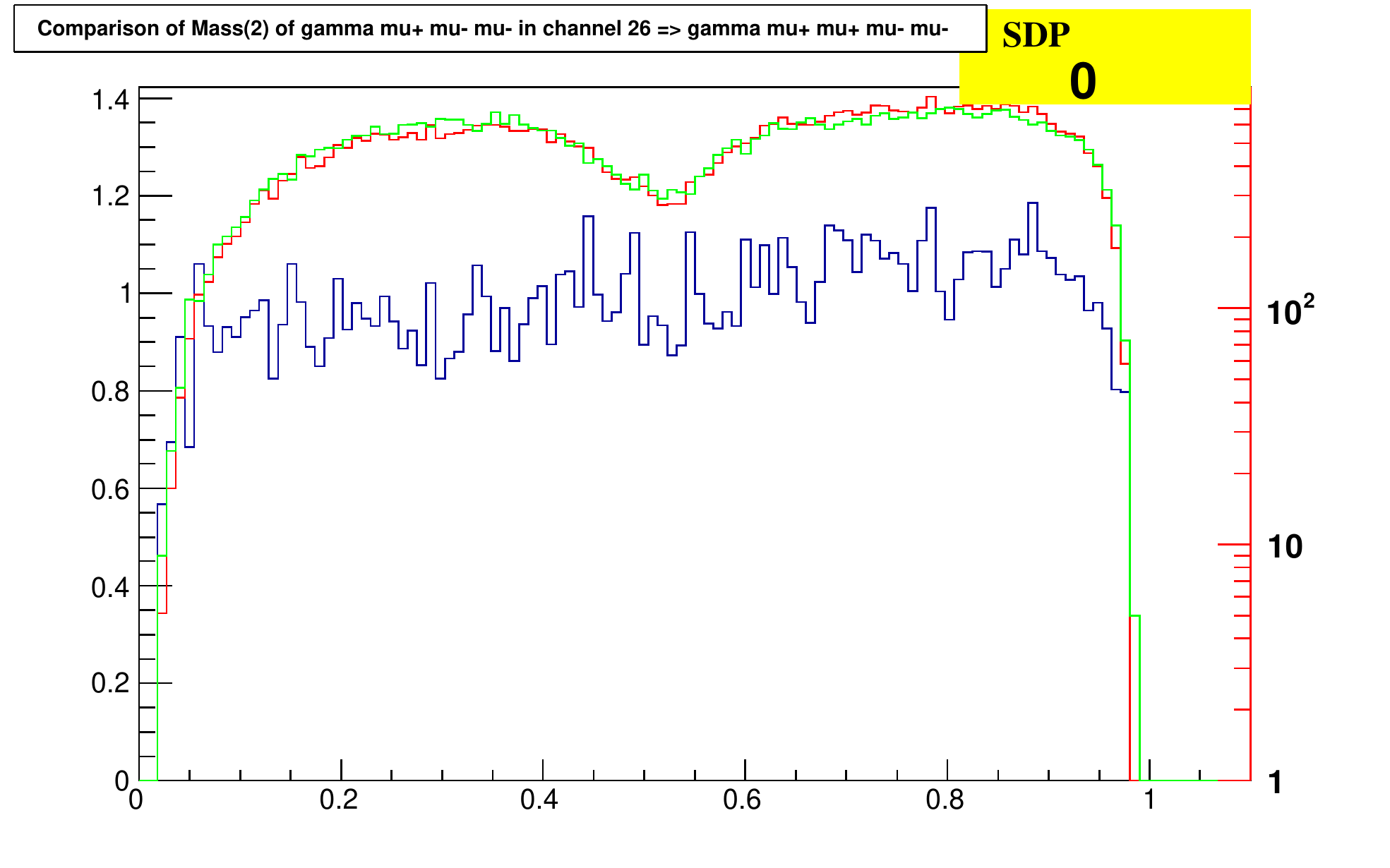}} }
{ \resizebox*{0.49\textwidth}{!}{\includegraphics{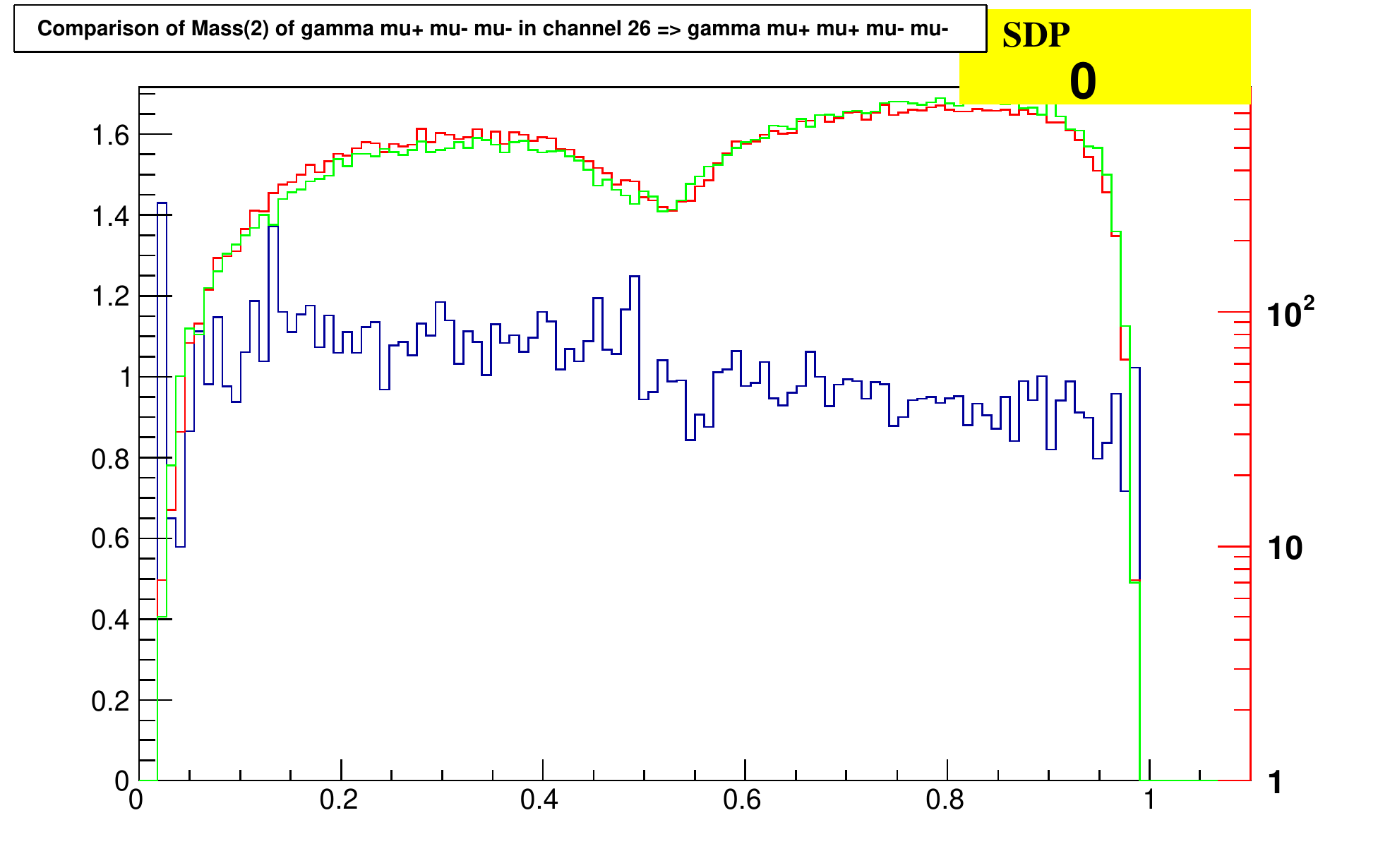}} }
{ \resizebox*{0.49\textwidth}{!}{\includegraphics{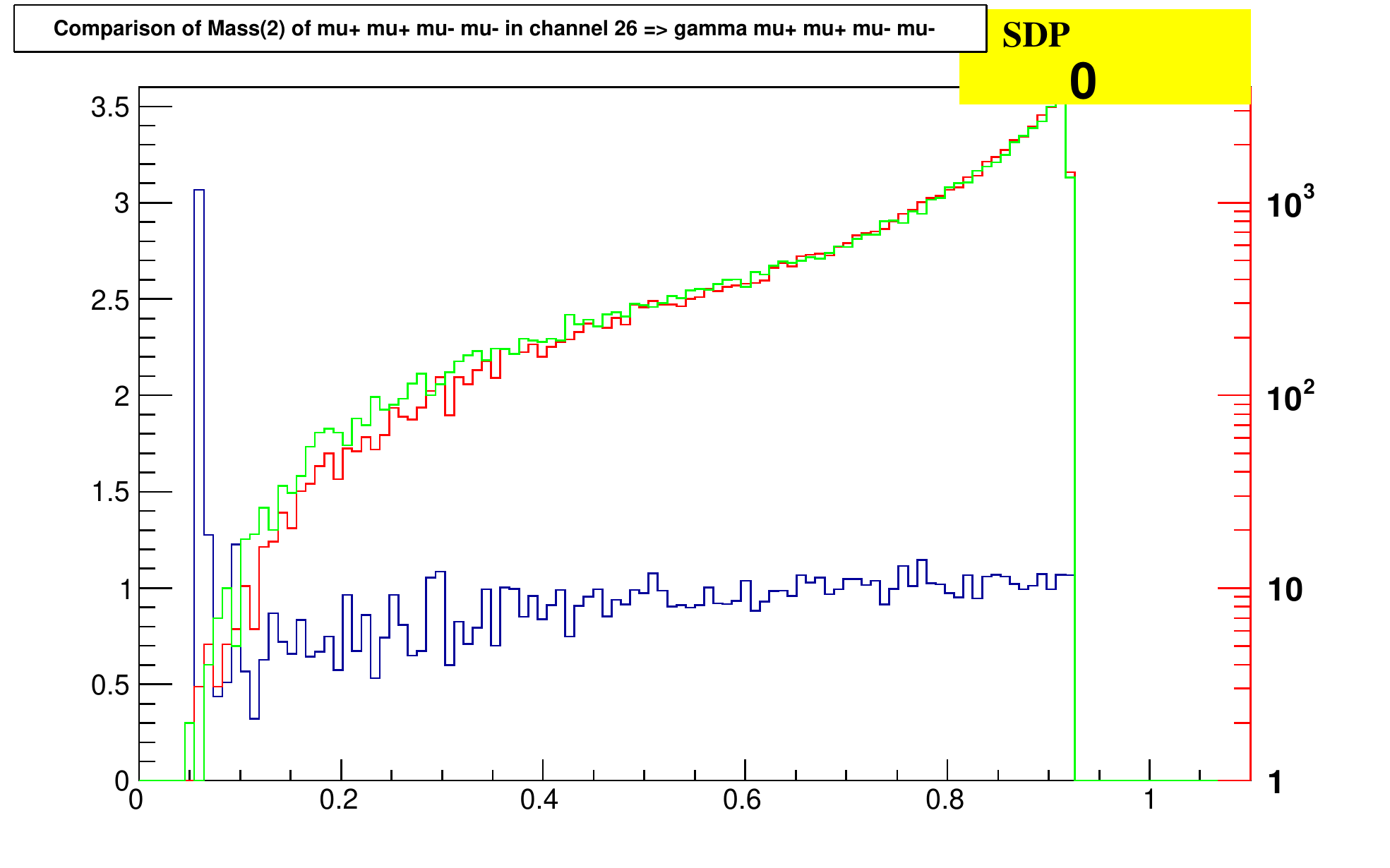}} }
{ \resizebox*{0.49\textwidth}{!}{\includegraphics{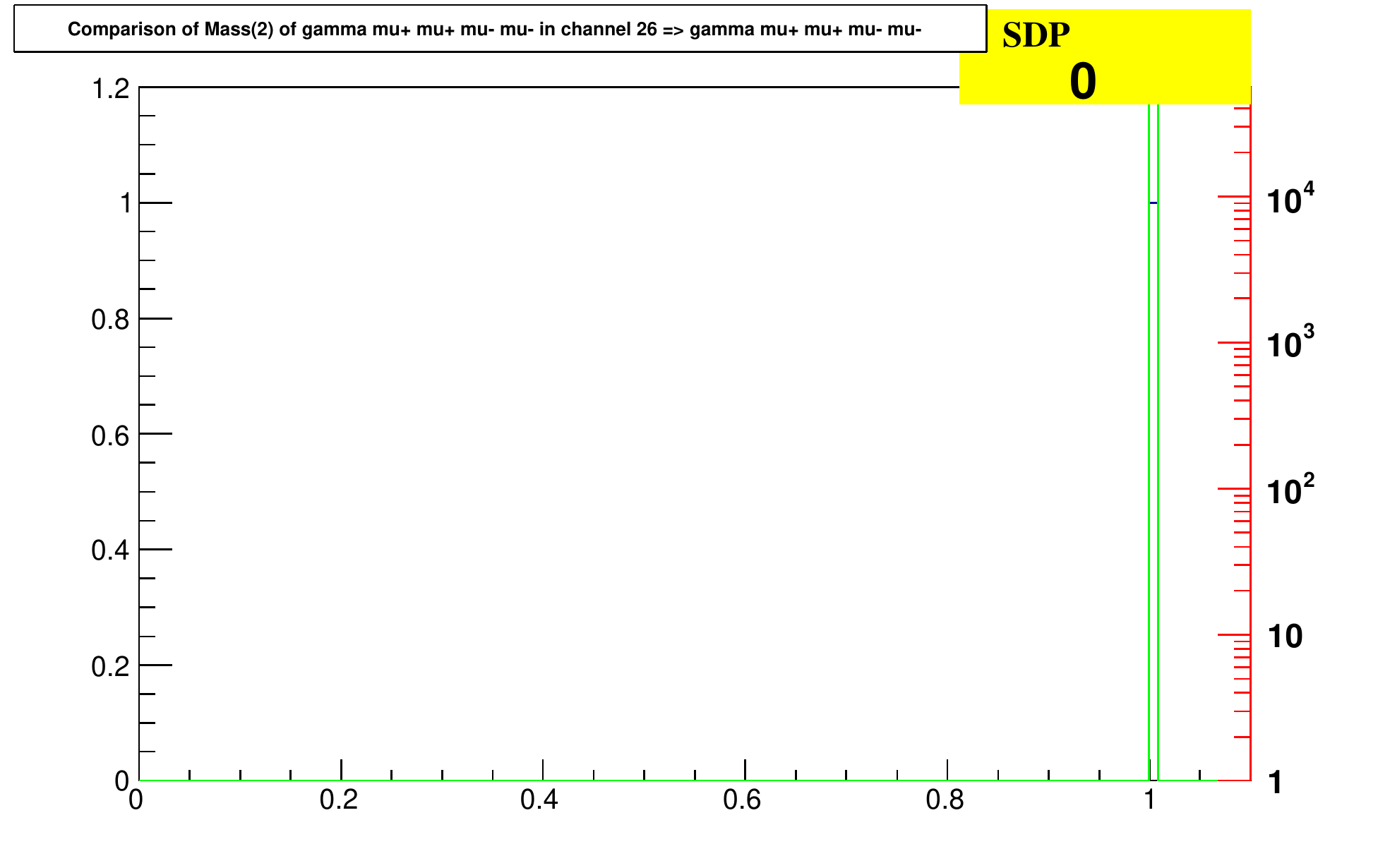}} }

%==================================
%\section{{\tt MC-tester} run for $\sqrt{s}=150$ GeV}
\subsection{{\tt MC-tester}: $q \bar q \rightarrow \gamma \mu^{+} \mu^{+} \mu^{-} \mu^{-}$ at $\sqrt{s}=150$ GeV}
\label{sec:mc-tester150}
%==================================
{ \resizebox*{0.49\textwidth}{!}{\includegraphics{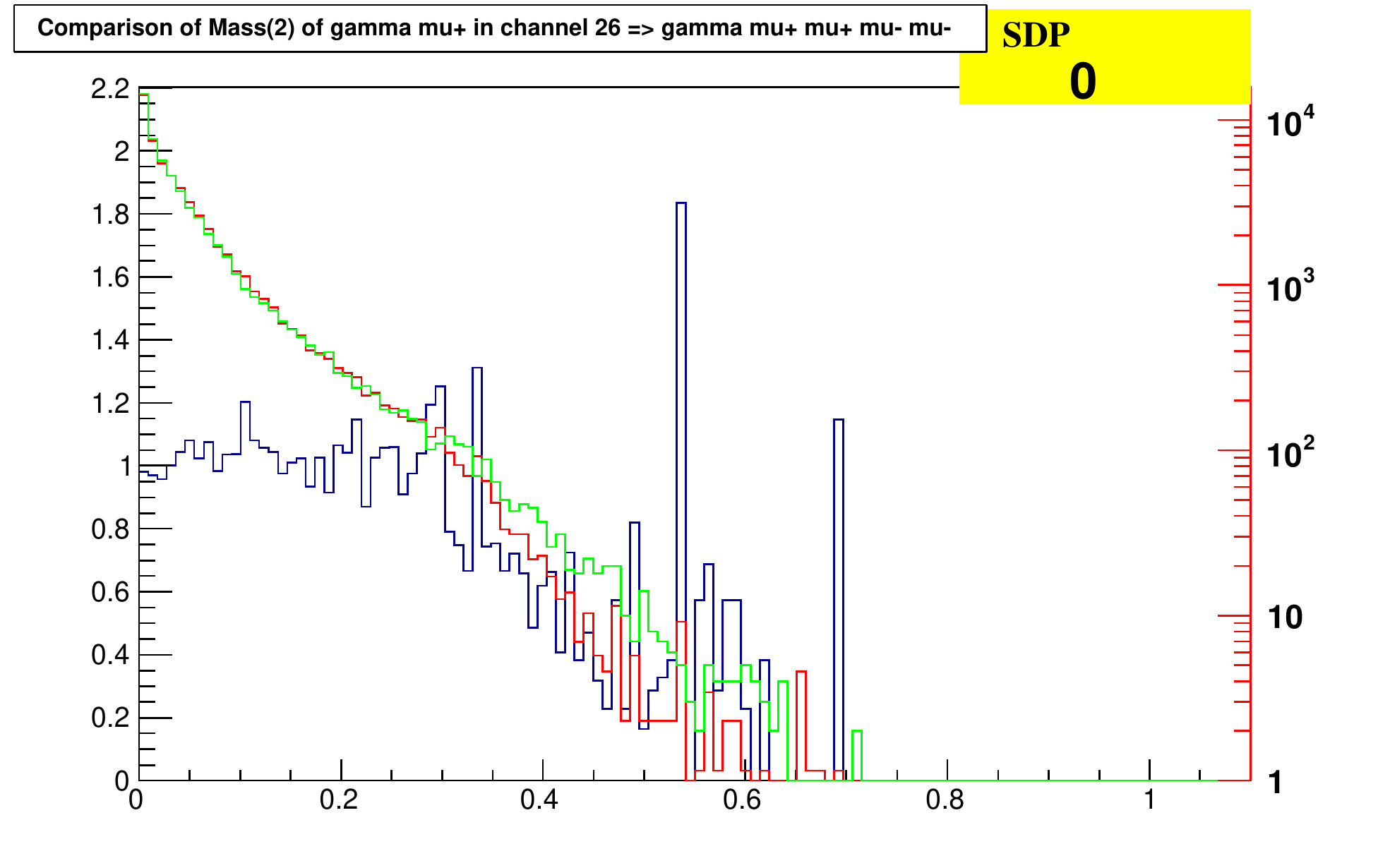}} }
{ \resizebox*{0.49\textwidth}{!}{\includegraphics{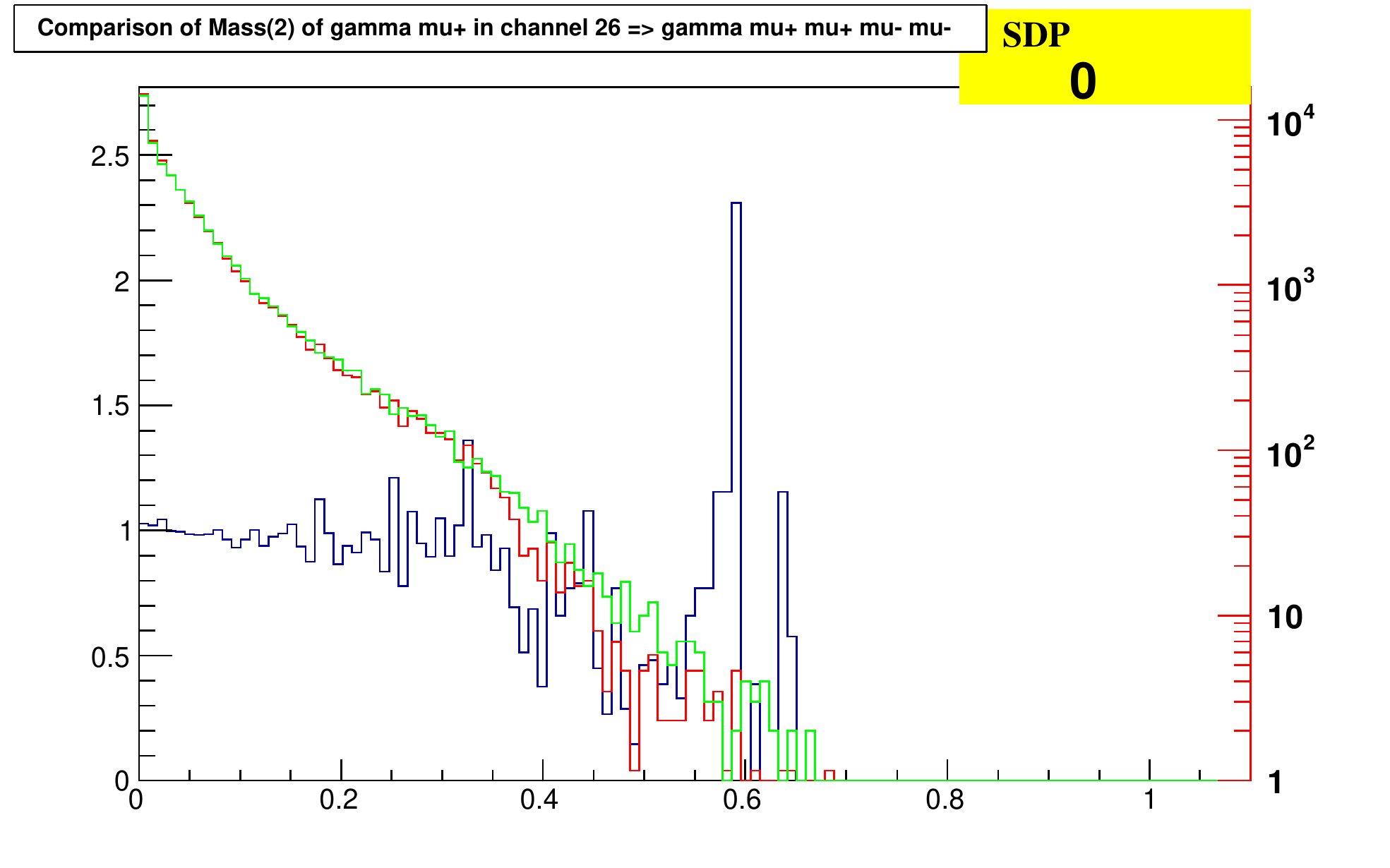}} }
{ \resizebox*{0.49\textwidth}{!}{\includegraphics{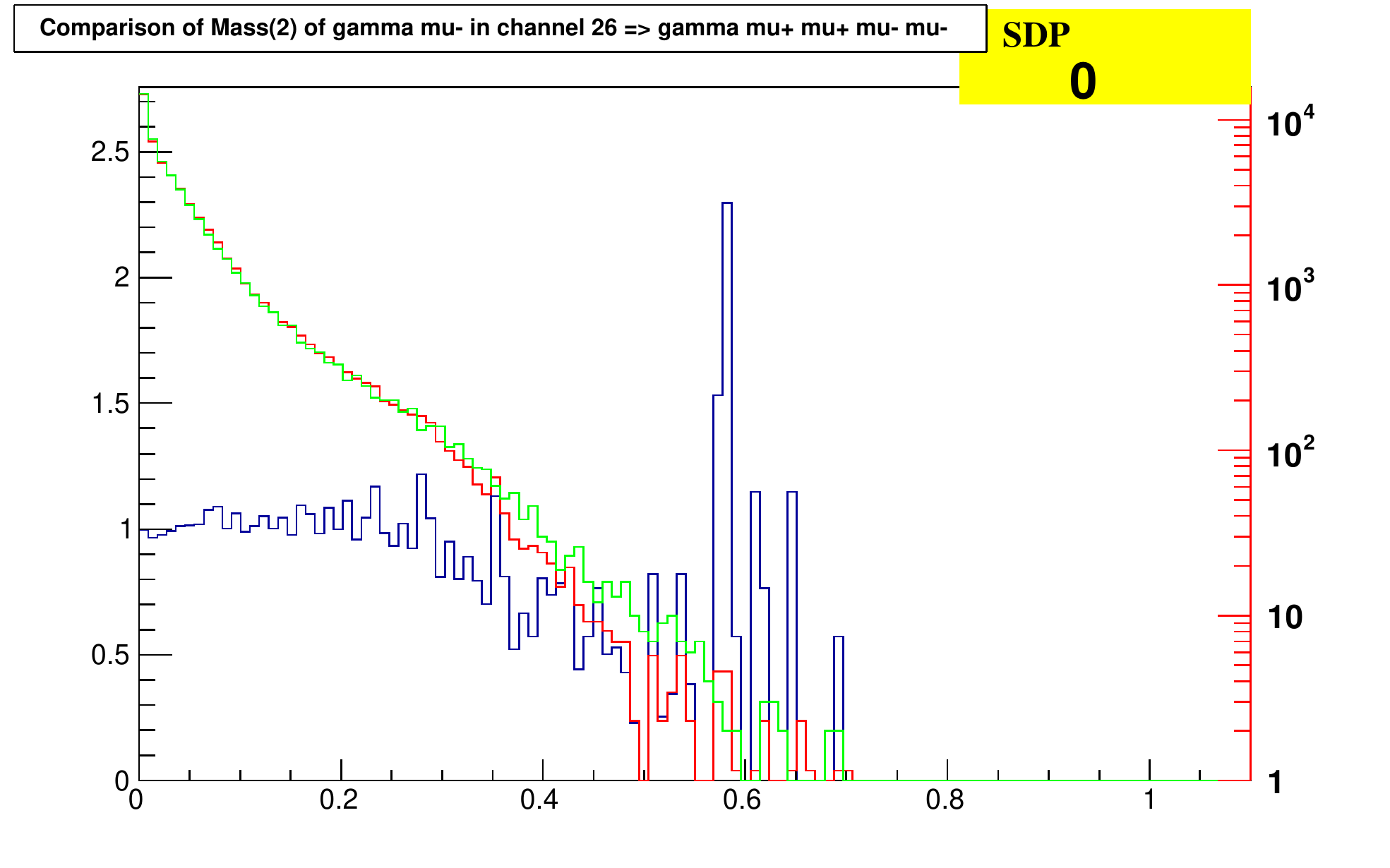}} }
{ \resizebox*{0.49\textwidth}{!}{\includegraphics{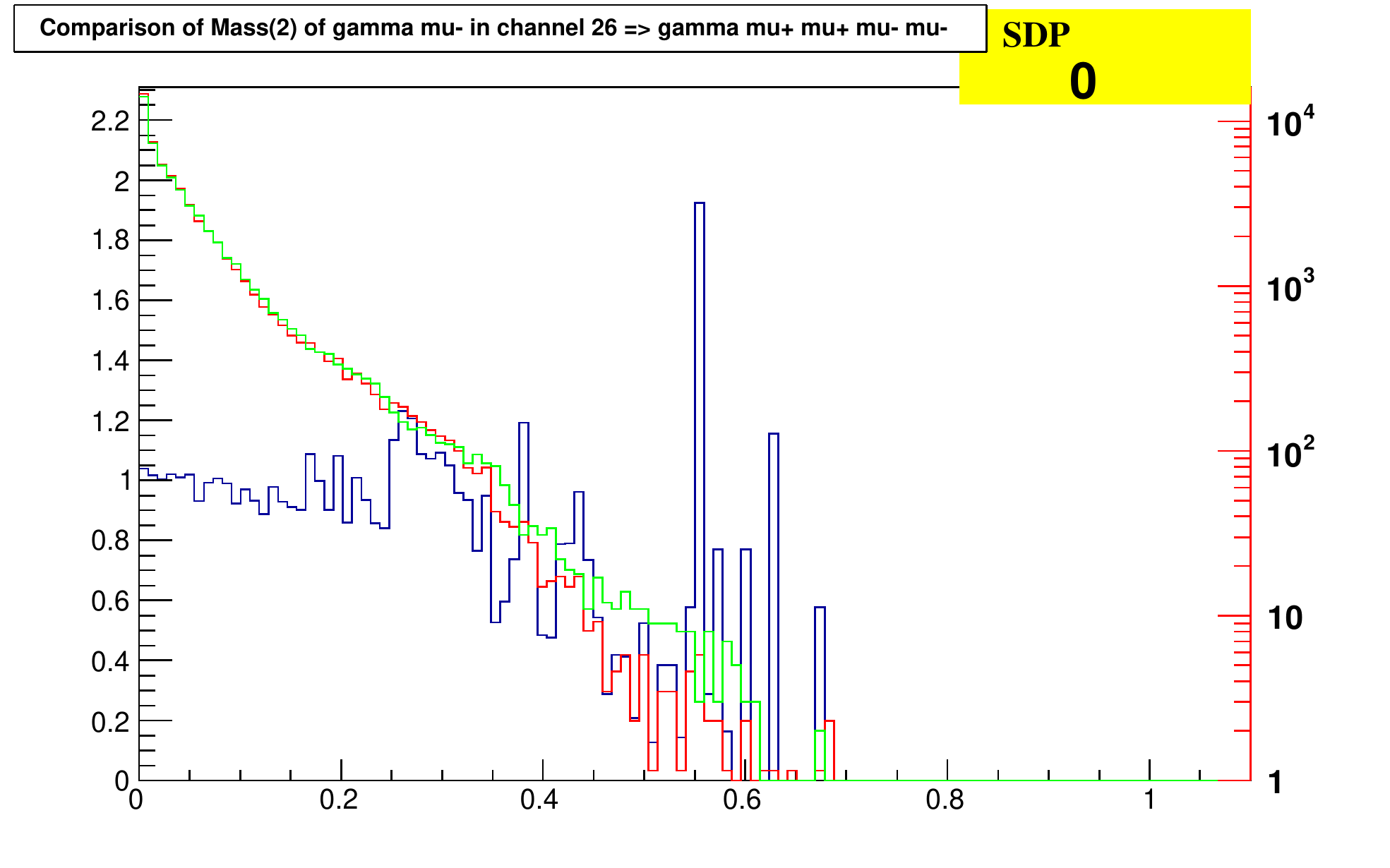}} }
{ \resizebox*{0.49\textwidth}{!}{\includegraphics{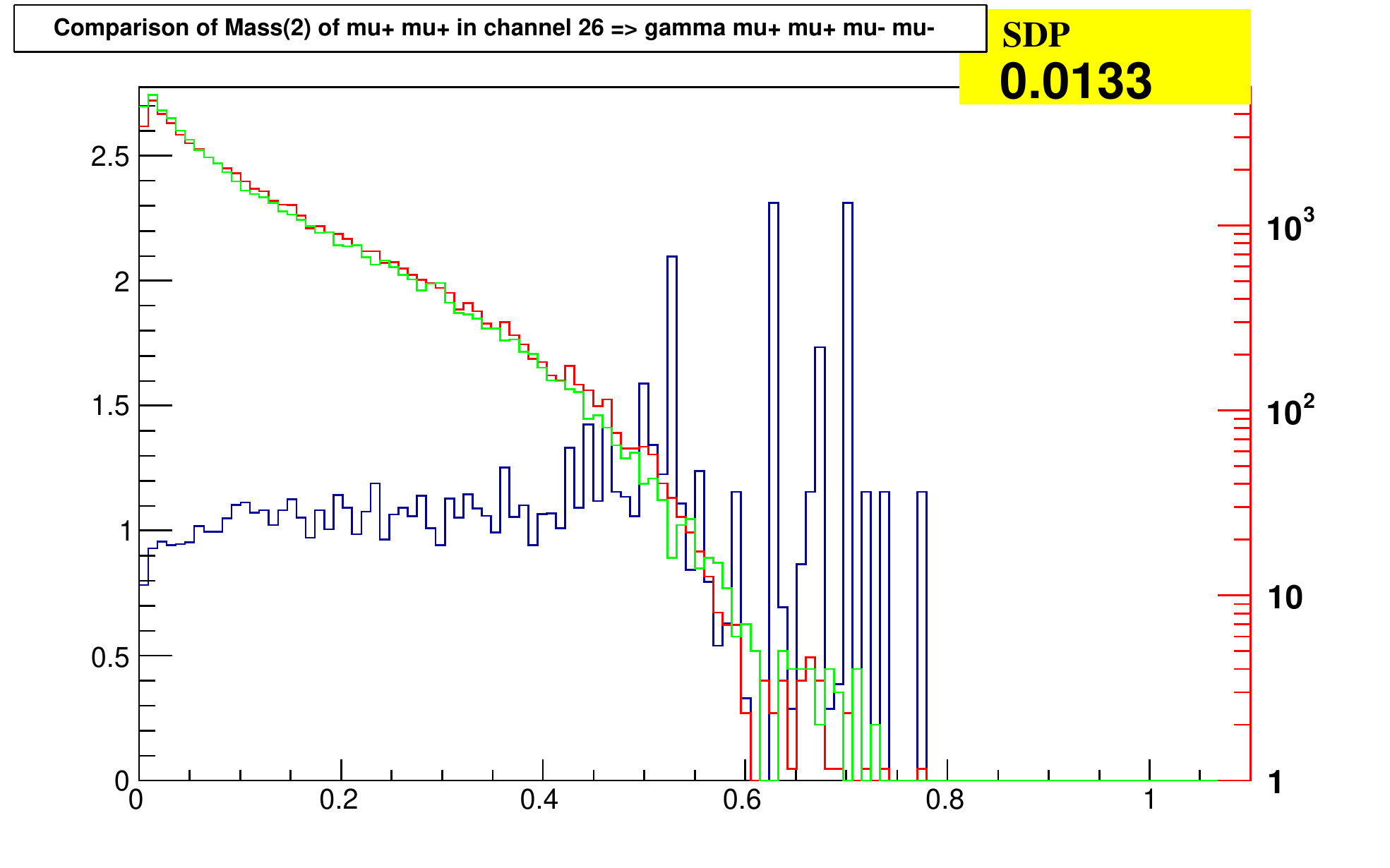}} }
{ \resizebox*{0.49\textwidth}{!}{\includegraphics{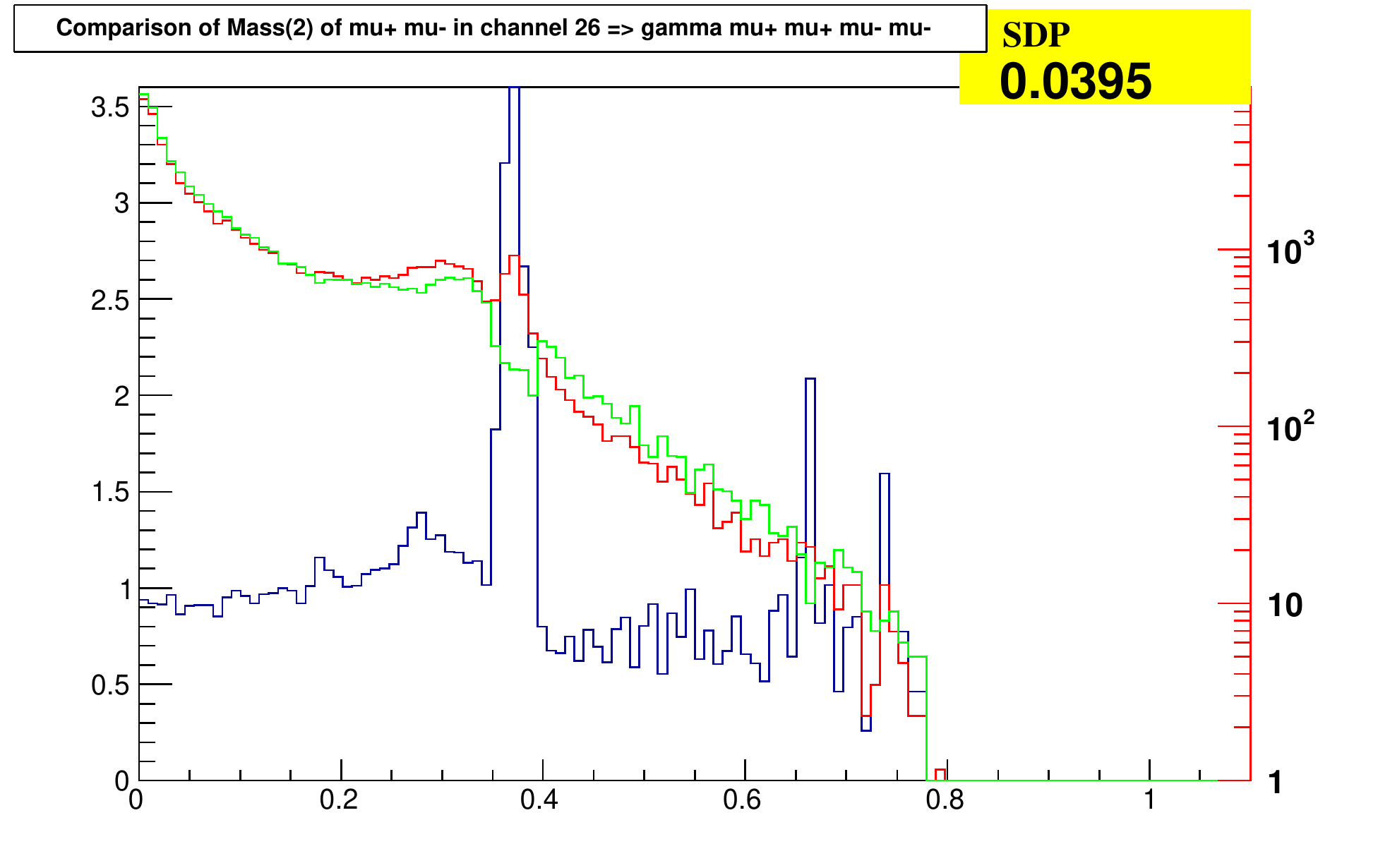}} }
{ \resizebox*{0.49\textwidth}{!}{\includegraphics{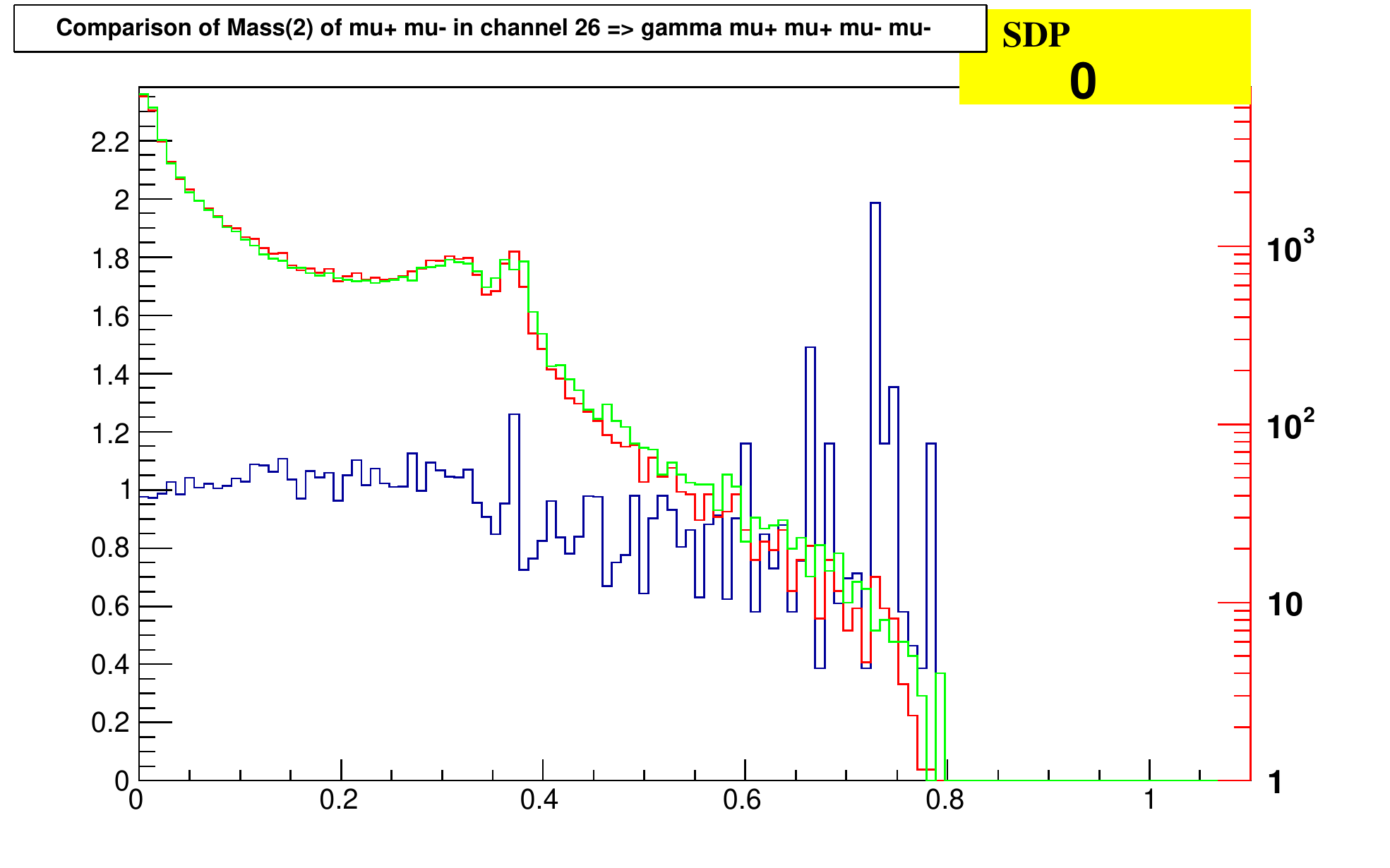}} }
{ \resizebox*{0.49\textwidth}{!}{\includegraphics{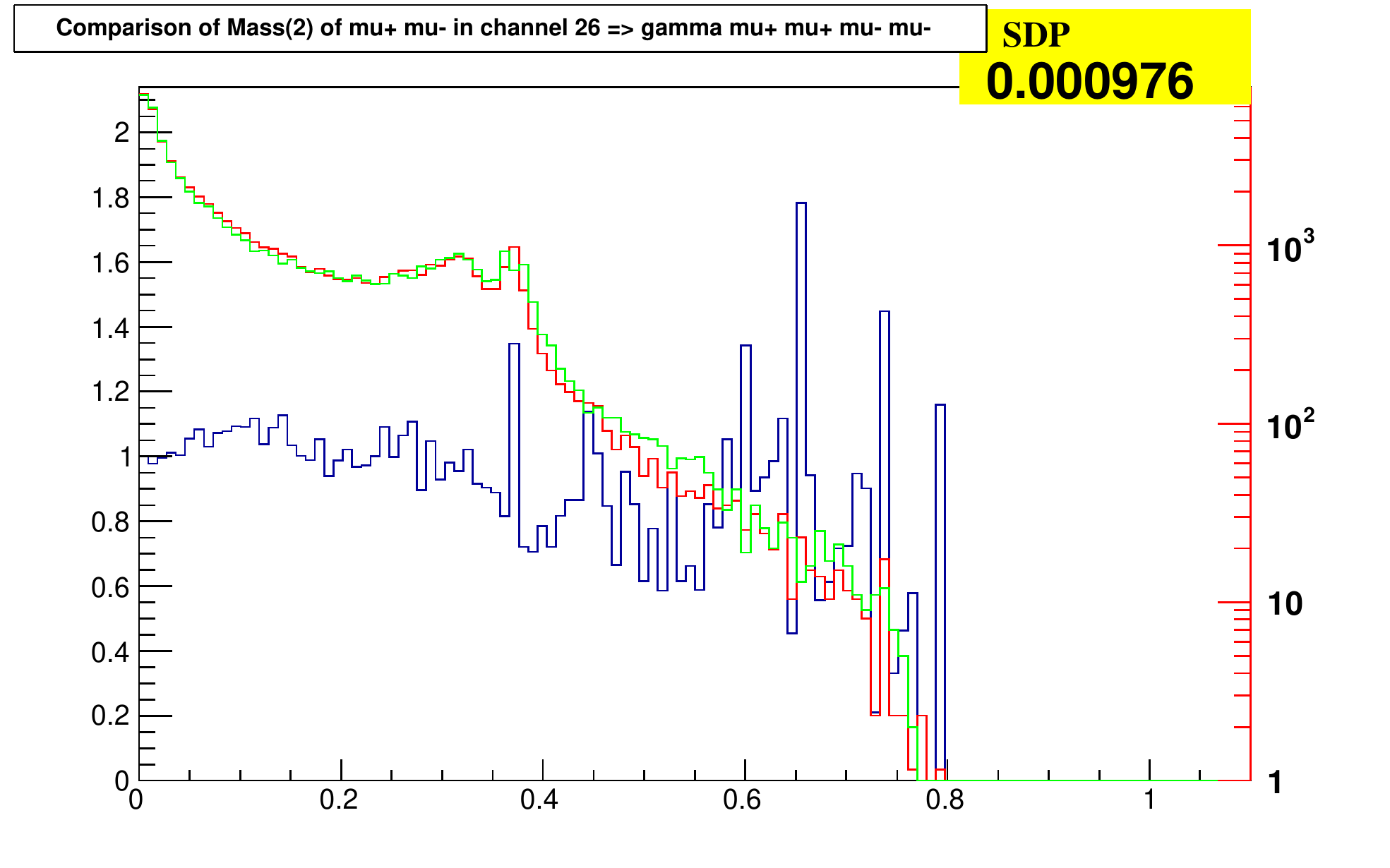}} }
{ \resizebox*{0.49\textwidth}{!}{\includegraphics{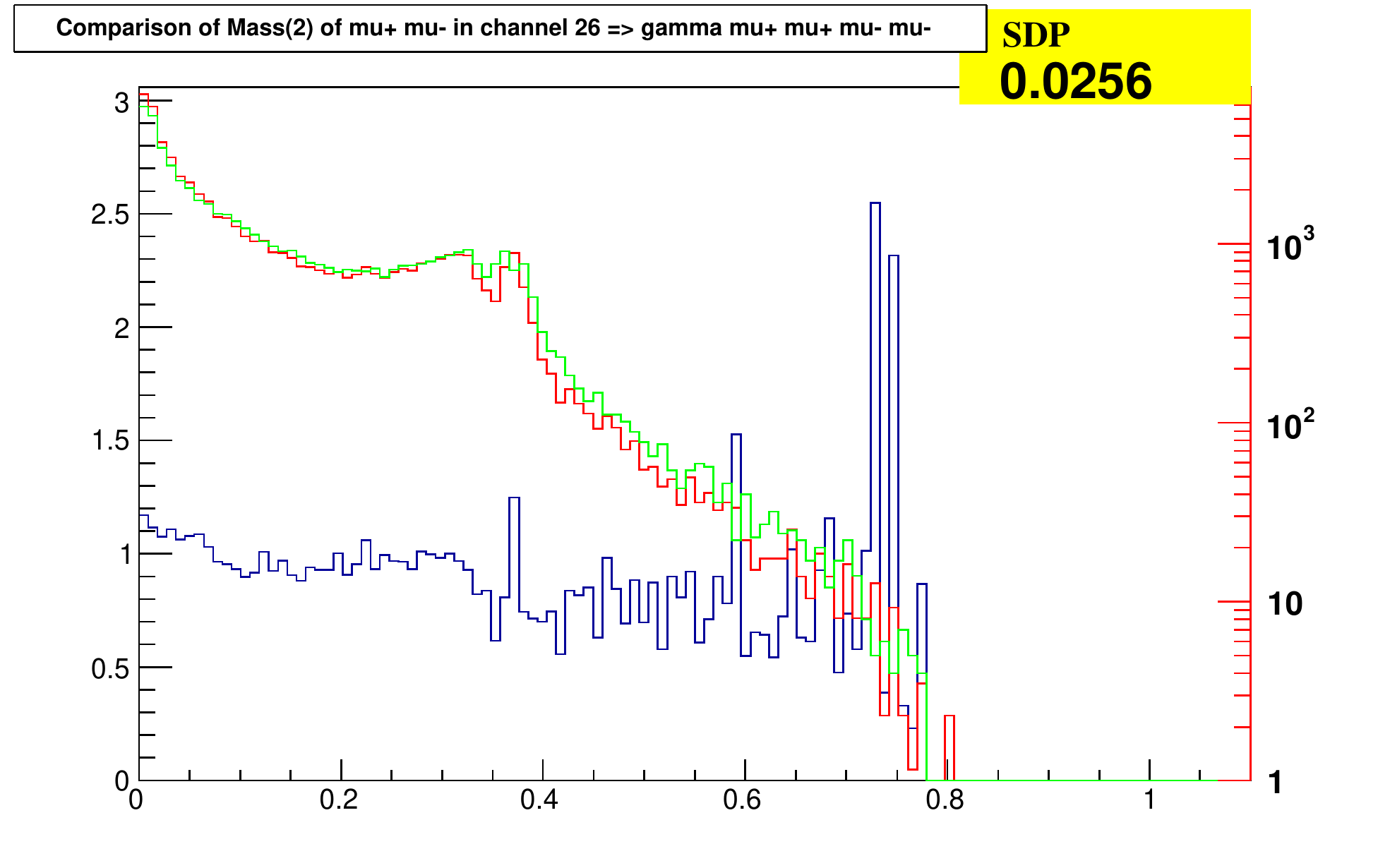}} }
{ \resizebox*{0.49\textwidth}{!}{\includegraphics{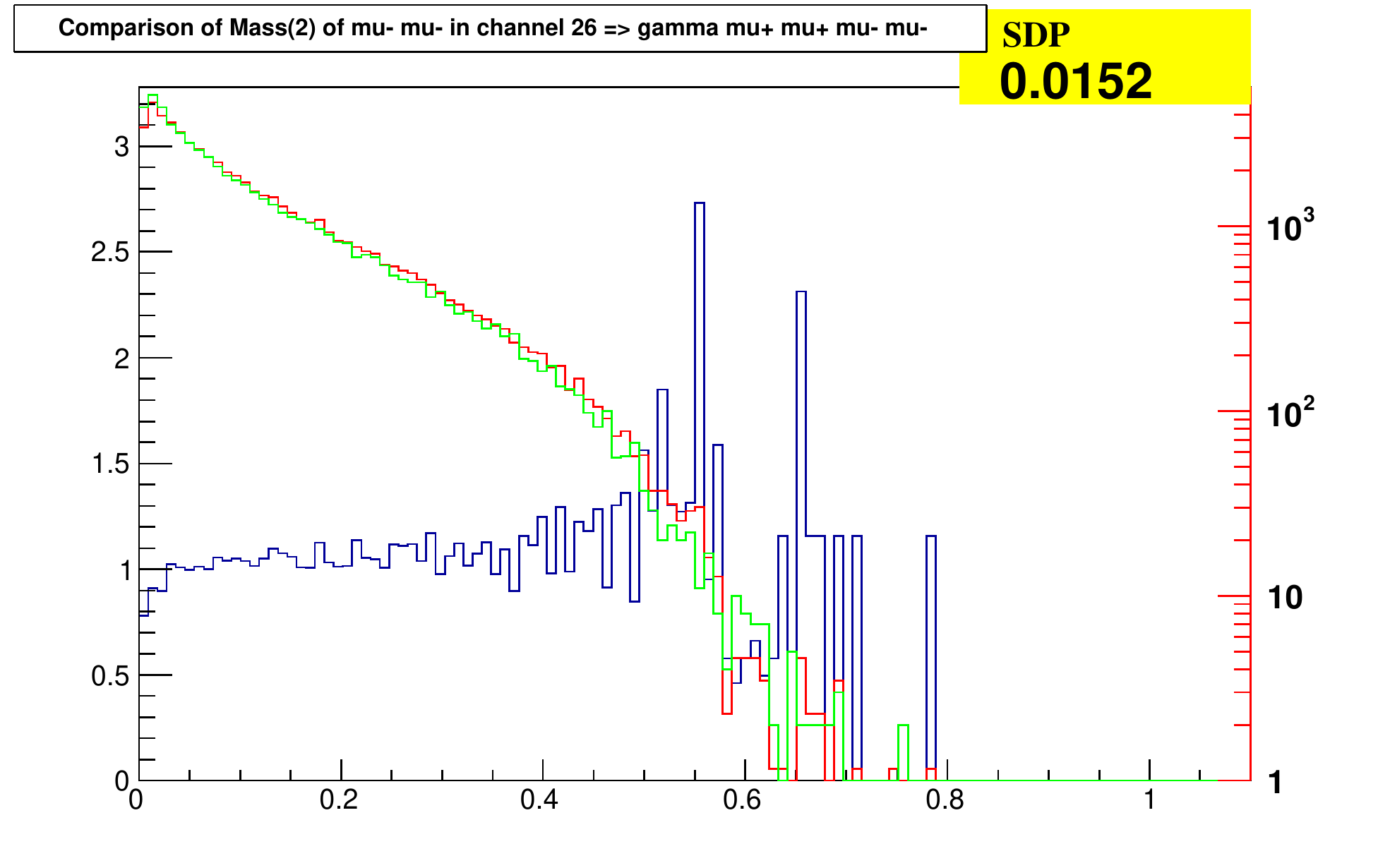}} }
{ \resizebox*{0.49\textwidth}{!}{\includegraphics{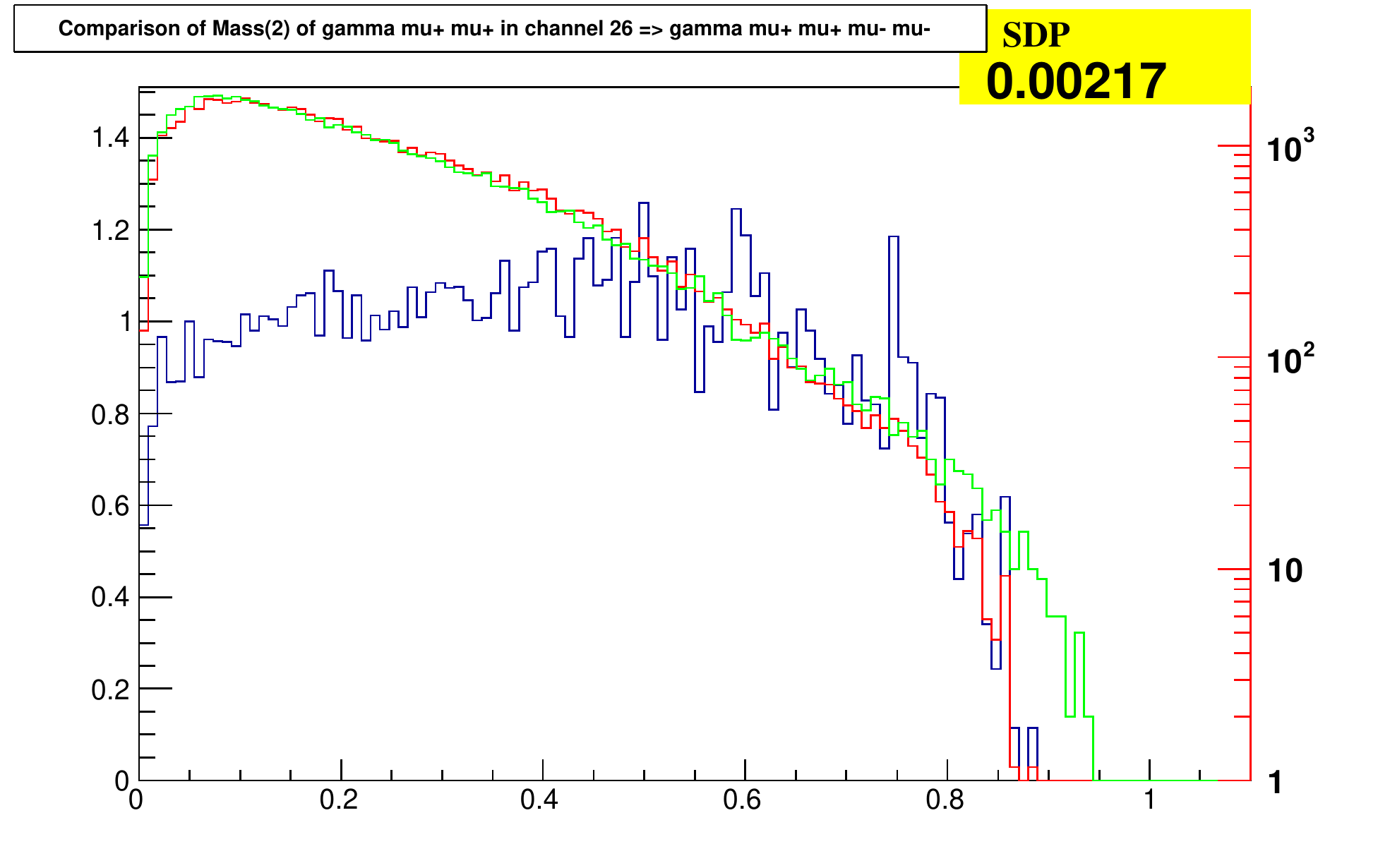}} }
{ \resizebox*{0.49\textwidth}{!}{\includegraphics{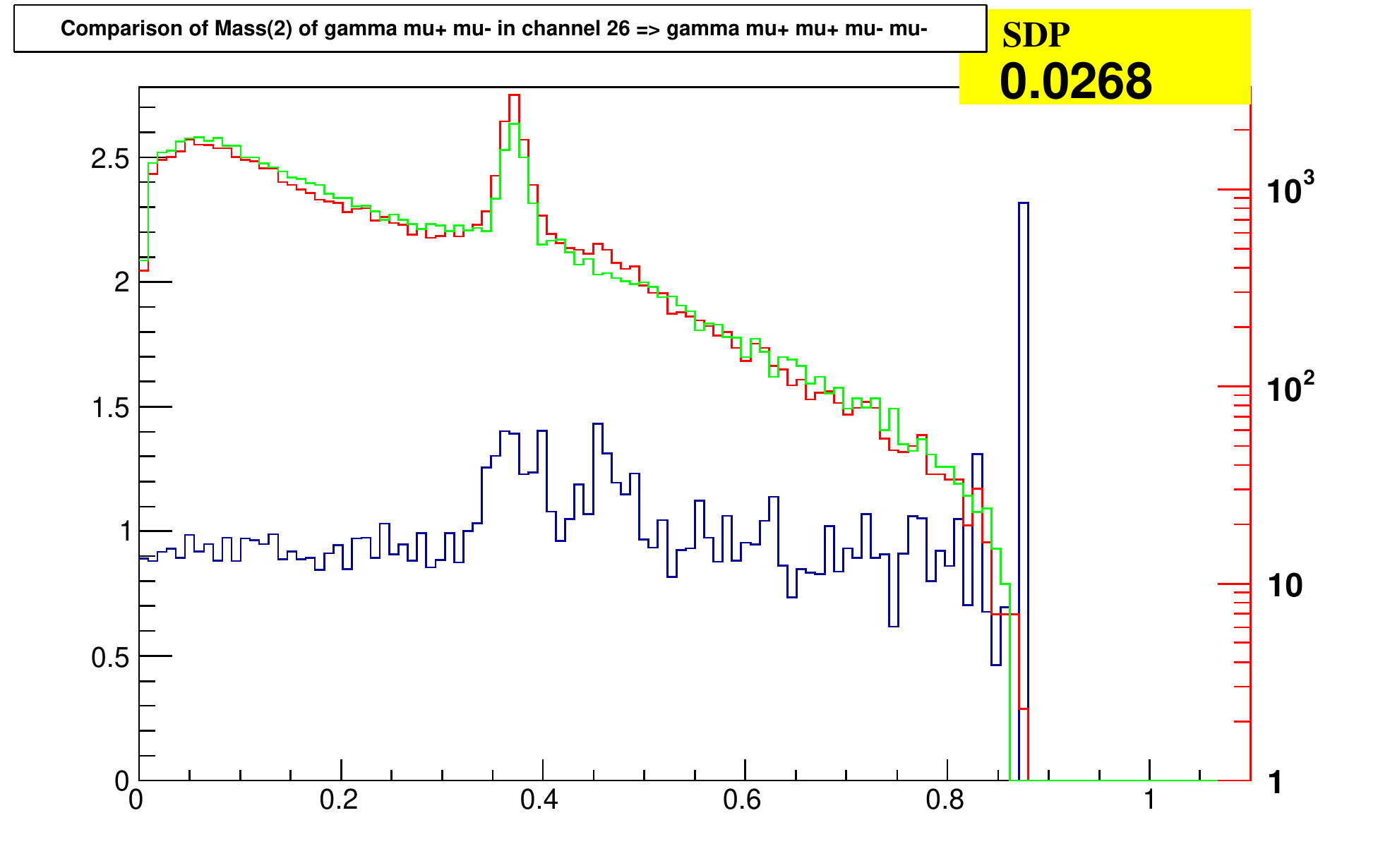}} }
{ \resizebox*{0.49\textwidth}{!}{\includegraphics{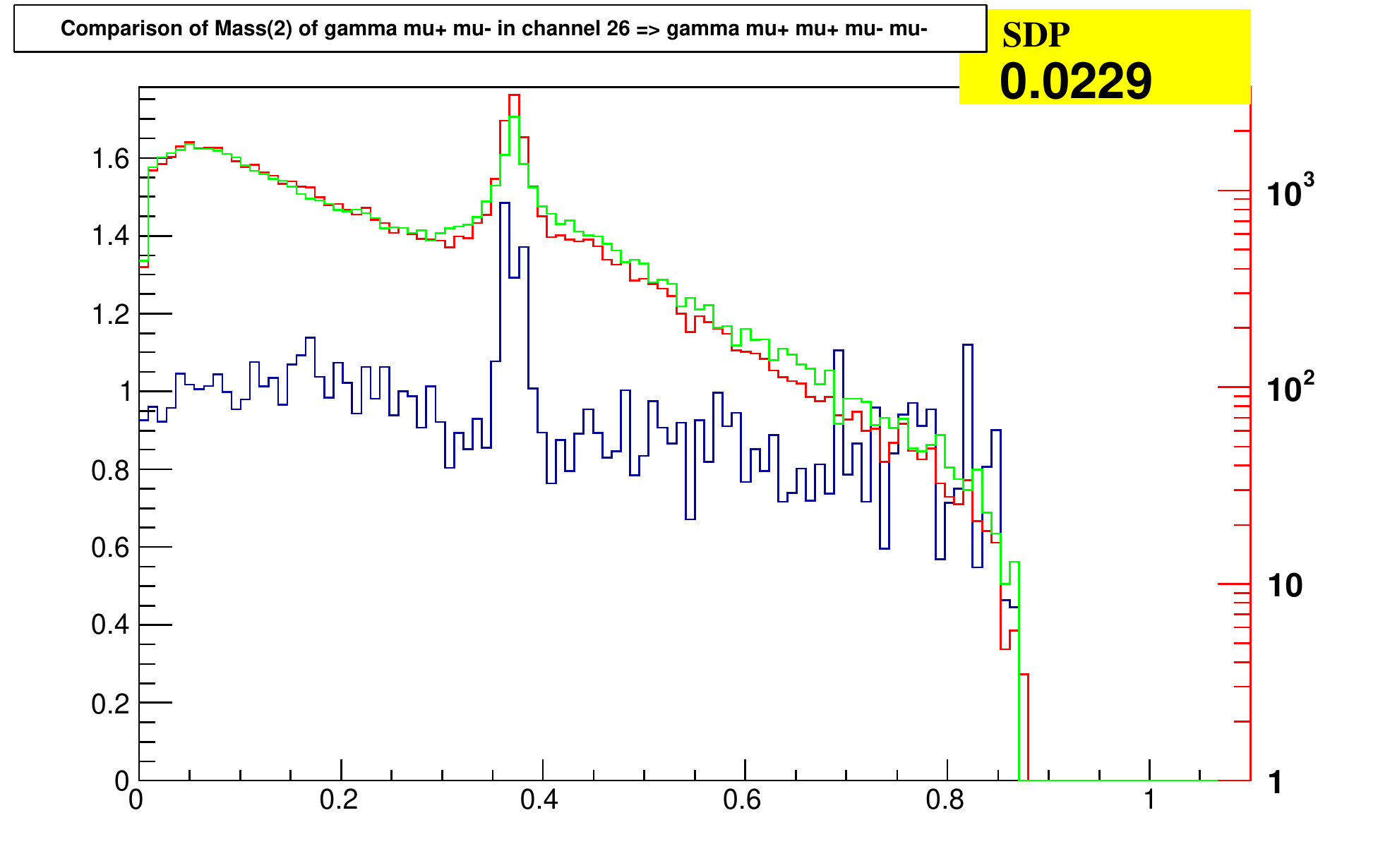}} }
{ \resizebox*{0.49\textwidth}{!}{\includegraphics{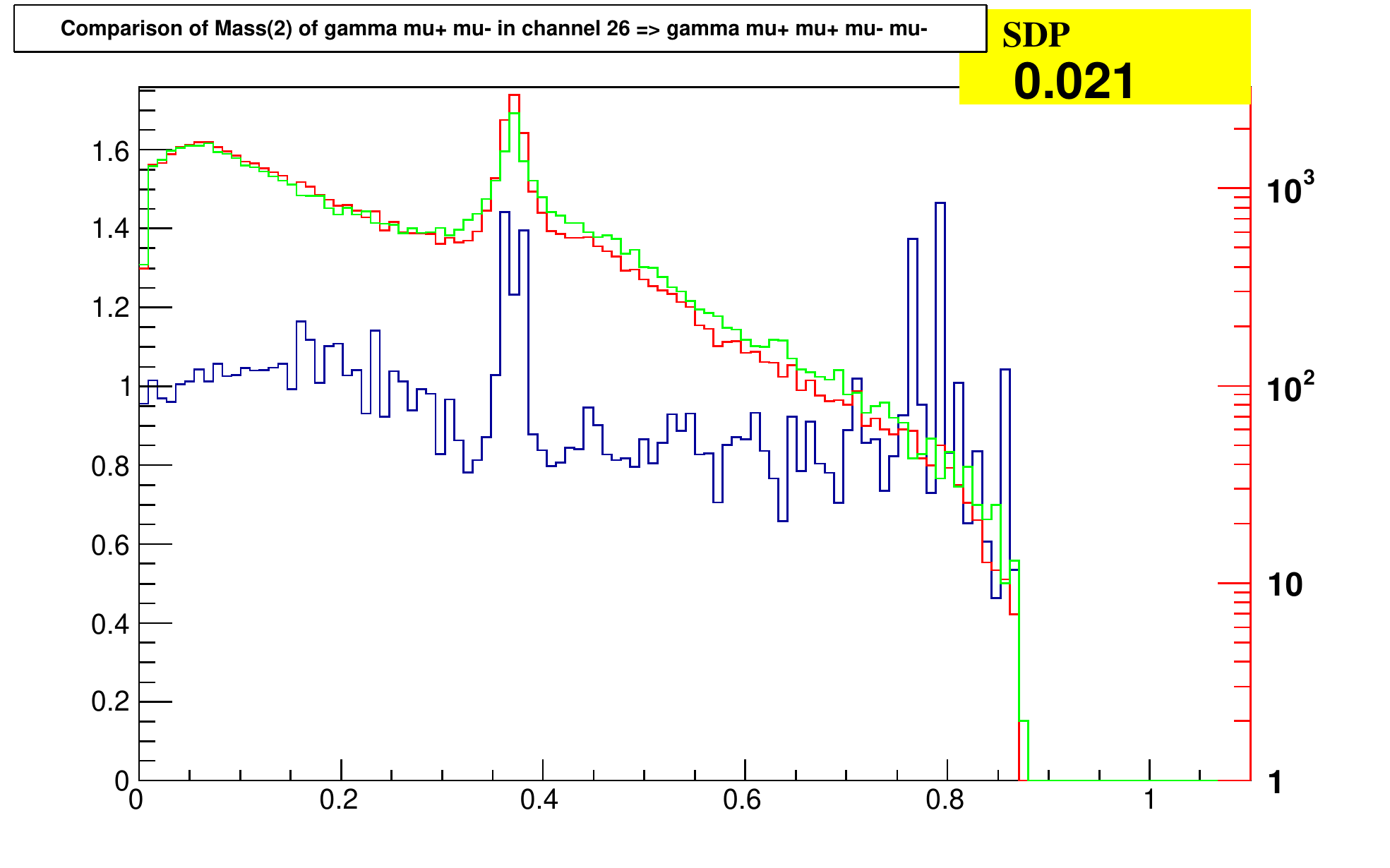}} }
{ \resizebox*{0.49\textwidth}{!}{\includegraphics{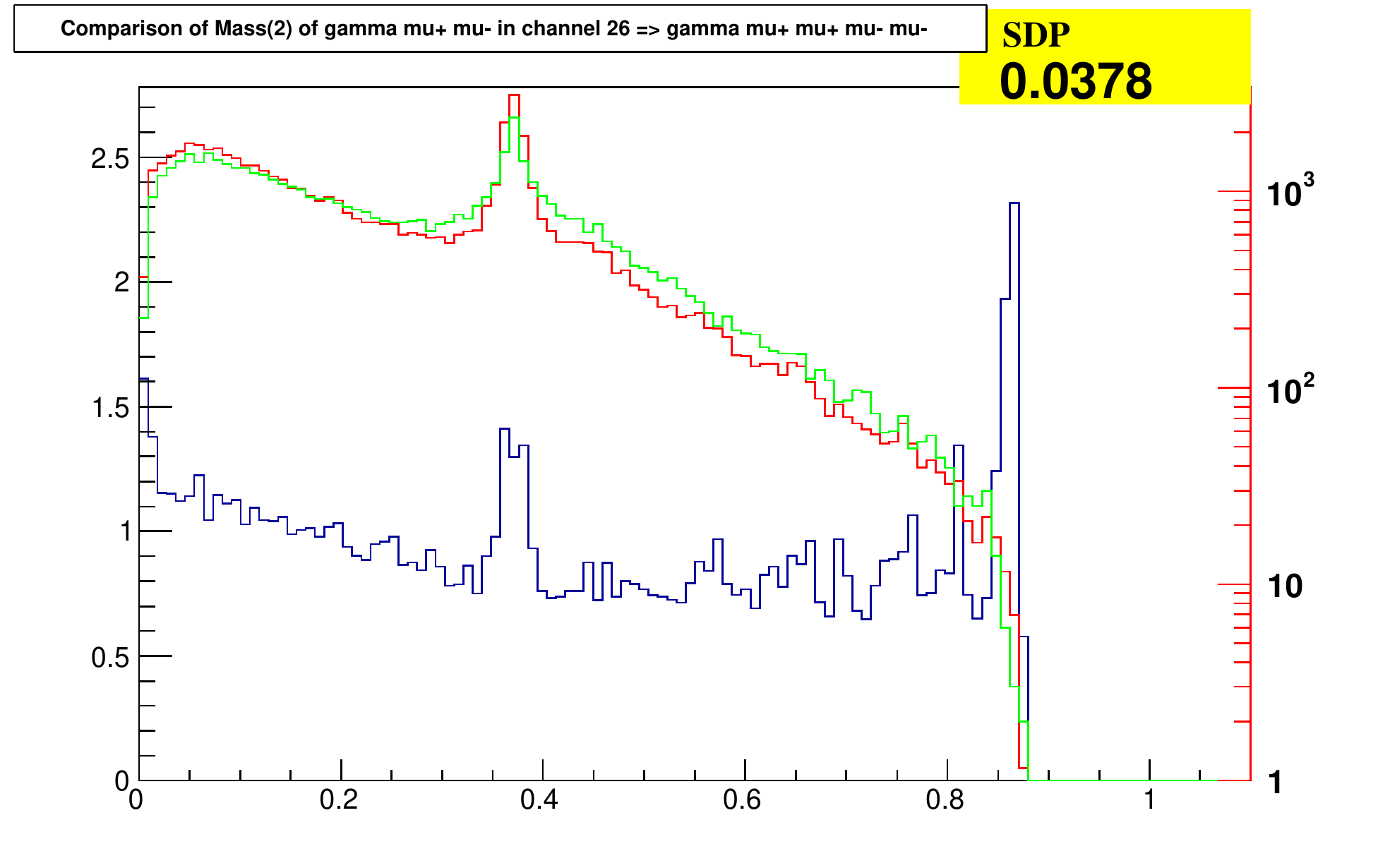}} }
{ \resizebox*{0.49\textwidth}{!}{\includegraphics{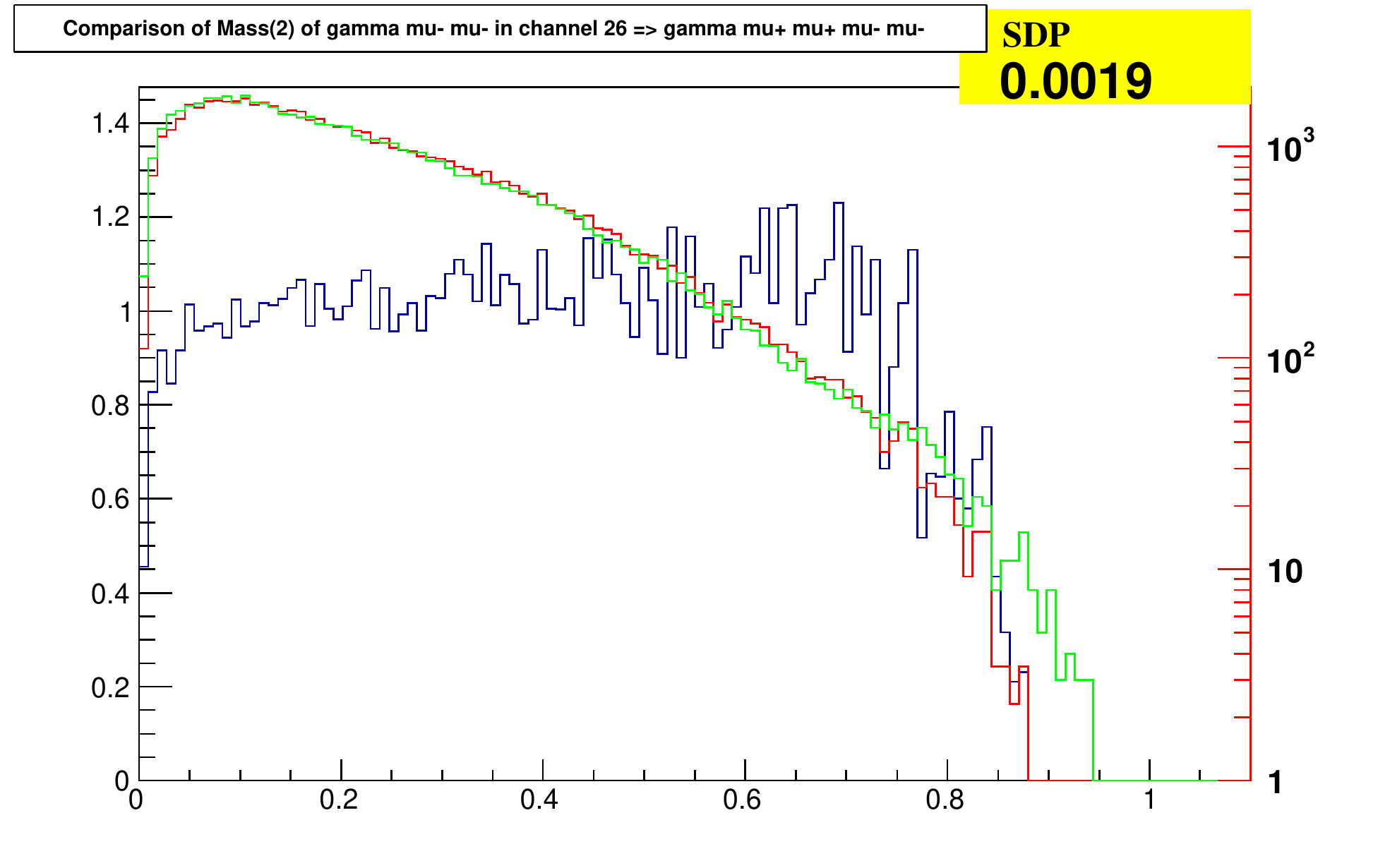}} }
{ \resizebox*{0.49\textwidth}{!}{\includegraphics{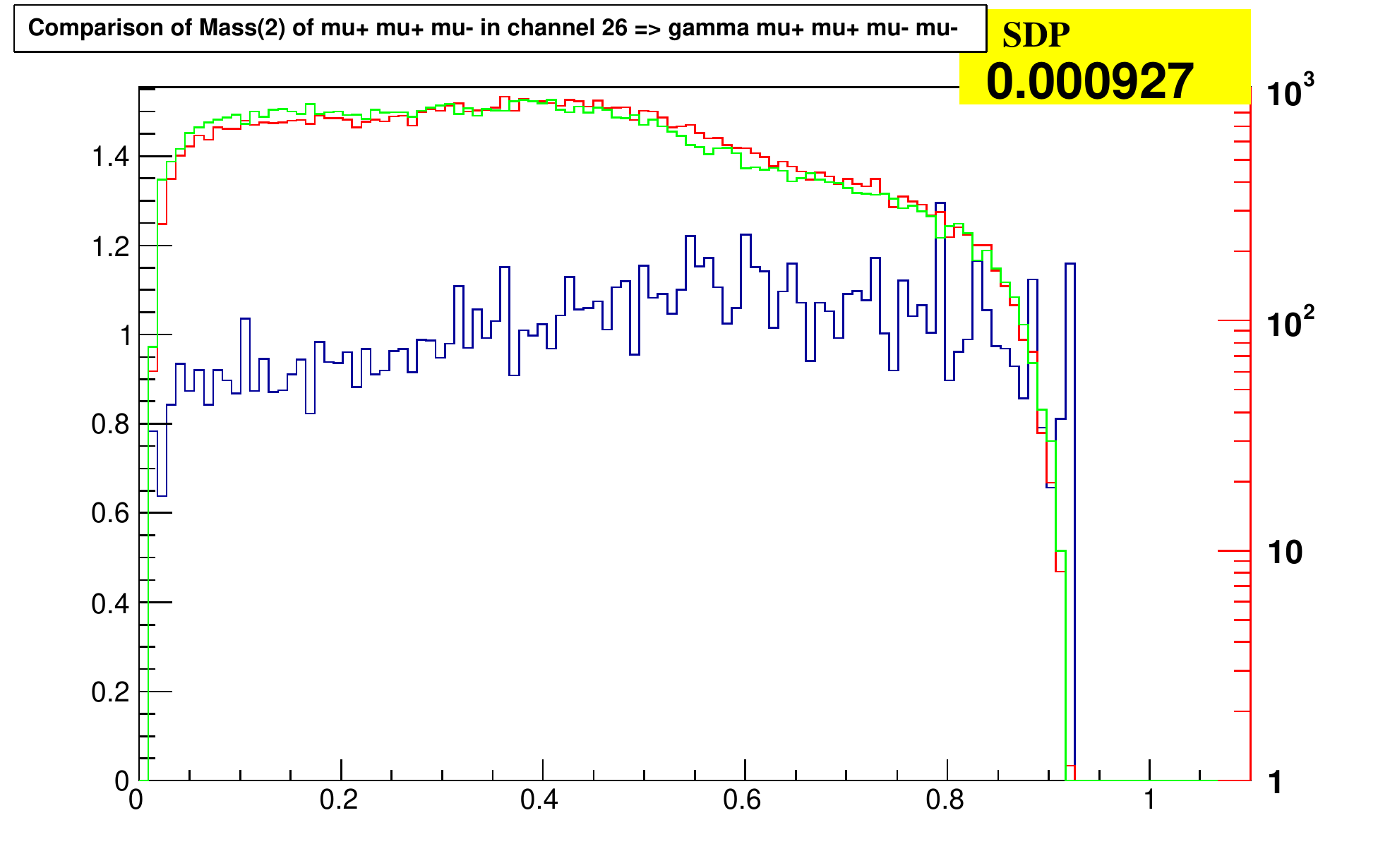}} }
{ \resizebox*{0.49\textwidth}{!}{\includegraphics{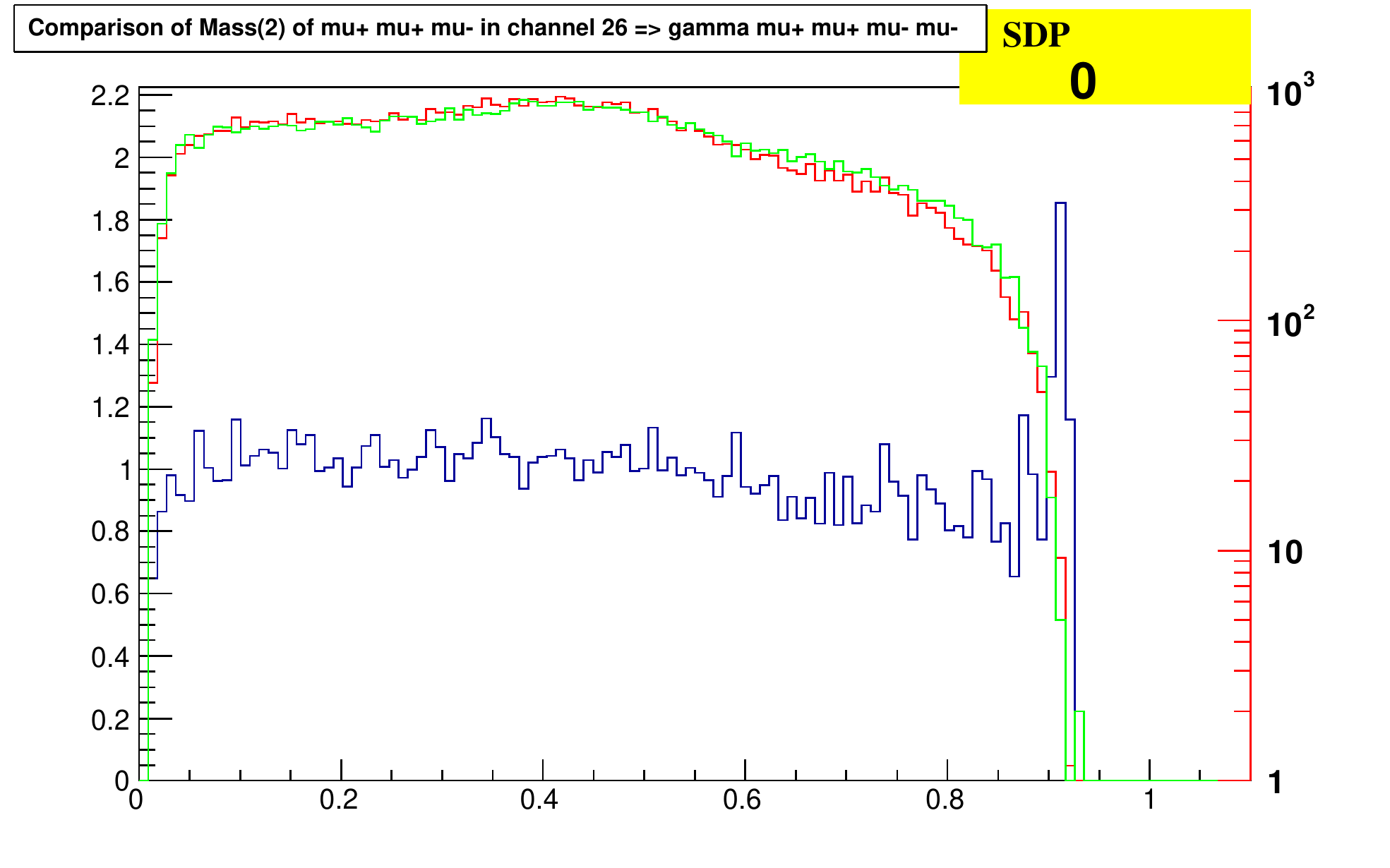}} }
{ \resizebox*{0.49\textwidth}{!}{\includegraphics{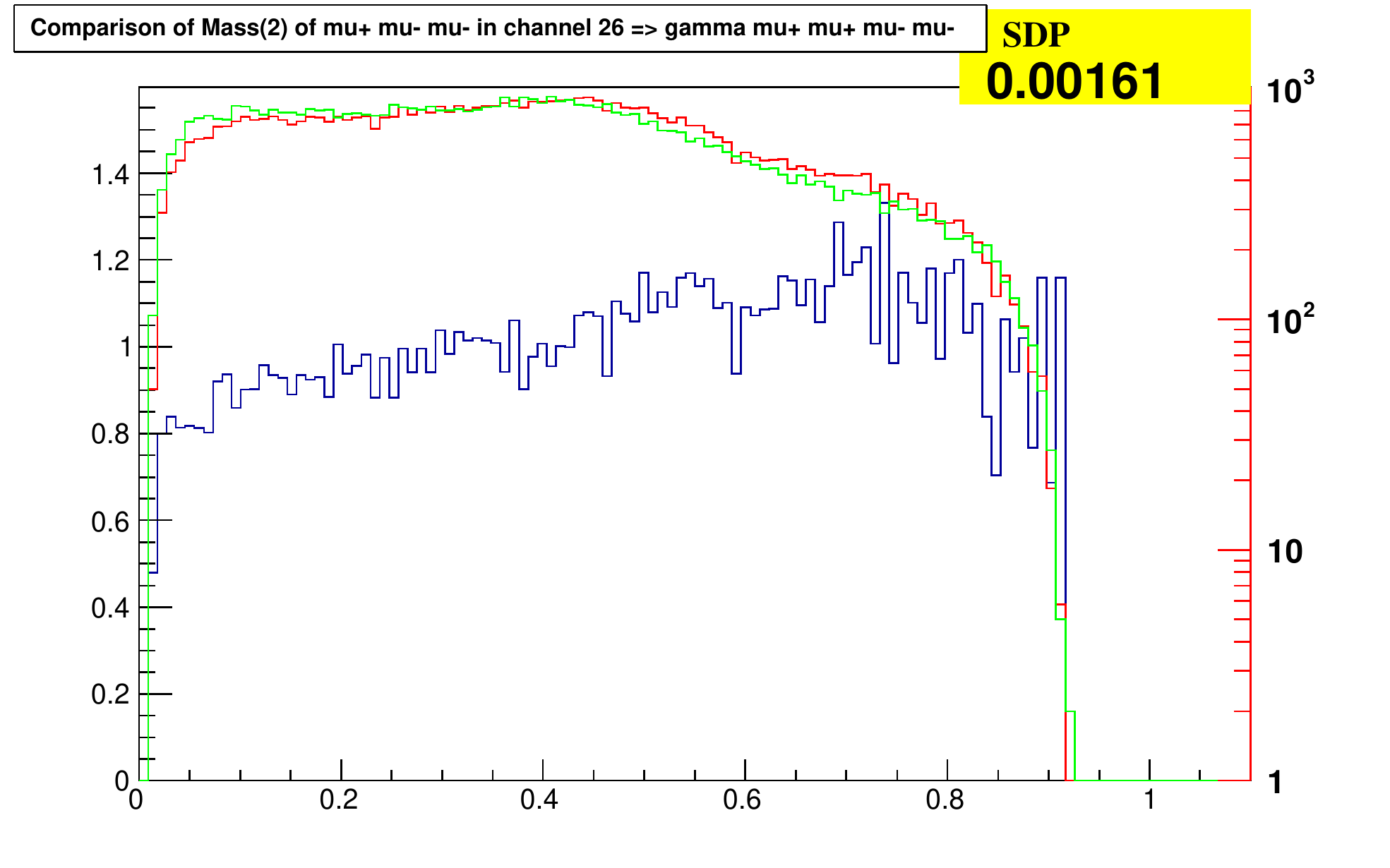}} }
{ \resizebox*{0.49\textwidth}{!}{\includegraphics{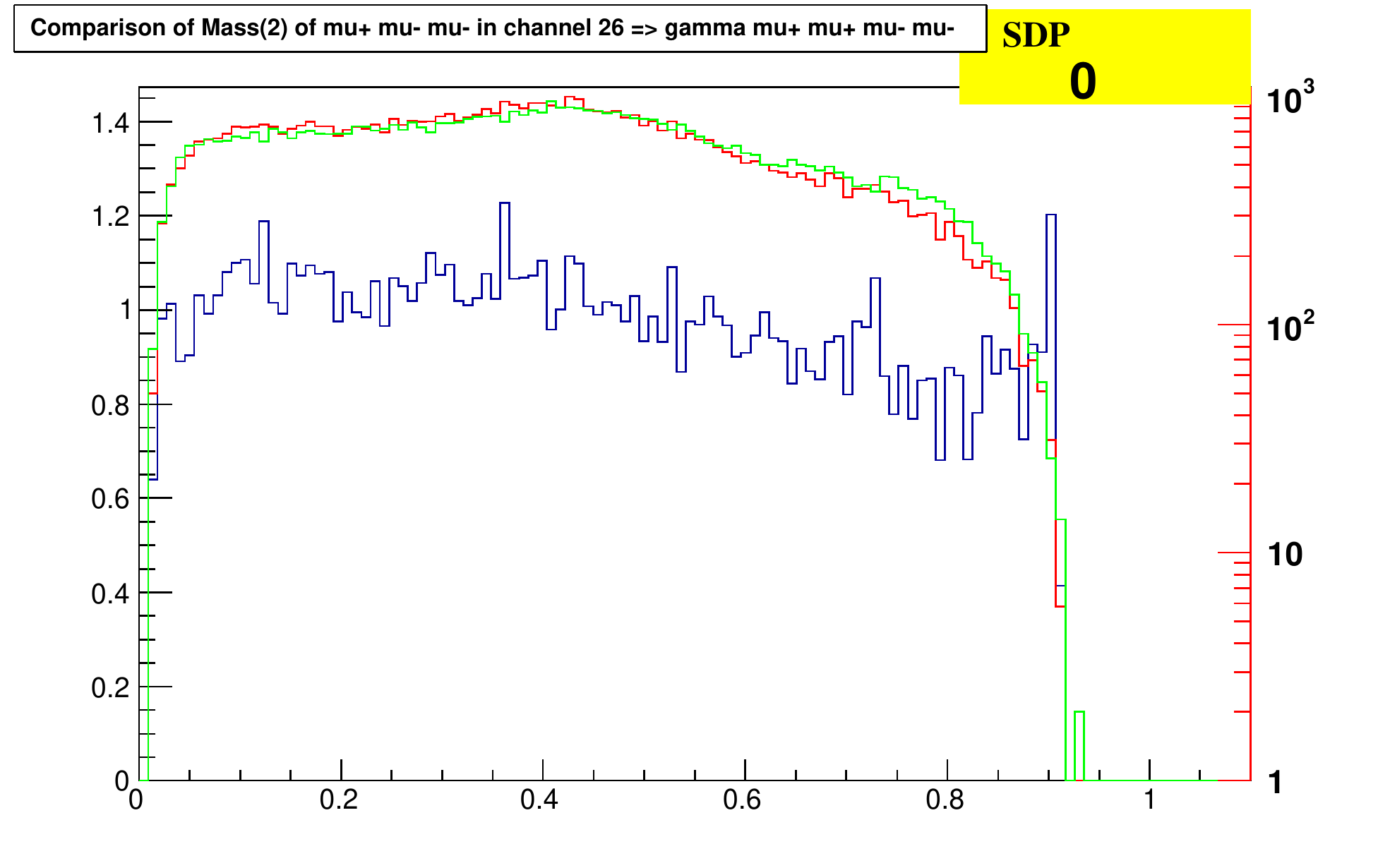}} }
{ \resizebox*{0.49\textwidth}{!}{\includegraphics{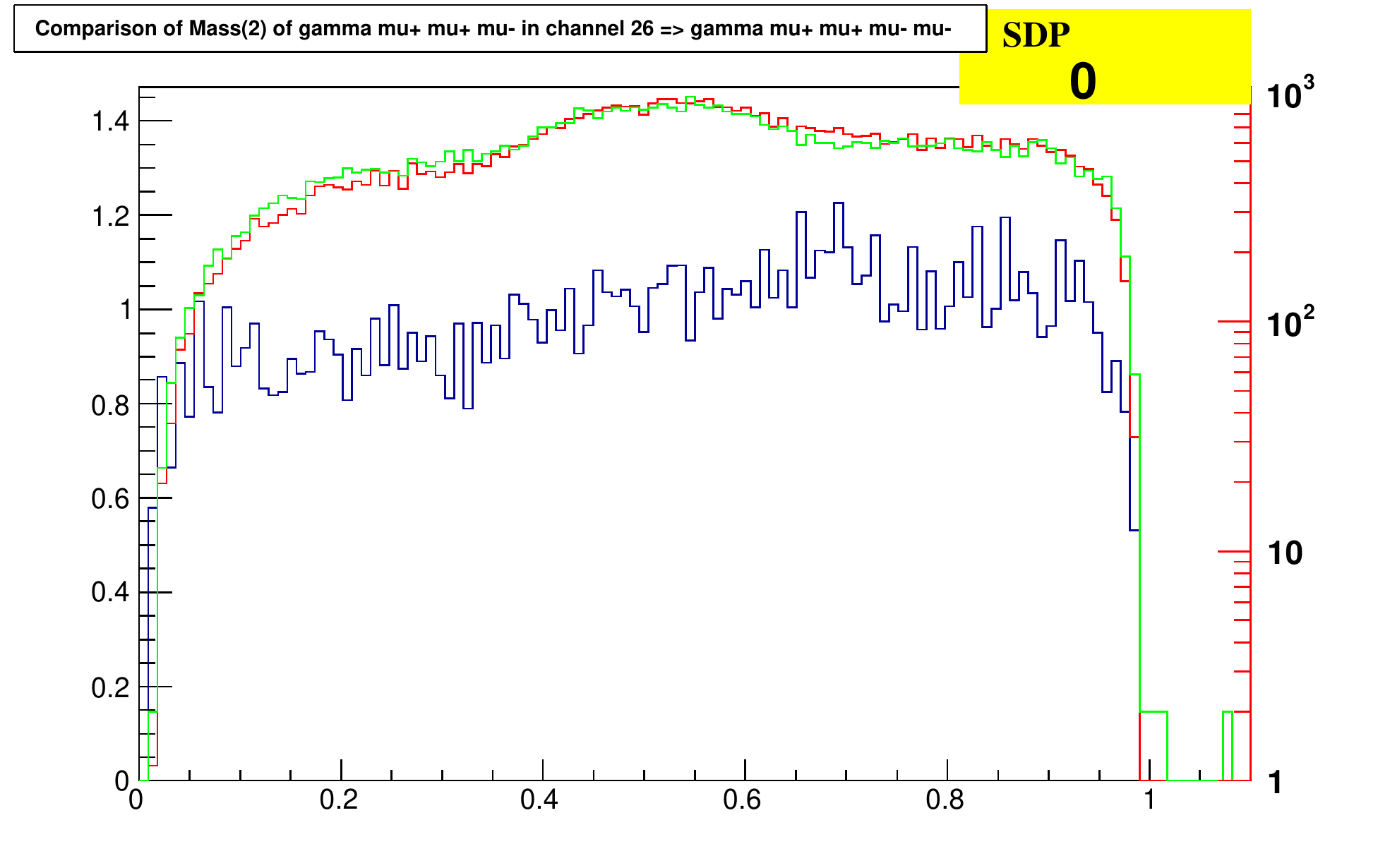}} }
{ \resizebox*{0.49\textwidth}{!}{\includegraphics{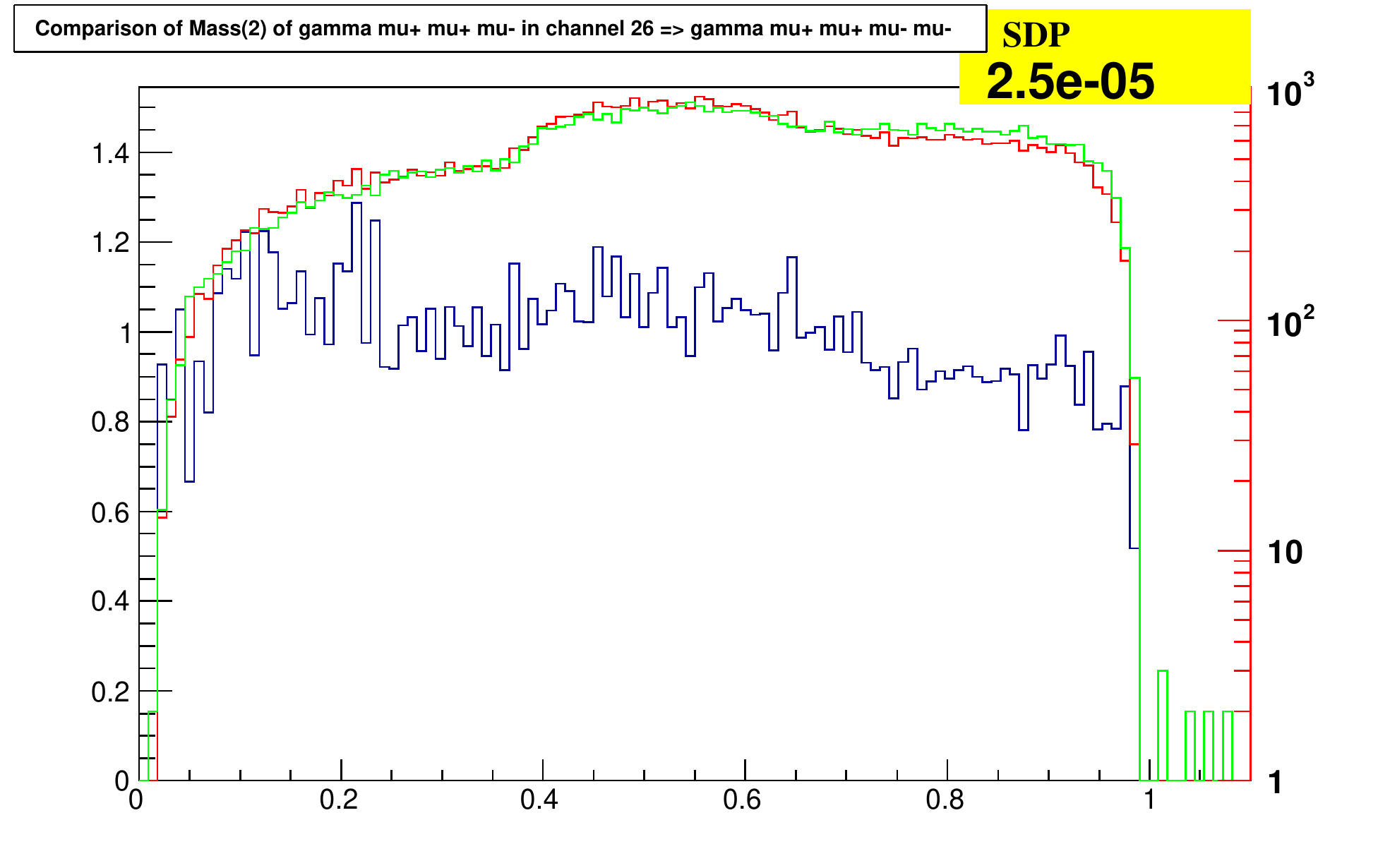}} }
{ \resizebox*{0.49\textwidth}{!}{\includegraphics{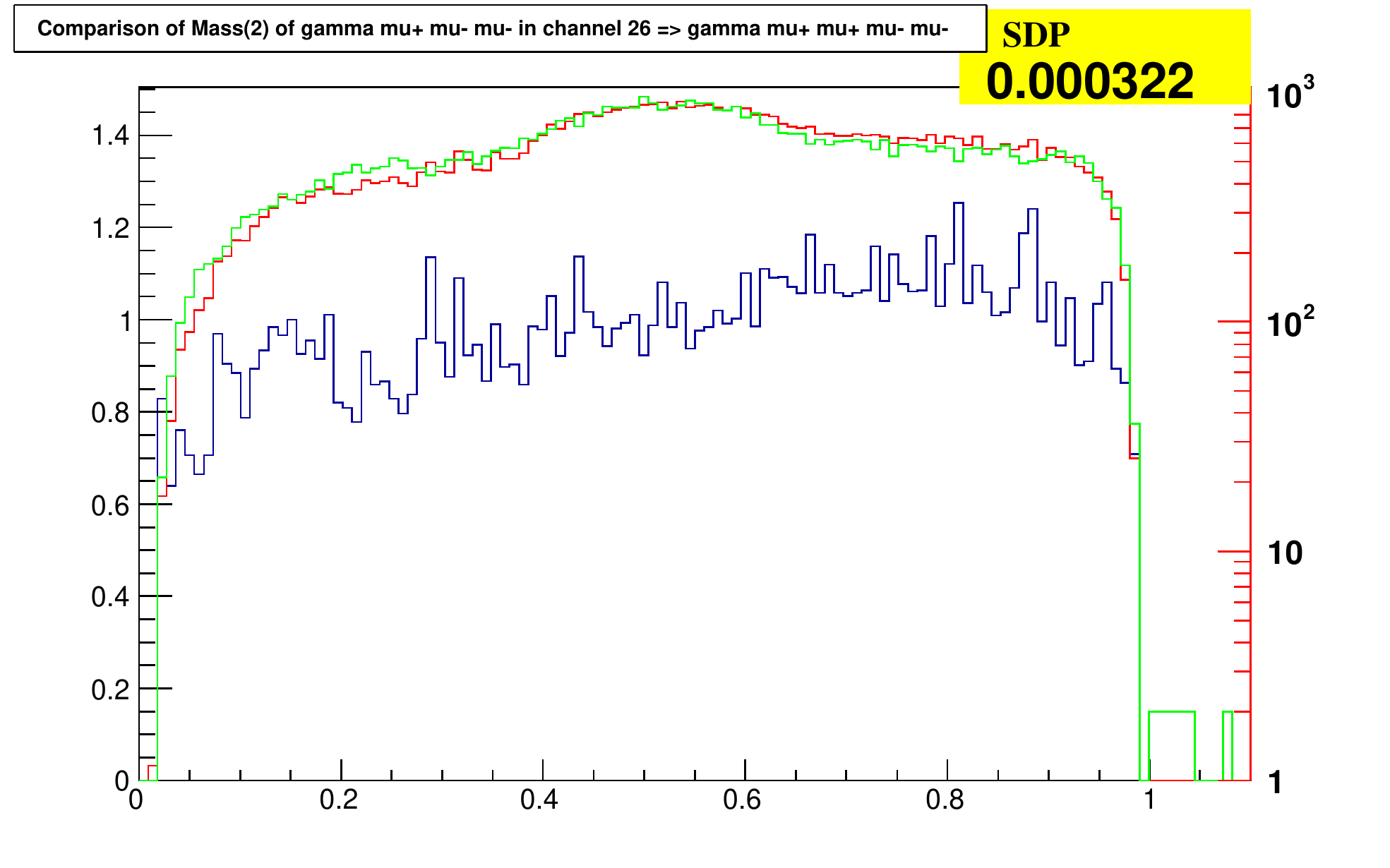}} }
{ \resizebox*{0.49\textwidth}{!}{\includegraphics{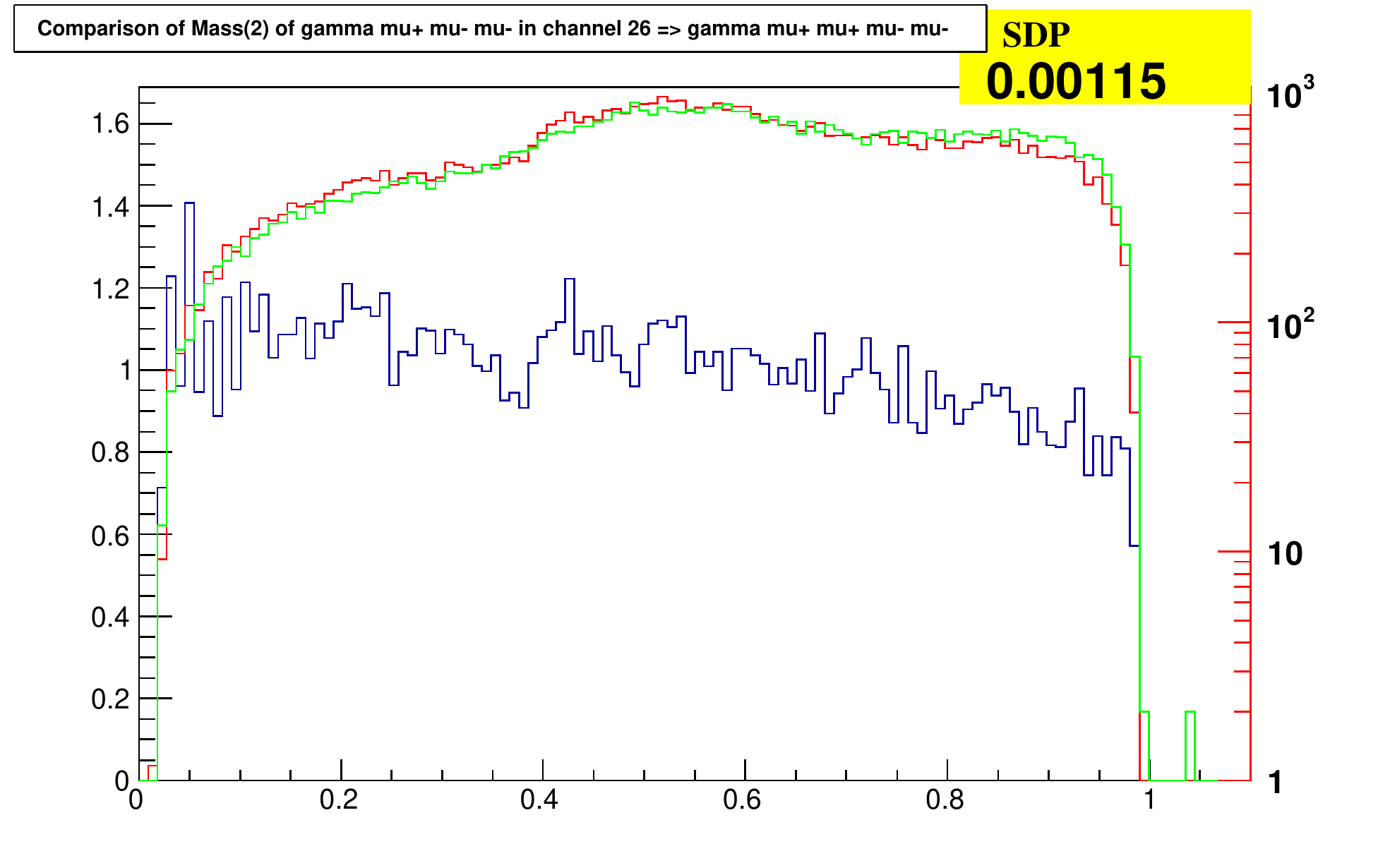}} }
{ \resizebox*{0.49\textwidth}{!}{\includegraphics{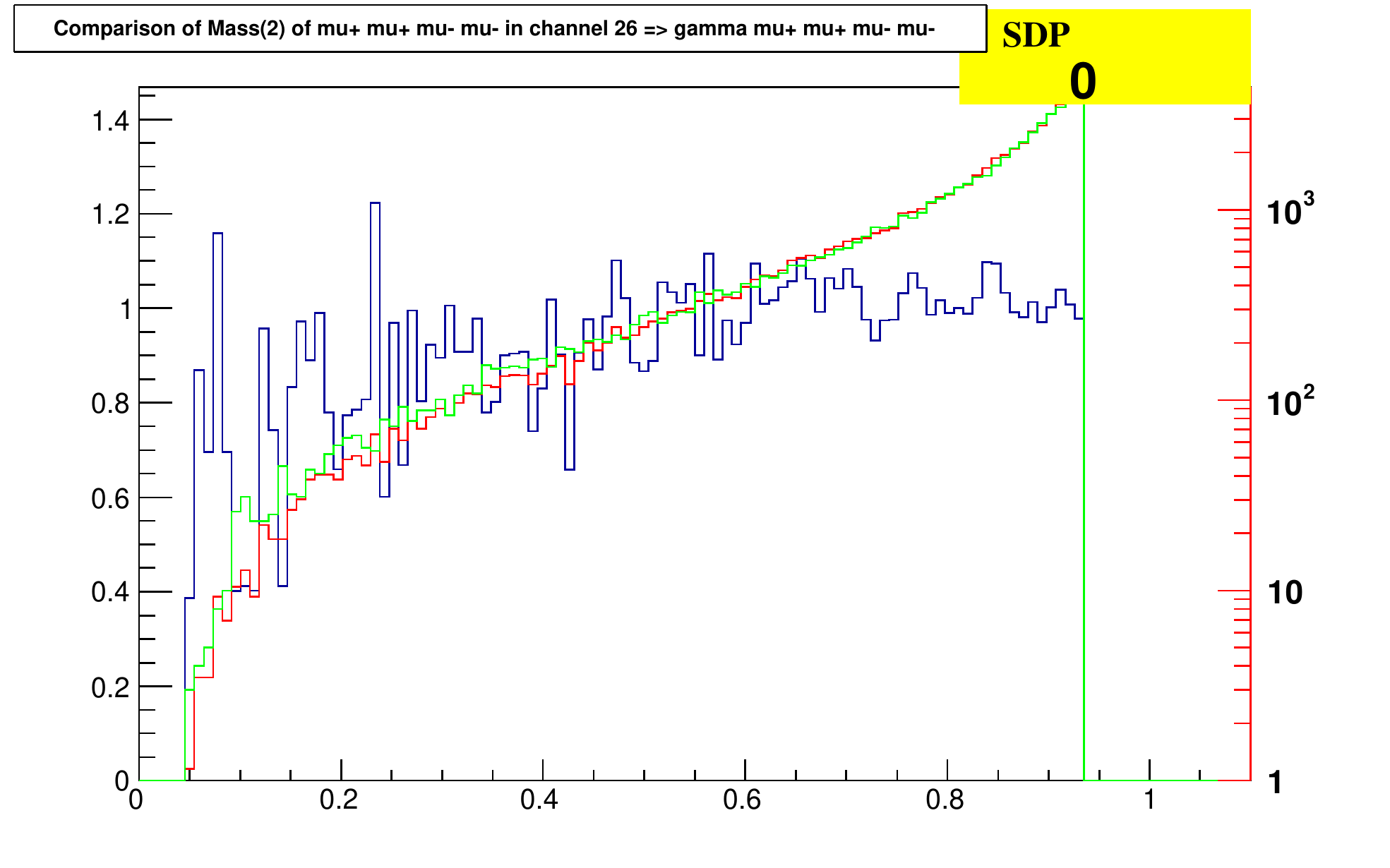}} }
{ \resizebox*{0.49\textwidth}{!}{\includegraphics{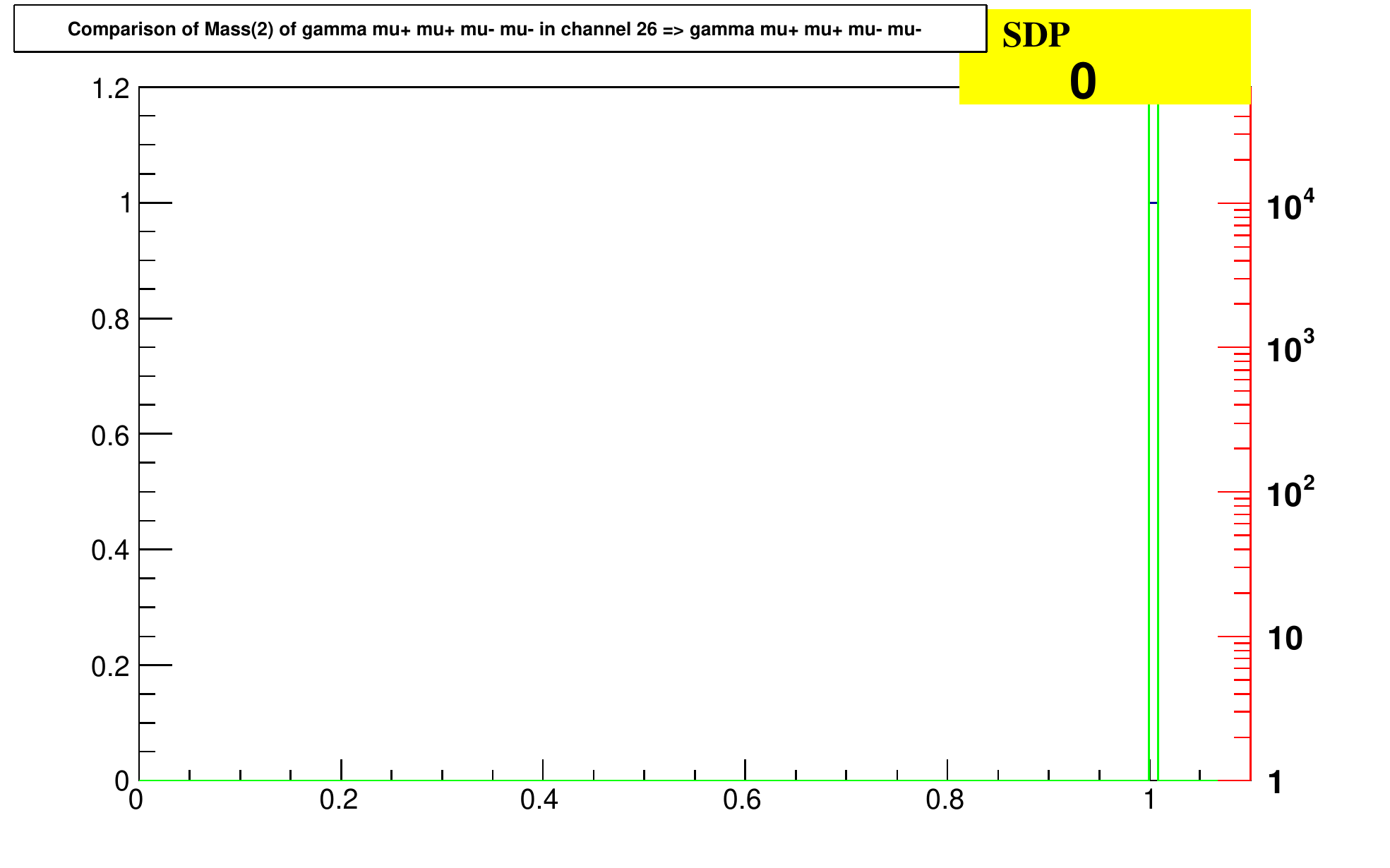}} }

%==================================
\subsection{{\tt MC-tester}: $q \bar q \rightarrow \gamma \mu^{+} \mu^{+} \mu^{-} \mu^{-}$ at $\sqrt{s}=240$ GeV}
\label{sec:mc-tester240}
%==================================
{ \resizebox*{0.49\textwidth}{!}{\includegraphics{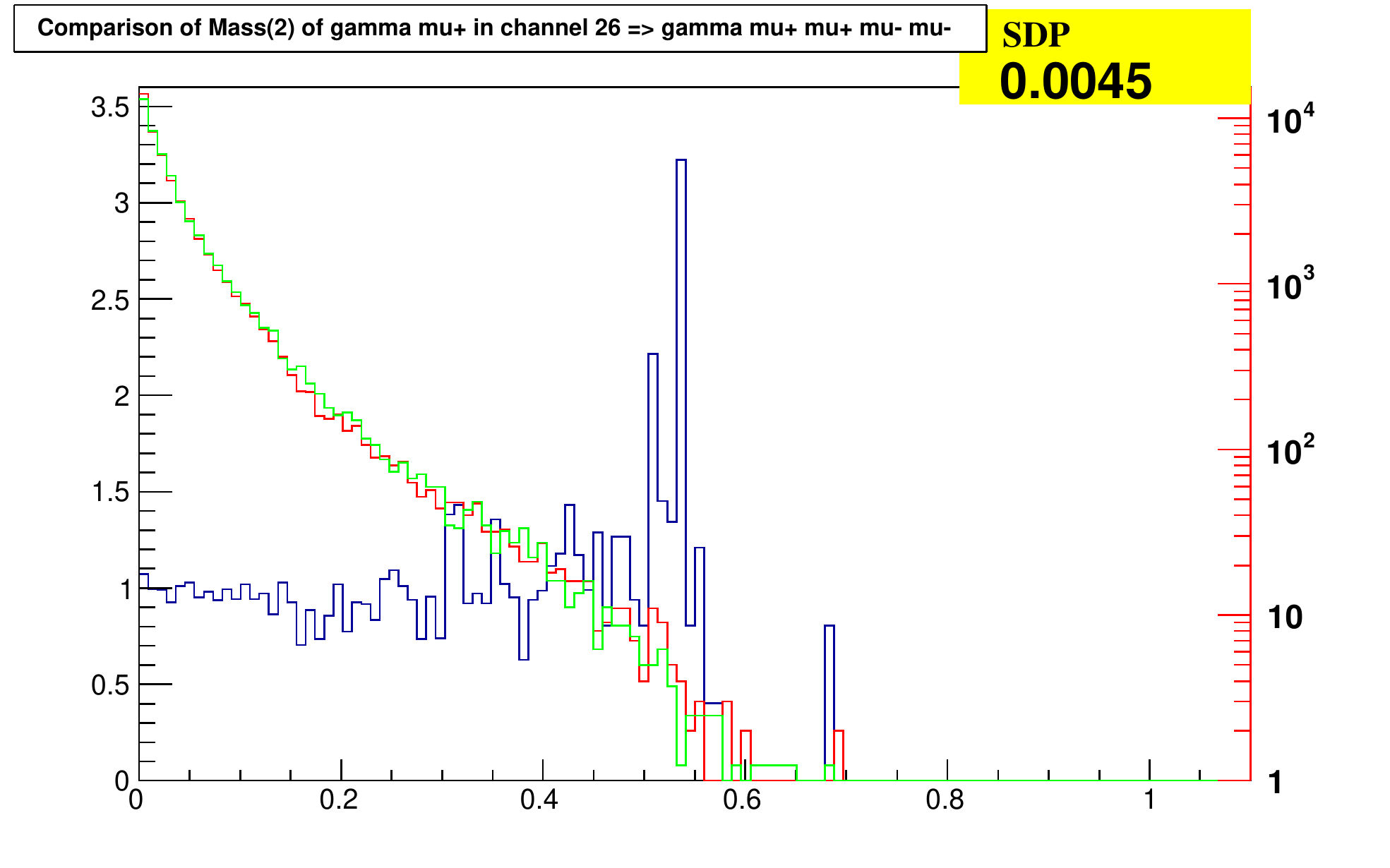}} }
{ \resizebox*{0.49\textwidth}{!}{\includegraphics{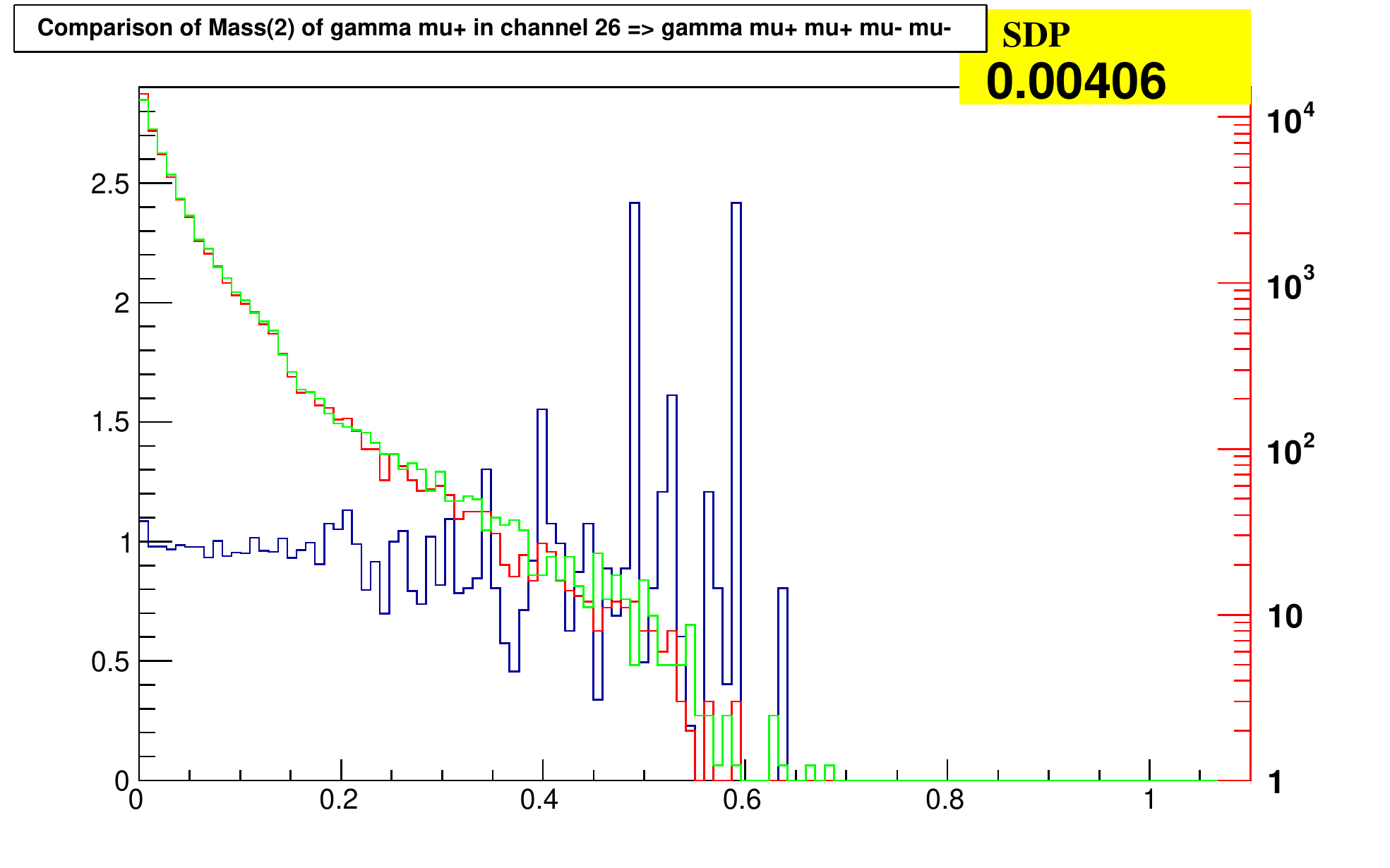}} }
{ \resizebox*{0.49\textwidth}{!}{\includegraphics{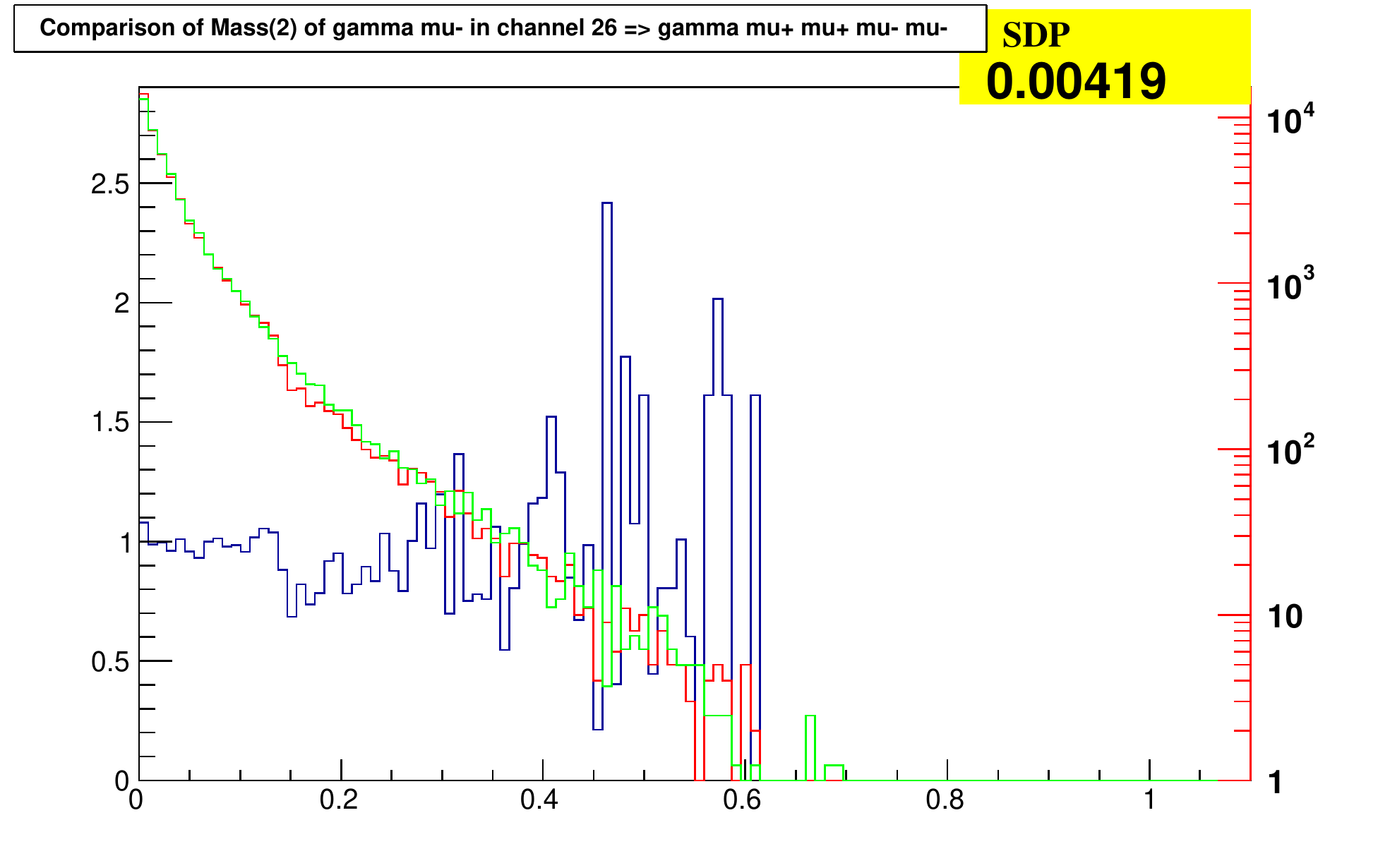}} }
{ \resizebox*{0.49\textwidth}{!}{\includegraphics{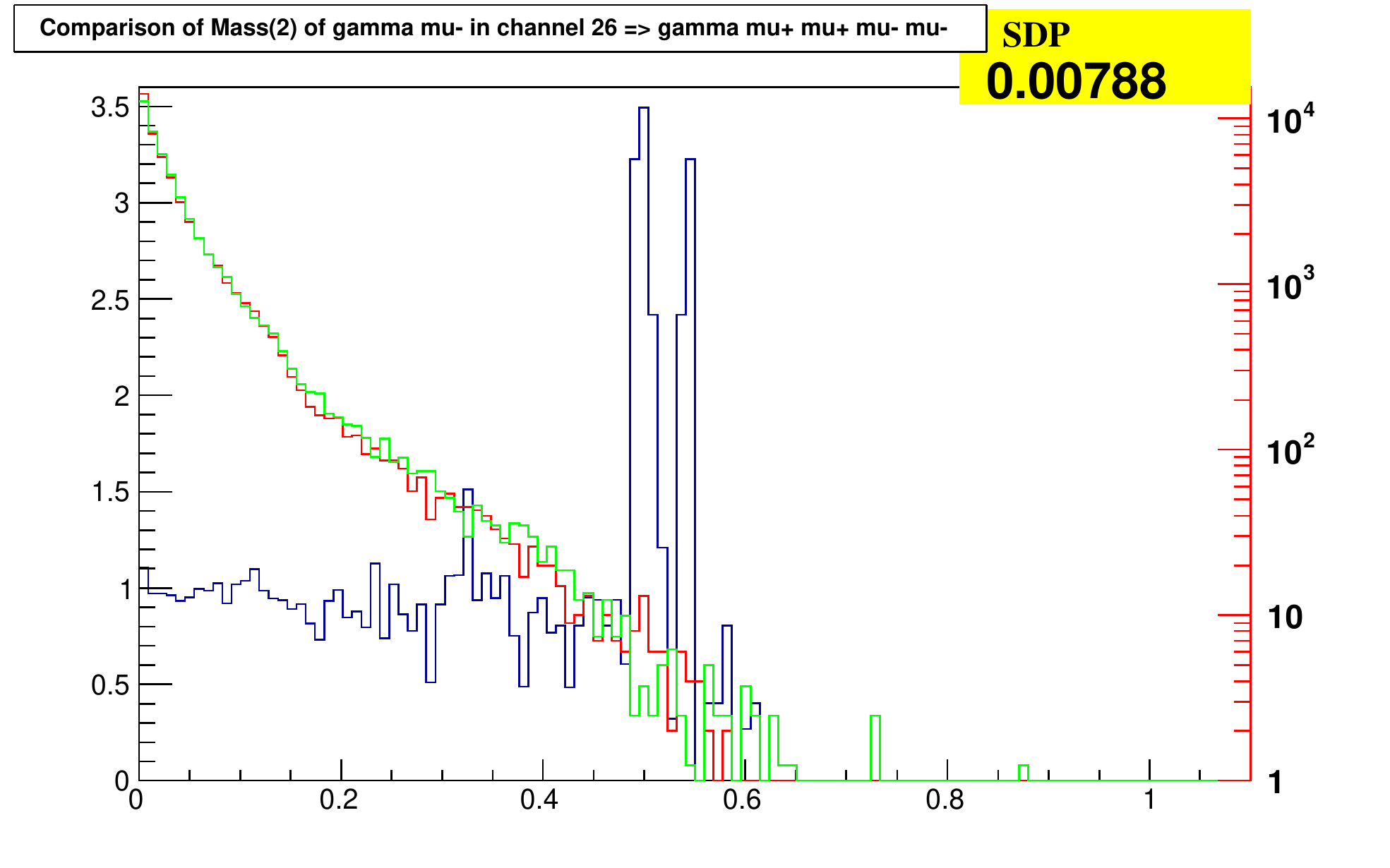}} }
{ \resizebox*{0.49\textwidth}{!}{\includegraphics{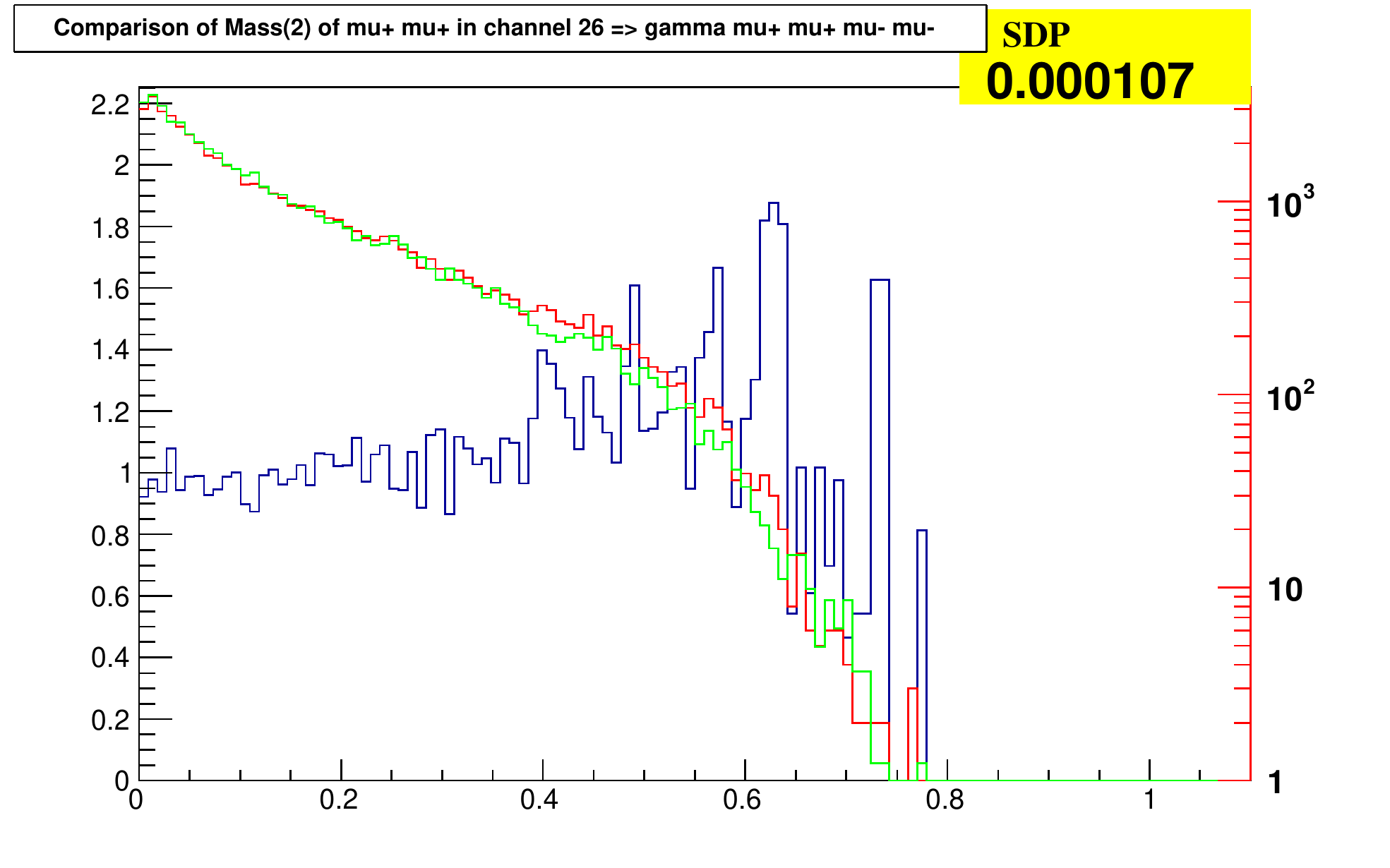}} }
{ \resizebox*{0.49\textwidth}{!}{\includegraphics{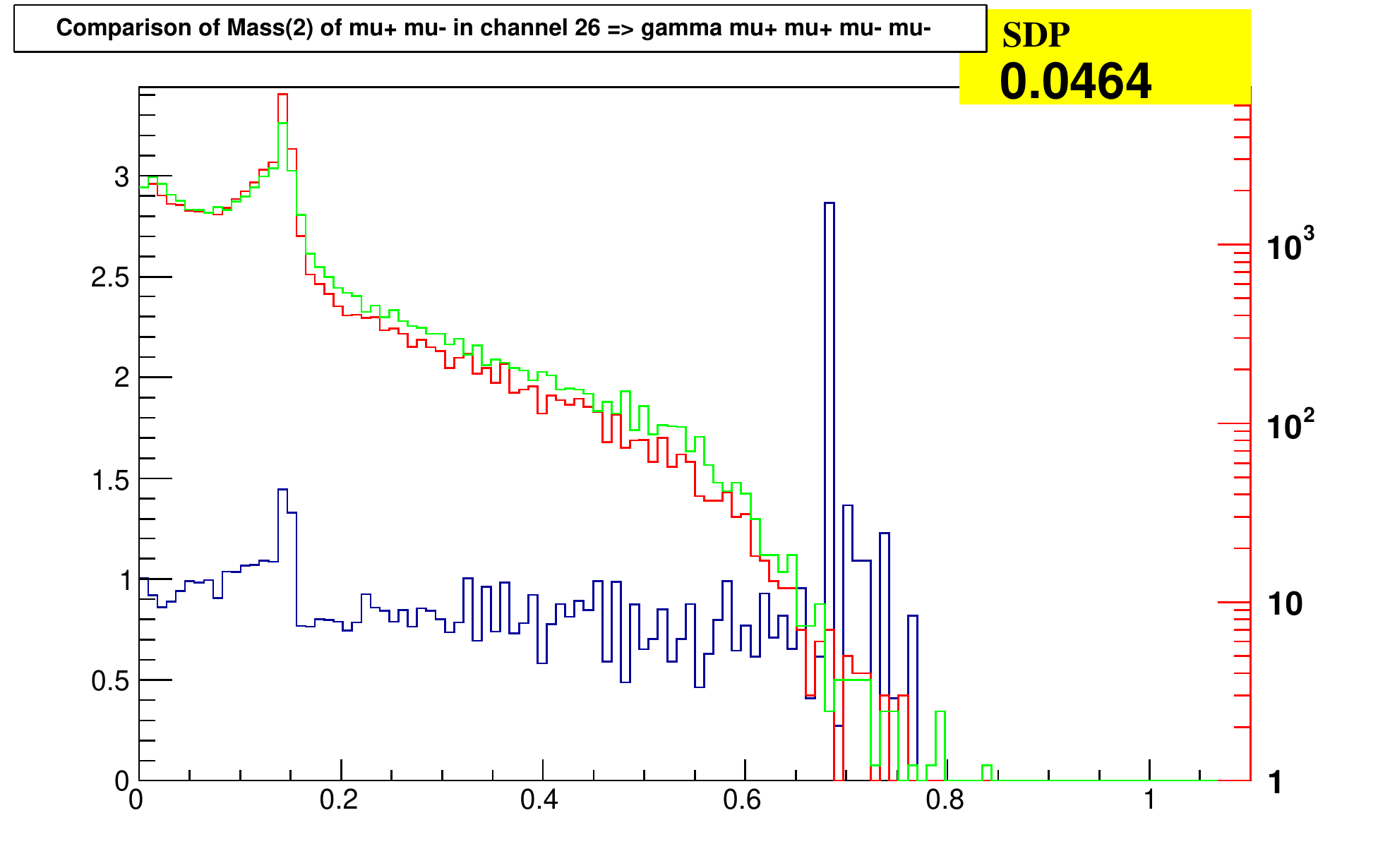}} }
{ \resizebox*{0.49\textwidth}{!}{\includegraphics{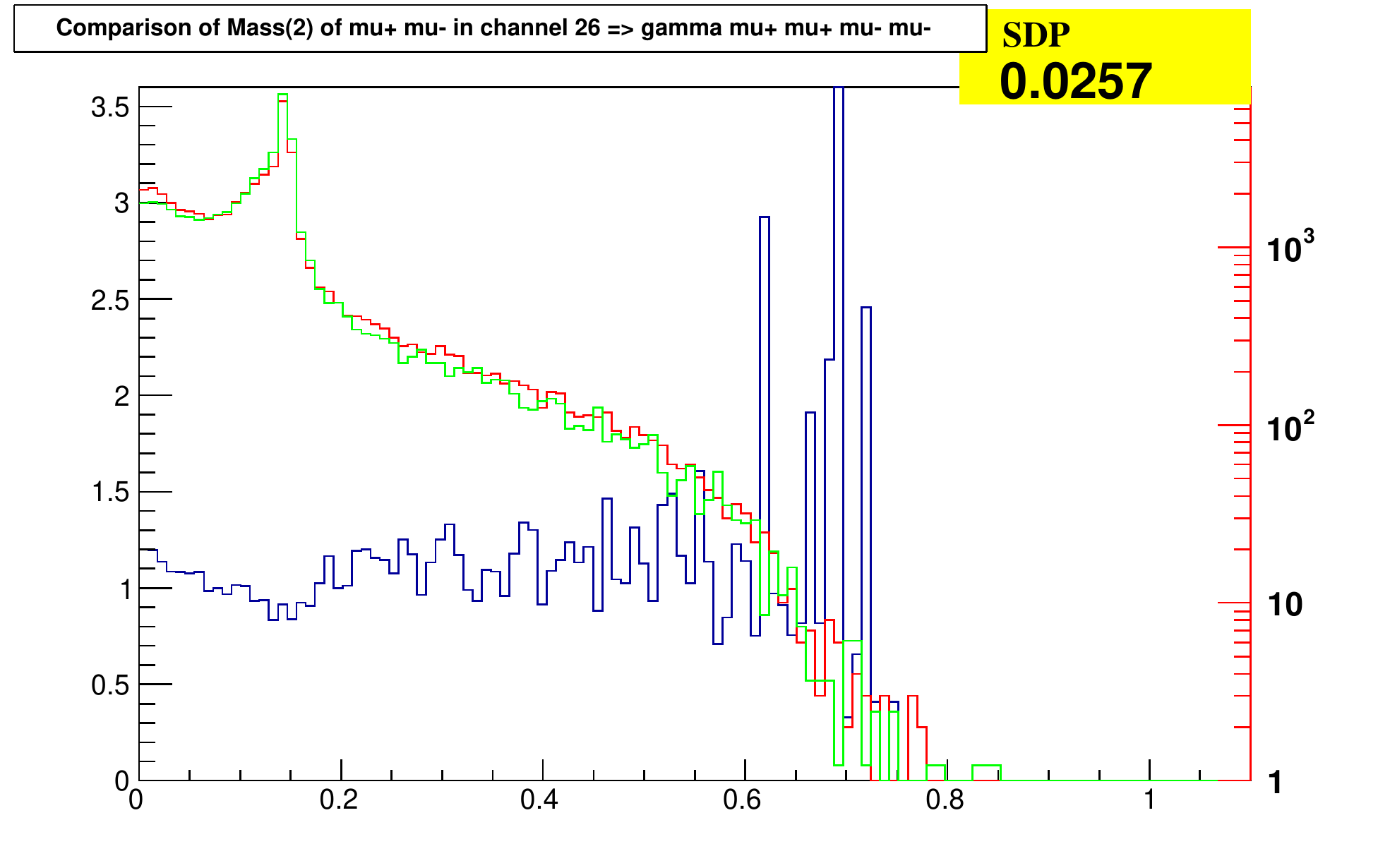}} }
{ \resizebox*{0.49\textwidth}{!}{\includegraphics{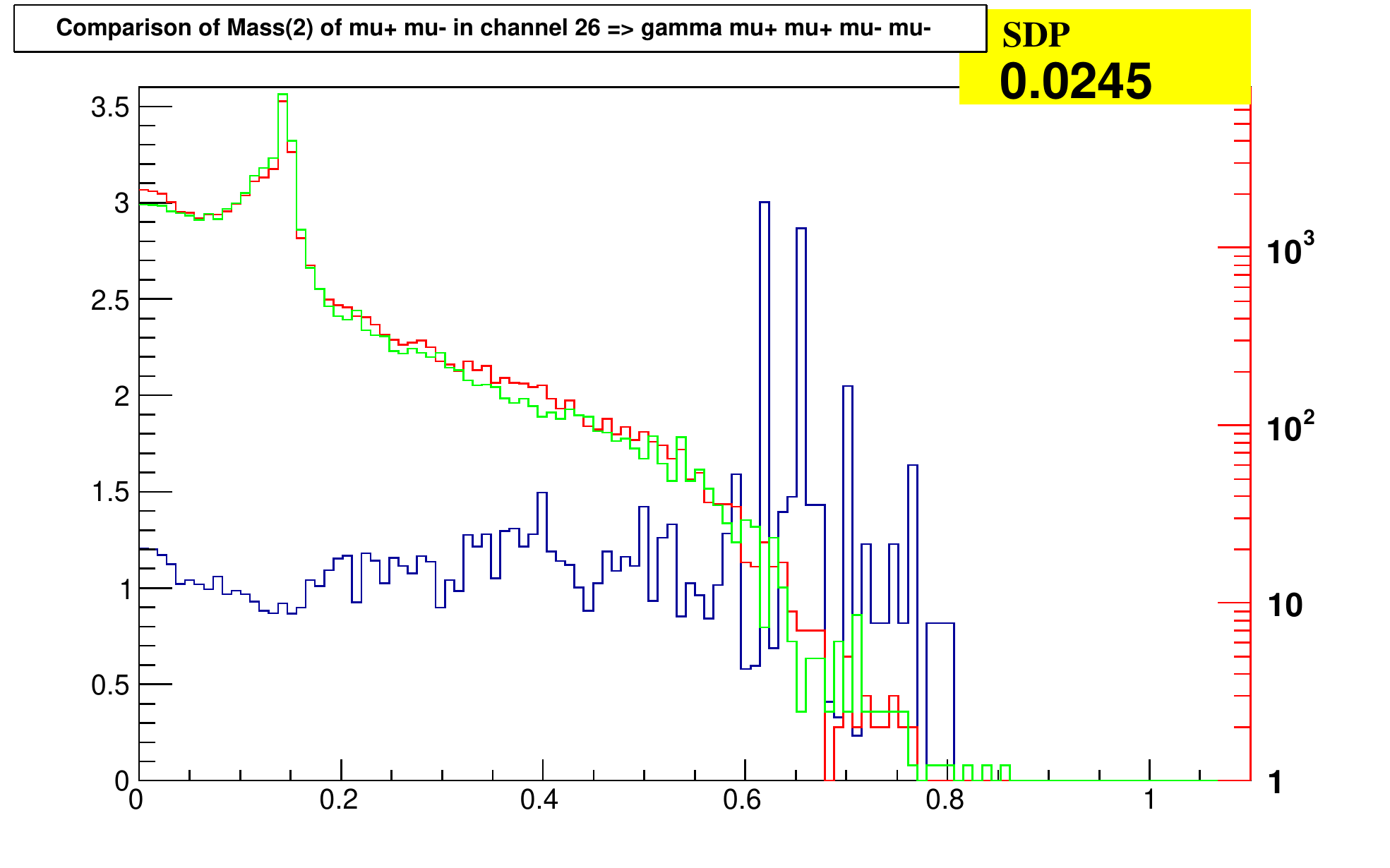}} }
{ \resizebox*{0.49\textwidth}{!}{\includegraphics{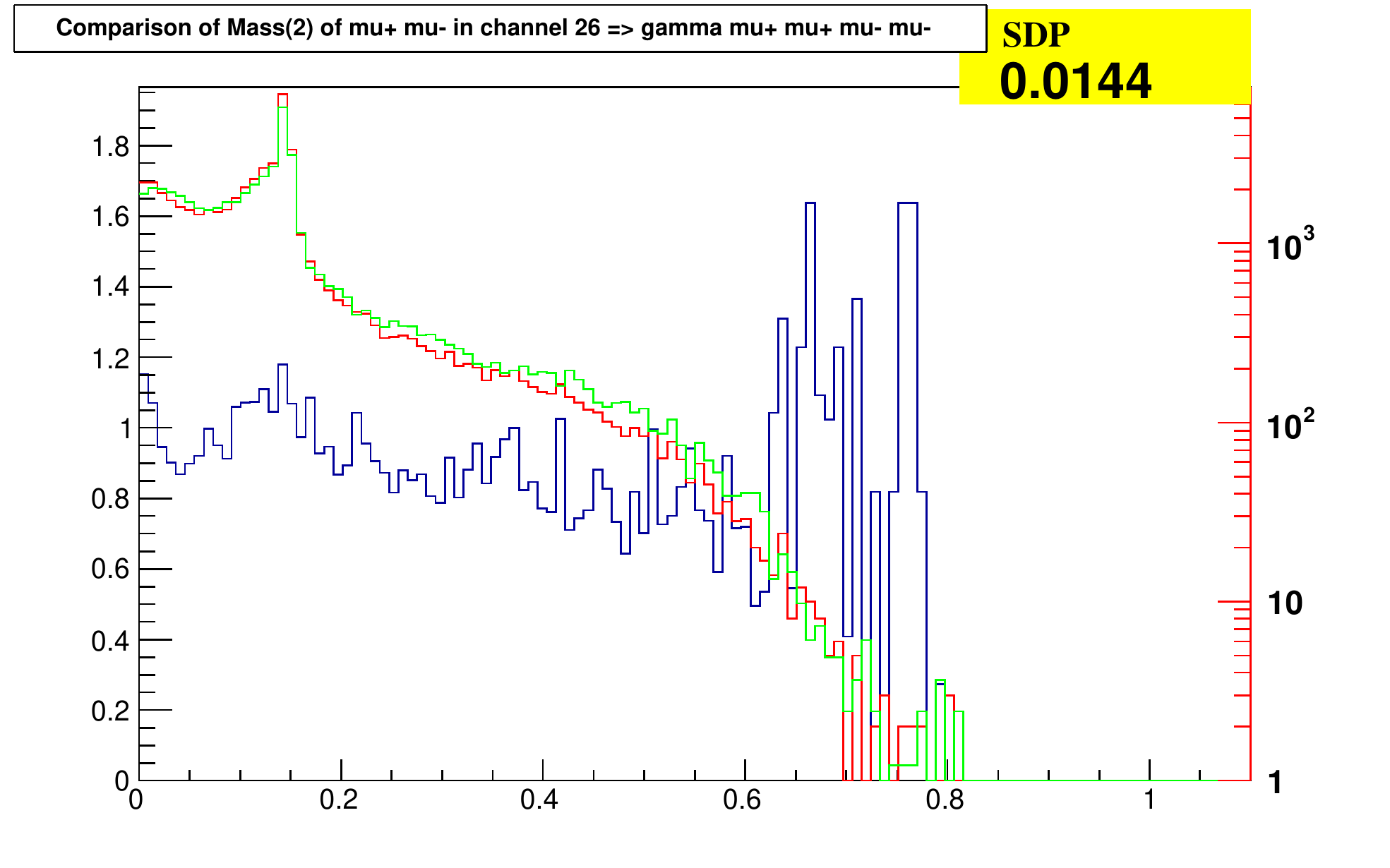}} }
{ \resizebox*{0.49\textwidth}{!}{\includegraphics{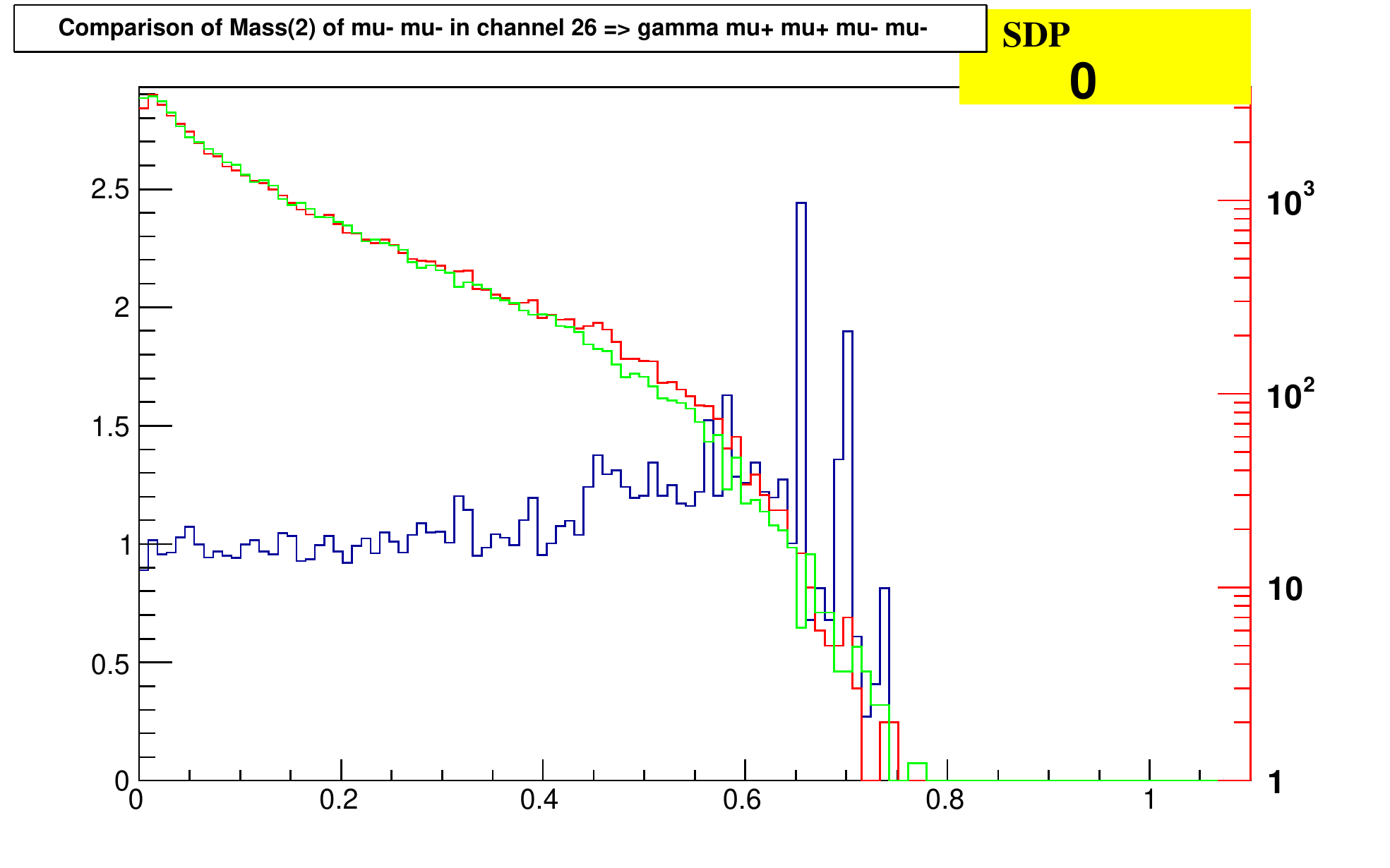}} }
{ \resizebox*{0.49\textwidth}{!}{\includegraphics{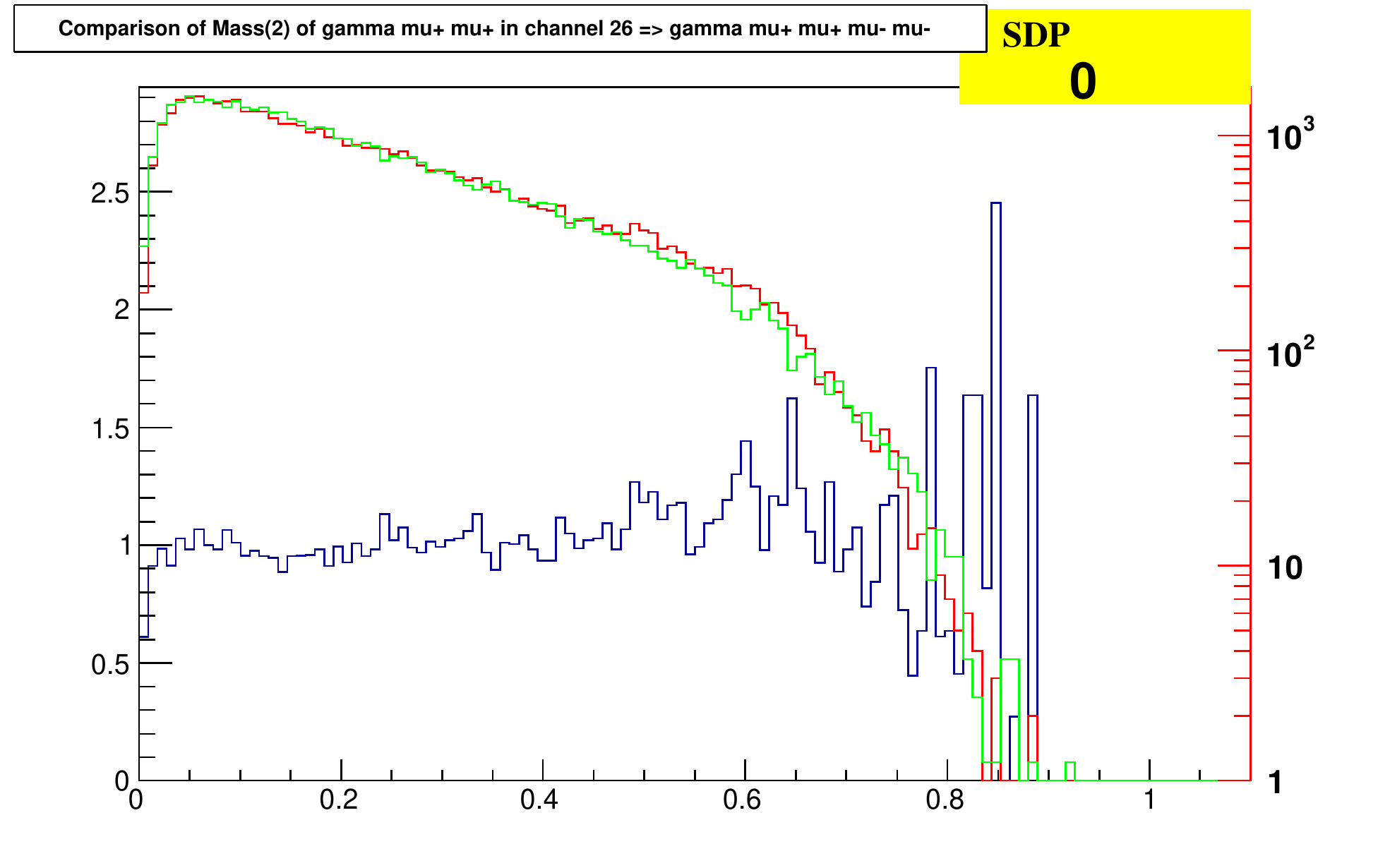}} }
{ \resizebox*{0.49\textwidth}{!}{\includegraphics{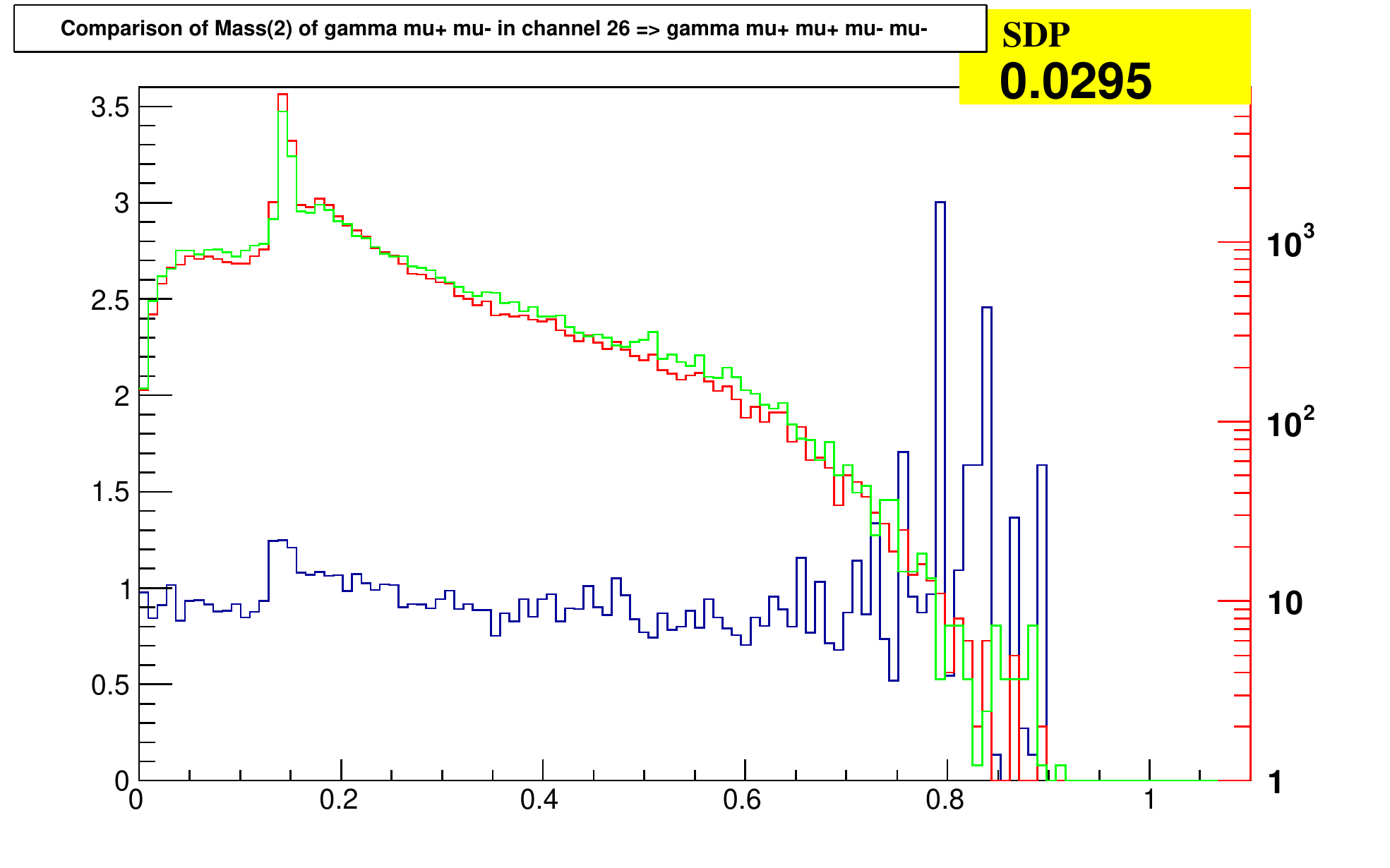}} }
{ \resizebox*{0.49\textwidth}{!}{\includegraphics{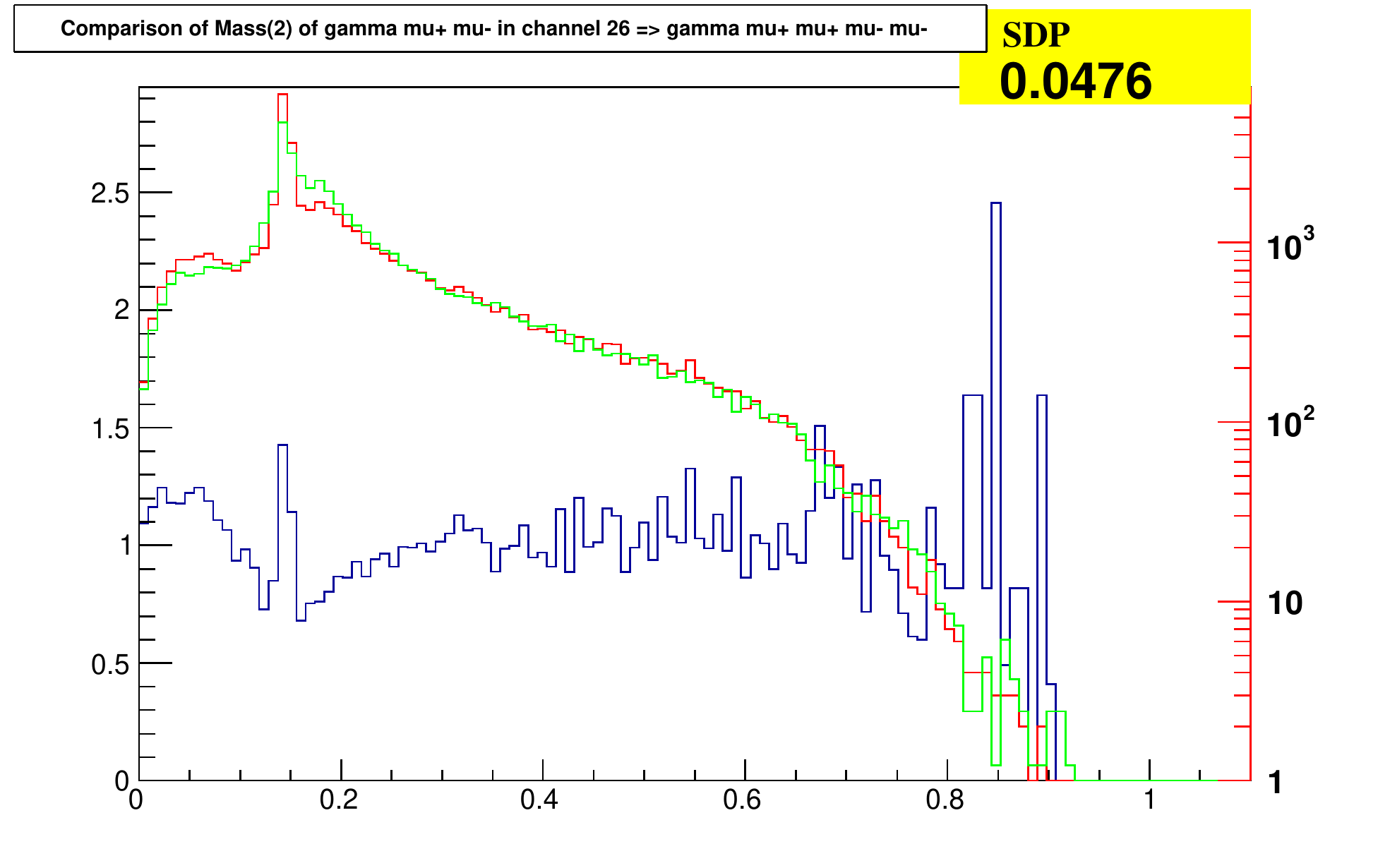}} }
{ \resizebox*{0.49\textwidth}{!}{\includegraphics{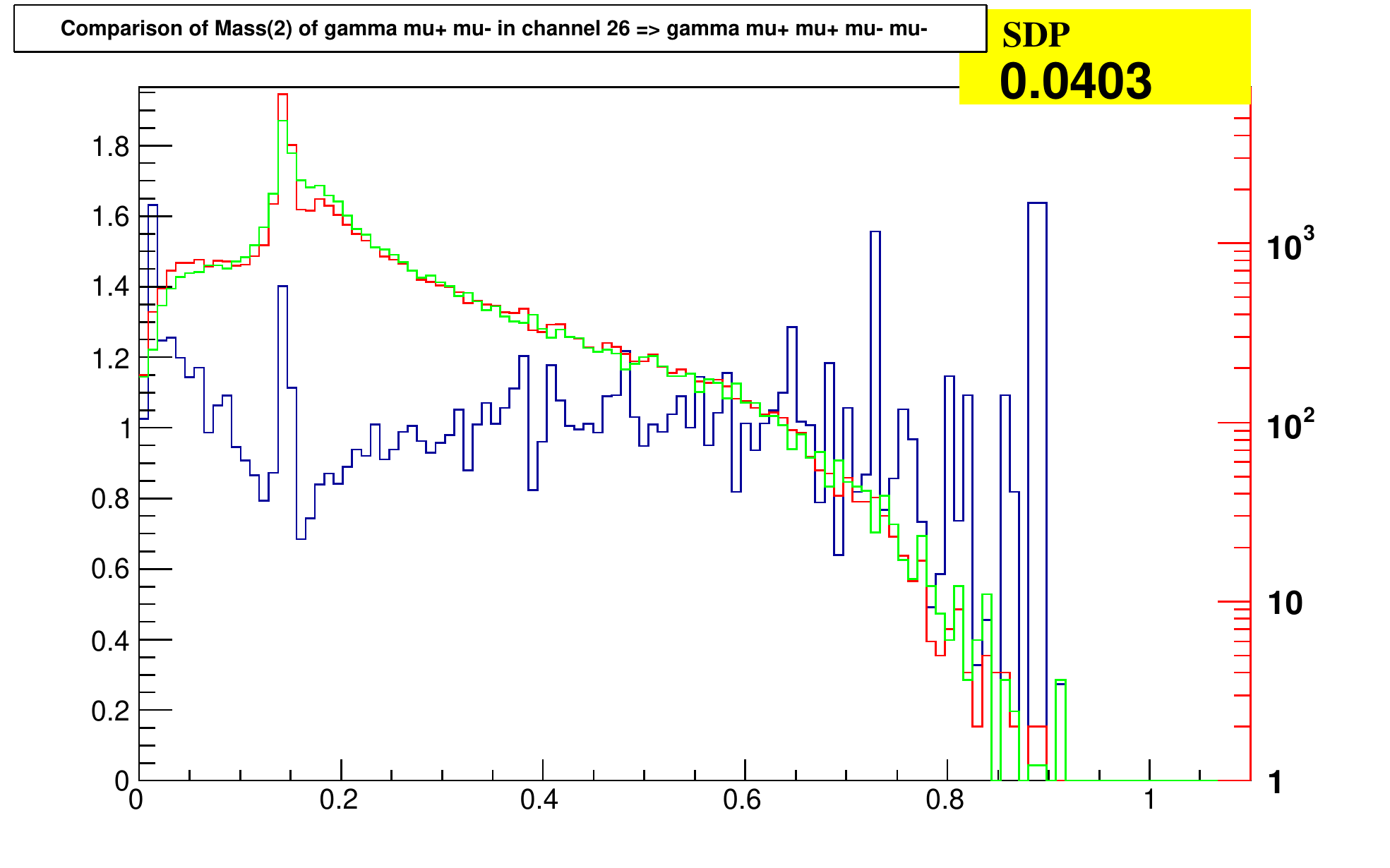}} }
{ \resizebox*{0.49\textwidth}{!}{\includegraphics{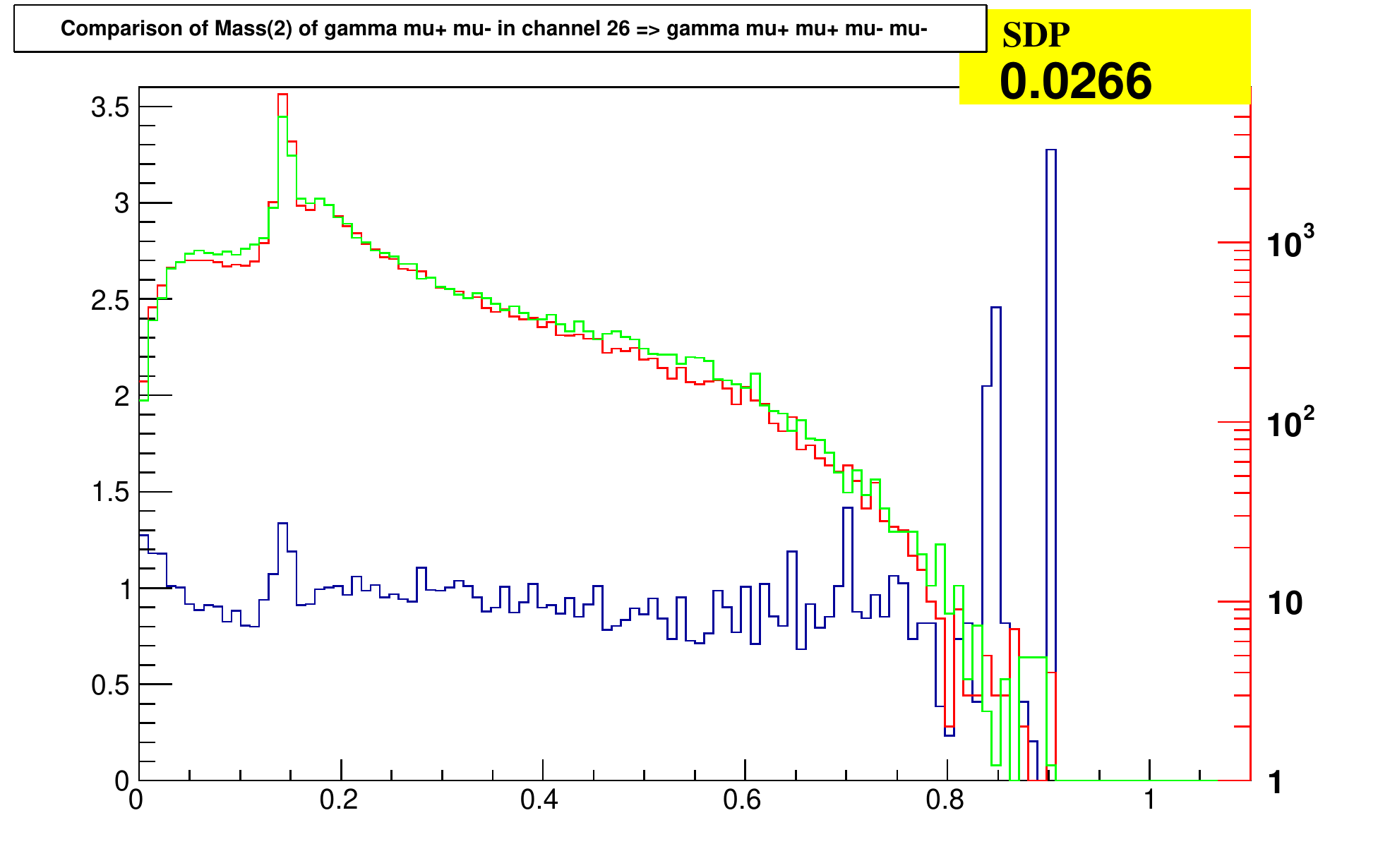}} }
{ \resizebox*{0.49\textwidth}{!}{\includegraphics{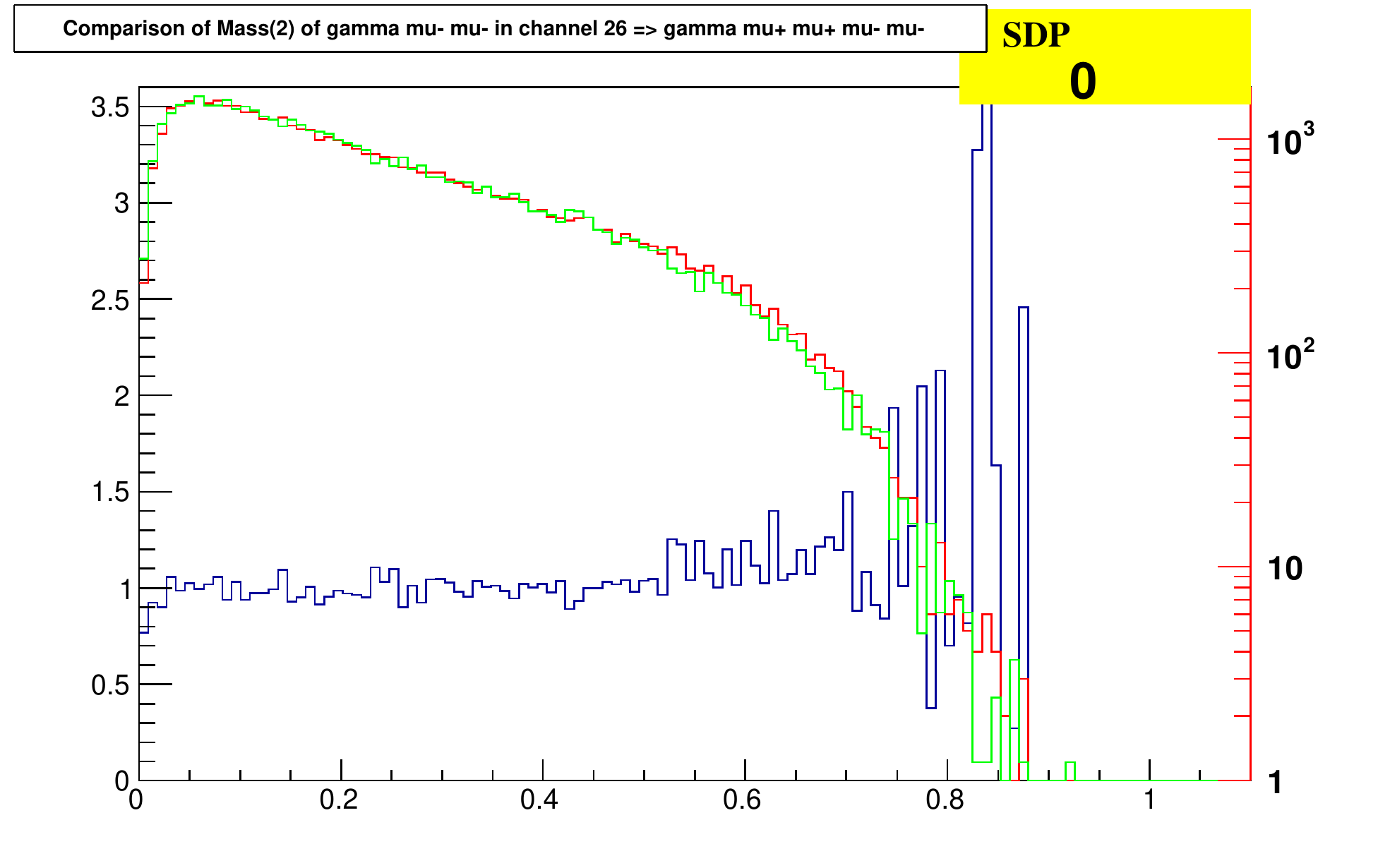}} }
{ \resizebox*{0.49\textwidth}{!}{\includegraphics{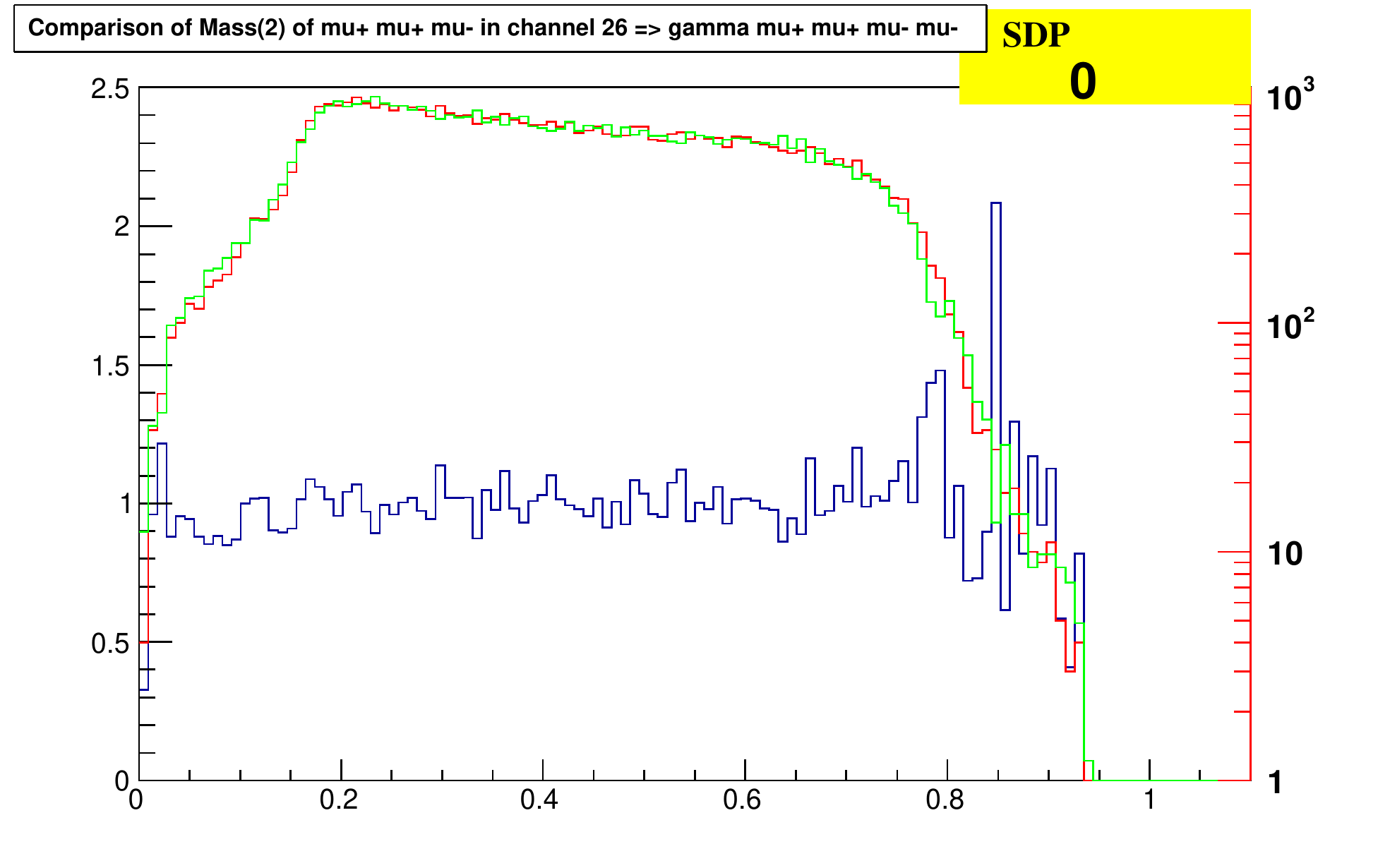}} }
{ \resizebox*{0.49\textwidth}{!}{\includegraphics{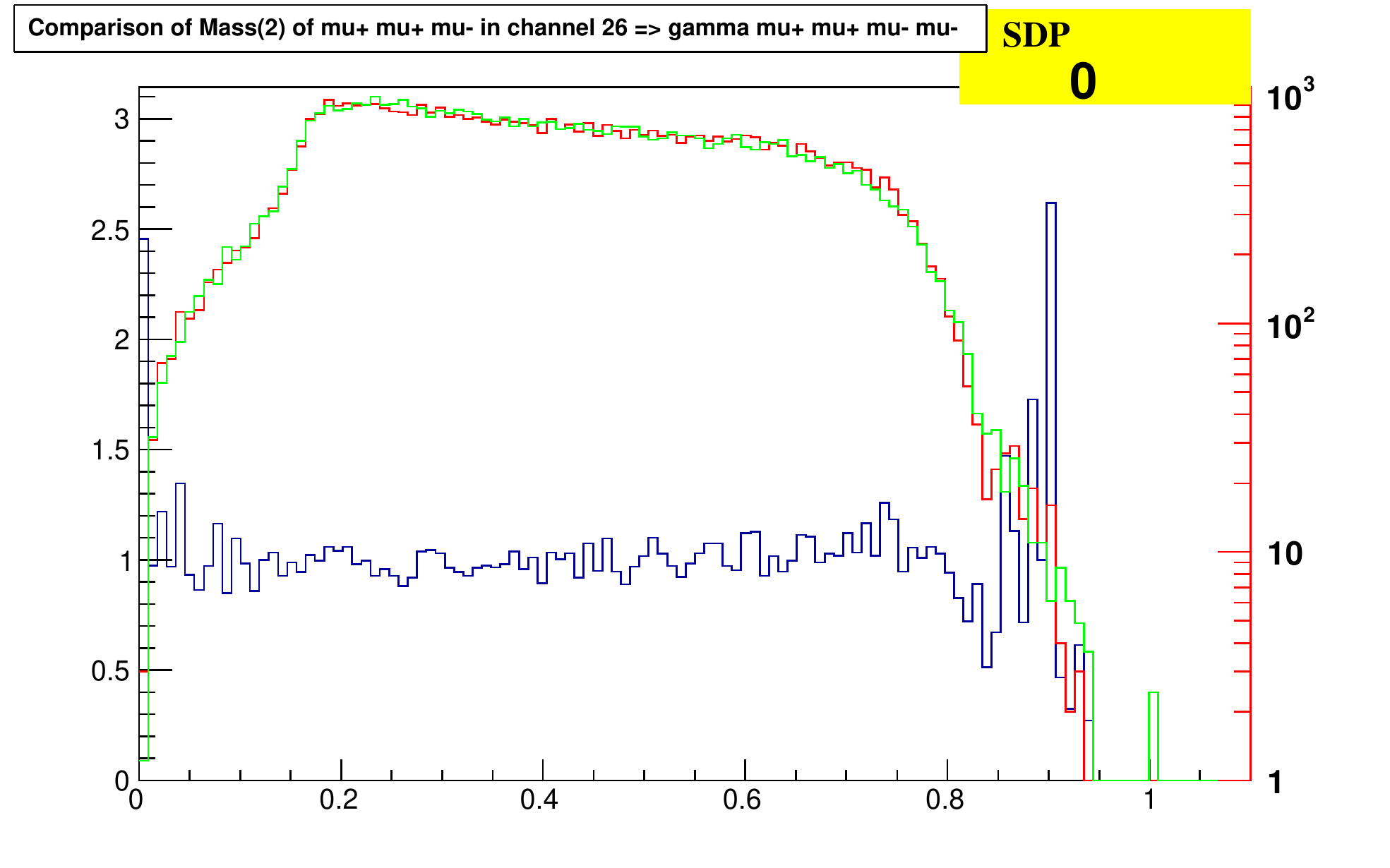}} }
{ \resizebox*{0.49\textwidth}{!}{\includegraphics{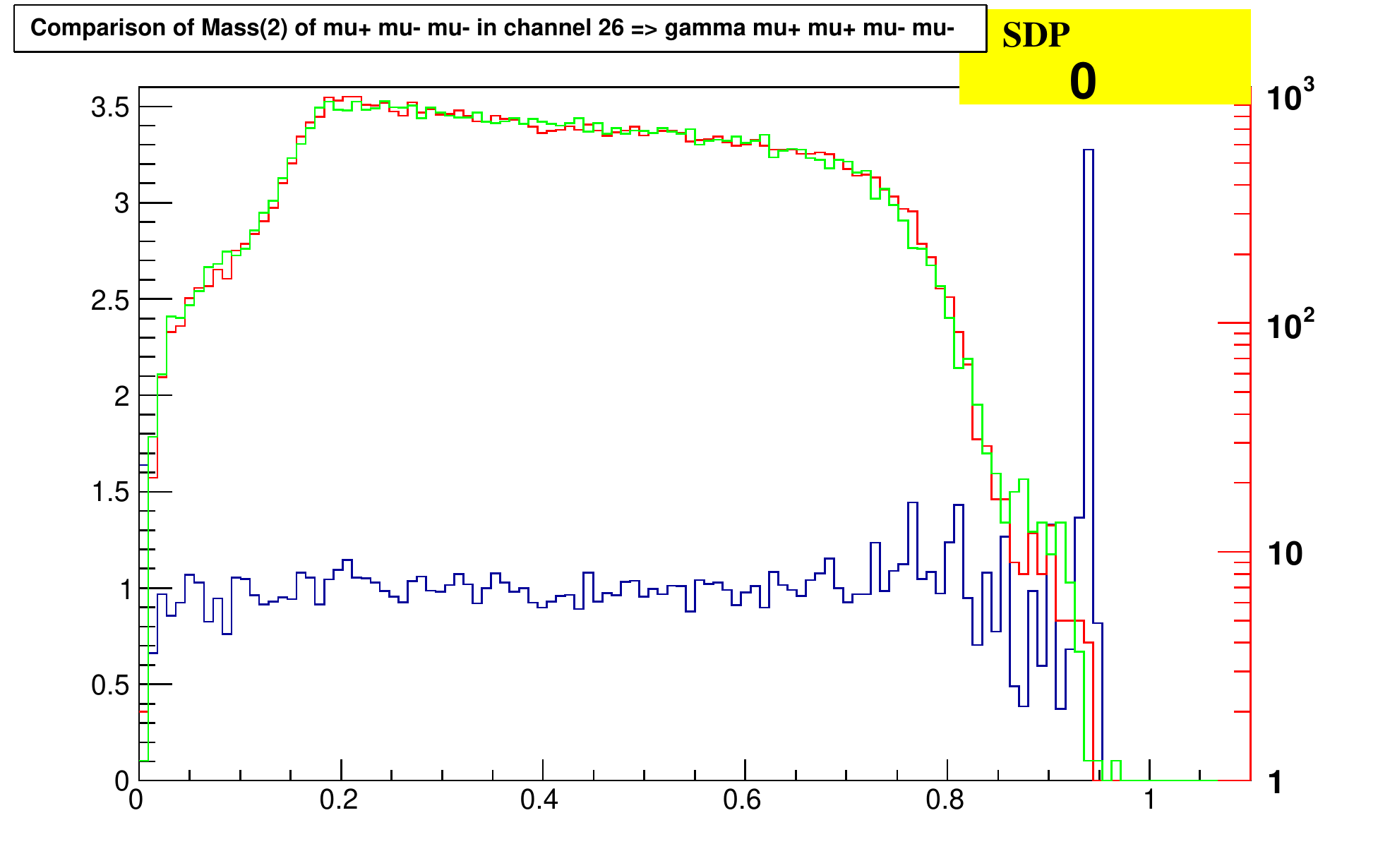}} }
{ \resizebox*{0.49\textwidth}{!}{\includegraphics{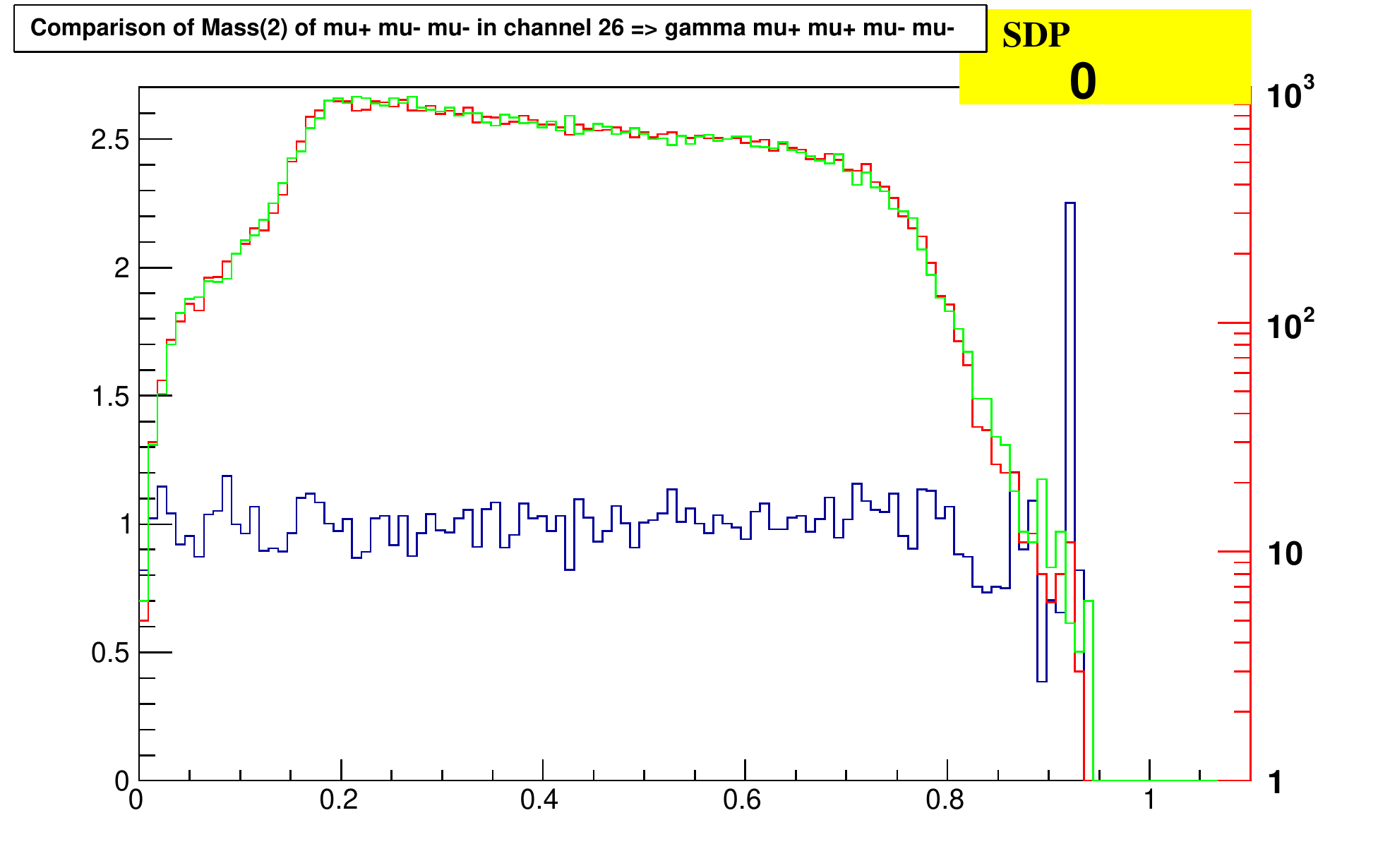}} }
{ \resizebox*{0.49\textwidth}{!}{\includegraphics{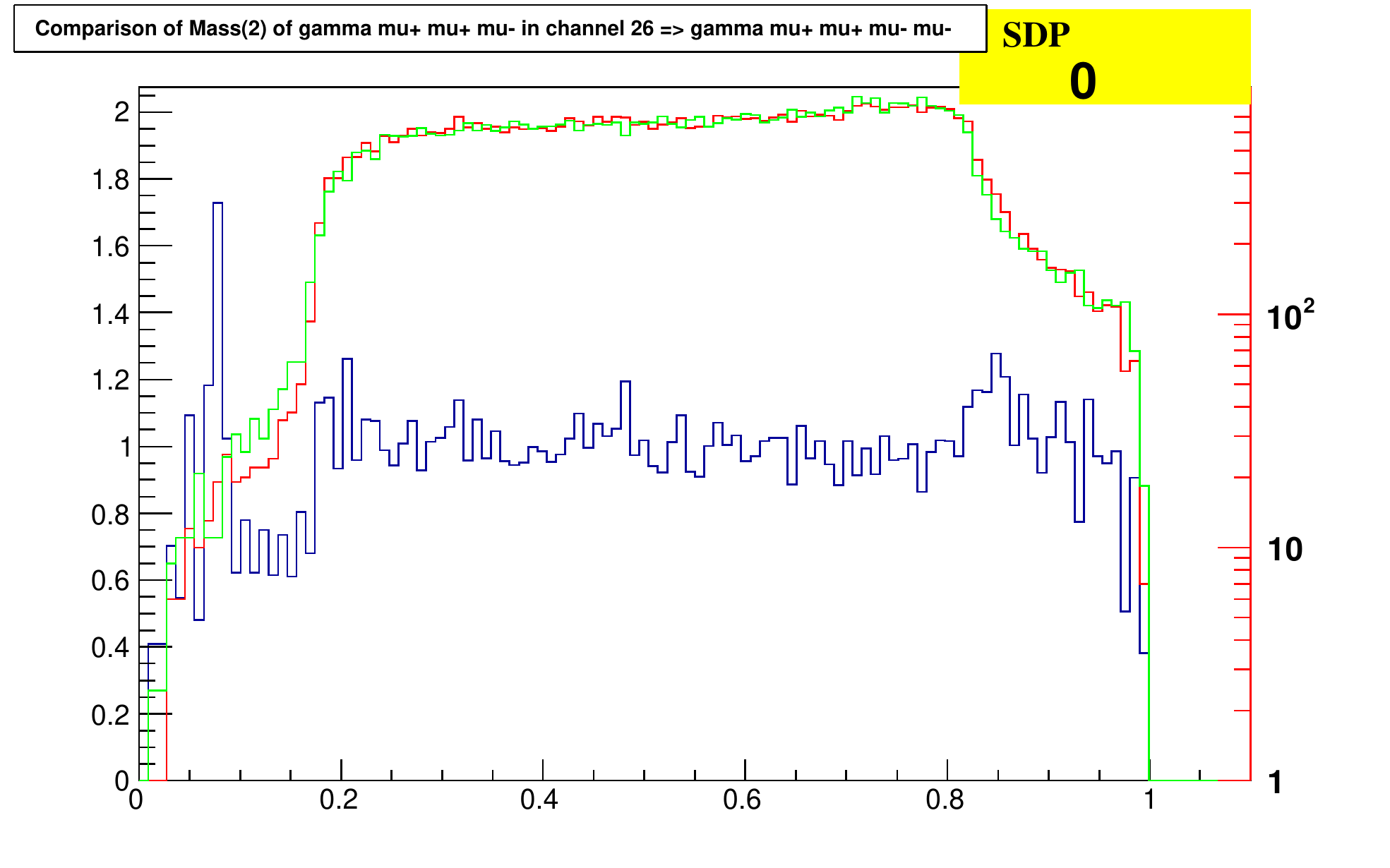}} }
{ \resizebox*{0.49\textwidth}{!}{\includegraphics{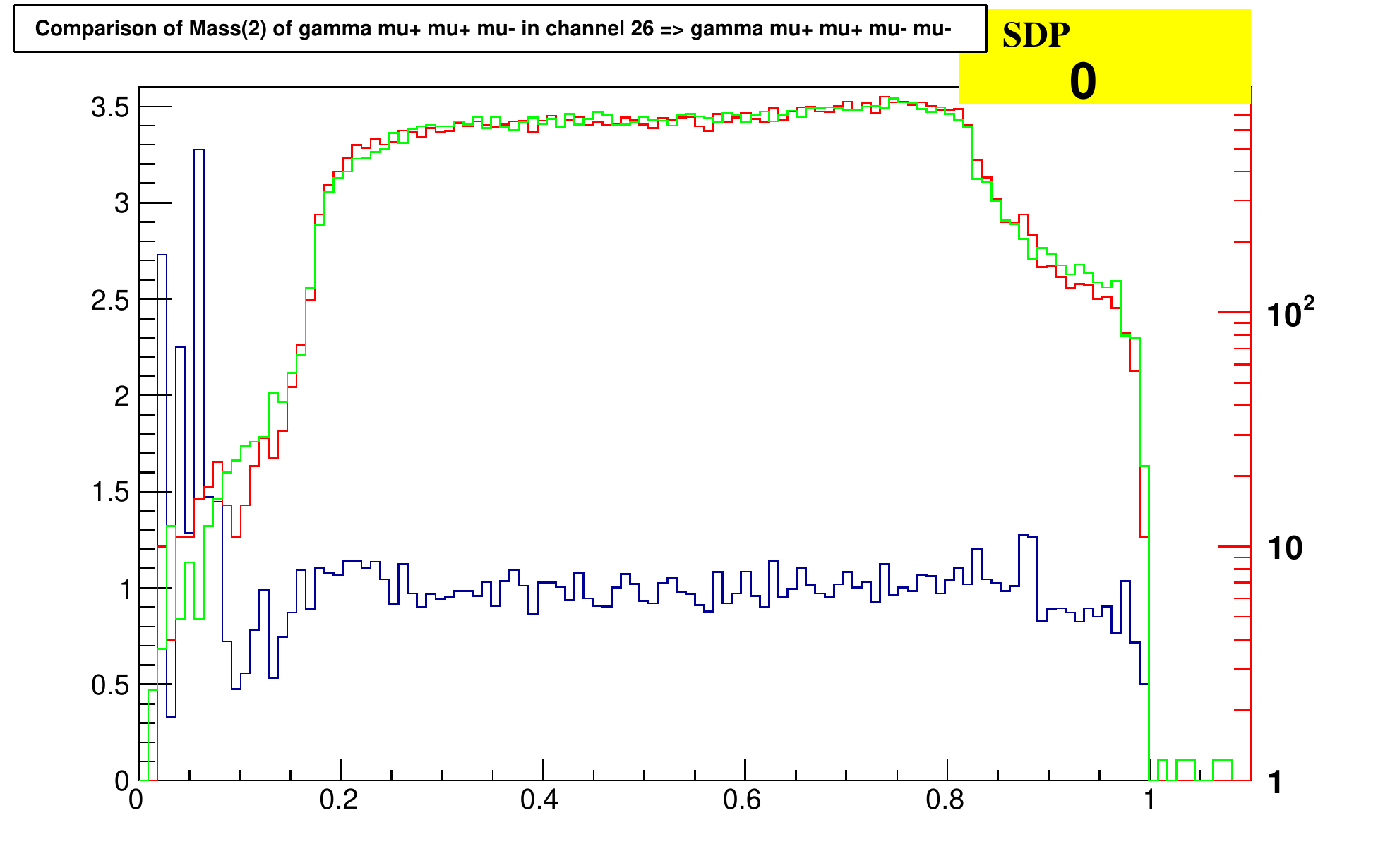}} }
{ \resizebox*{0.49\textwidth}{!}{\includegraphics{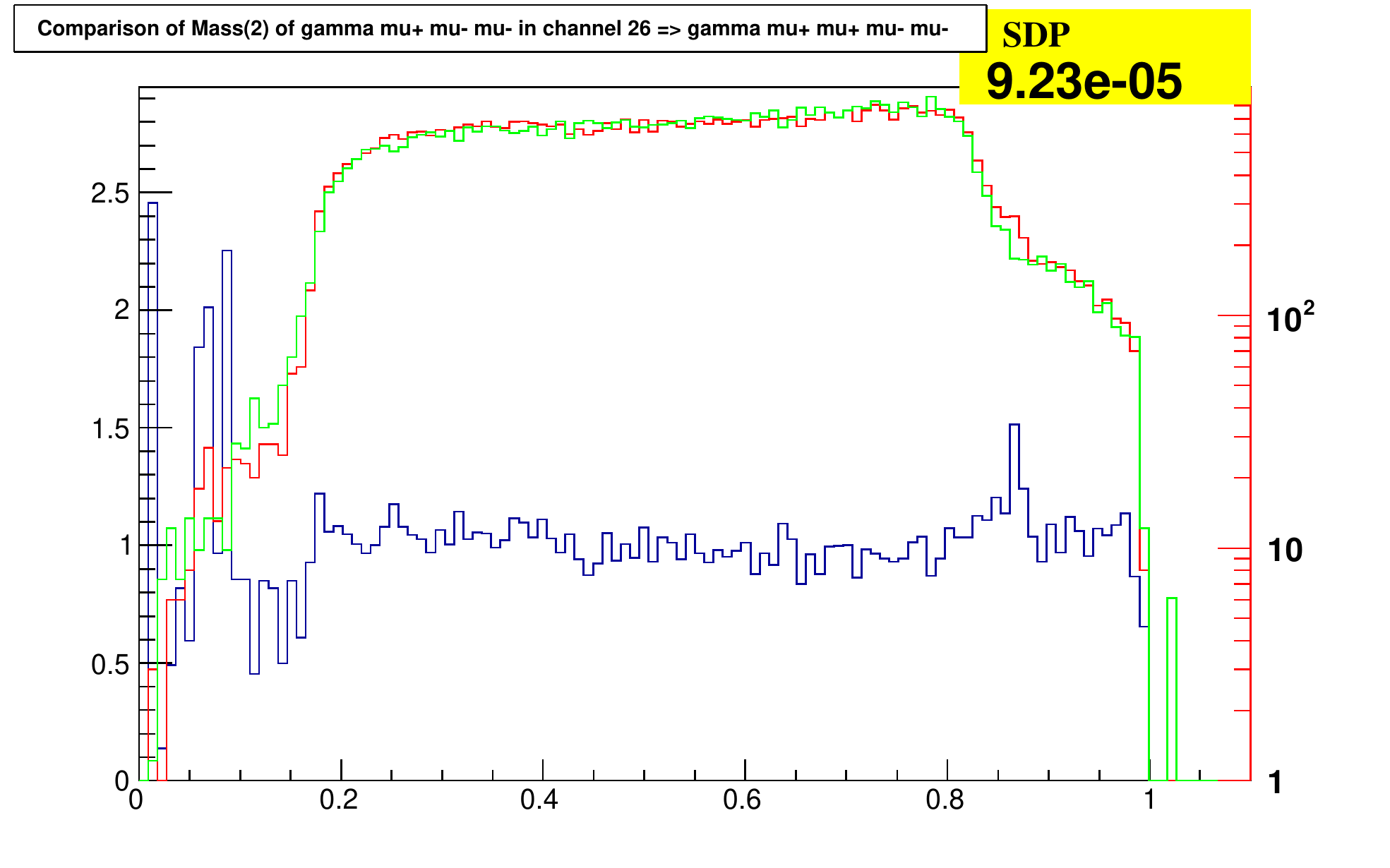}} }
{ \resizebox*{0.49\textwidth}{!}{\includegraphics{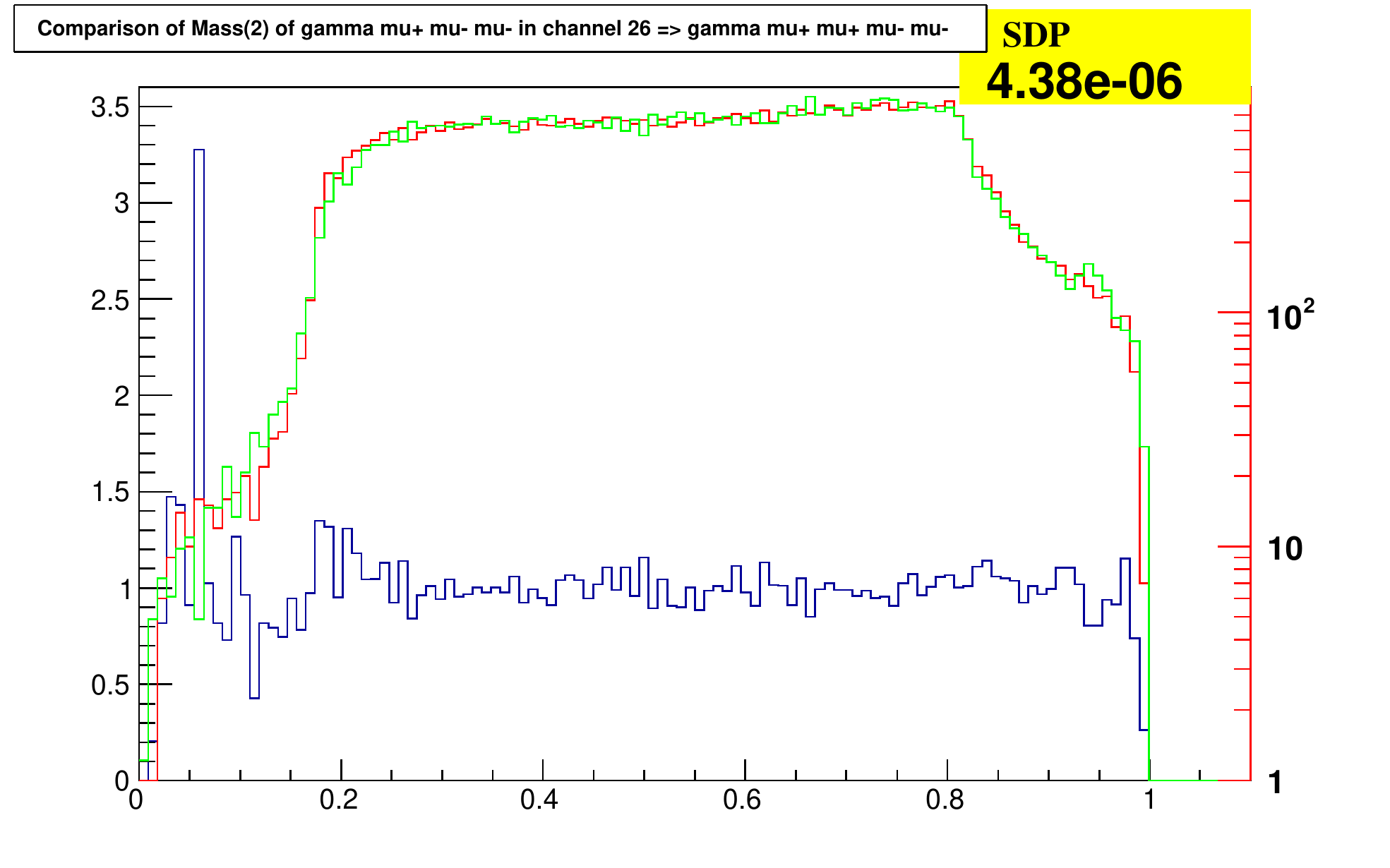}} }
{ \resizebox*{0.49\textwidth}{!}{\includegraphics{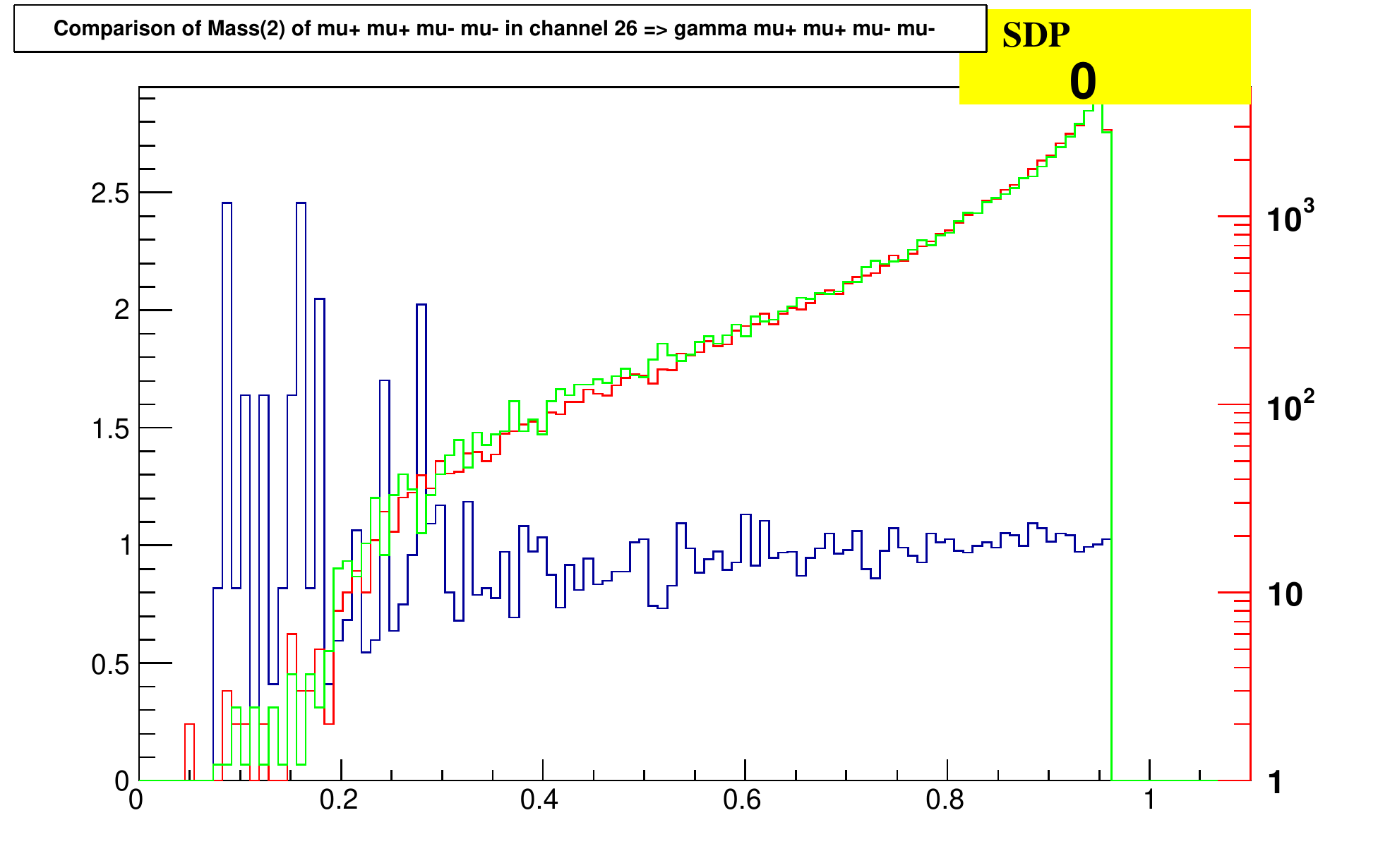}} }
{ \resizebox*{0.49\textwidth}{!}{\includegraphics{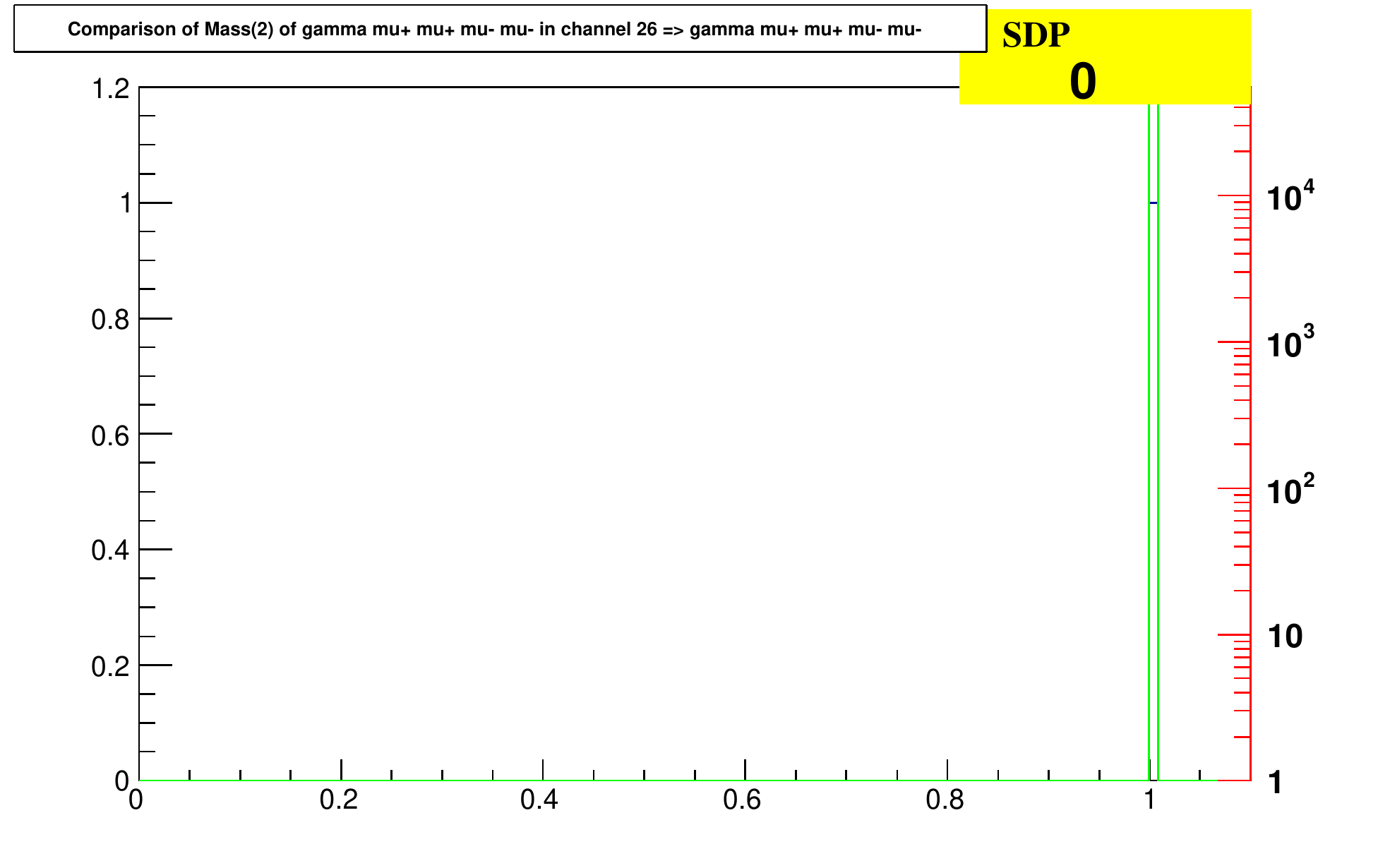}} }

%%%%%%%%%%%%%%%%%%%%%%%%%%%%%%%%%%%%%%%%%%%%%%%%%%%%%%%
%%%%%%%%%%%%%%%%%%%%%%%%%%%%%%%%%%%%%%%%%%%%%%%%%%%%%%%
\end{document}